\documentclass[11pt]{article} 

\makeatletter\renewcommand{\section}{\@startsection
	{section}{1}{\z@}{-3.5ex plus -1ex minus
		-.2ex}{2.3ex plus .2ex}{\bf }}

\makeatletter\renewcommand{\subsection}{\@startsection{subsection}{2}{\z@}{-3.25ex
    plus -1ex minus
    -.2ex}{1.5ex plus .2ex}{\it }}
\makeatletter\renewcommand{\subsubsection}{\@startsection{subsubsection}{3}{-2.45ex}{-3.25ex
    plus -1ex minus -.2ex}{1.5ex plus .2ex}{\it }}

\makeatletter \@addtoreset{equation}{section}

\usepackage{graphicx}
\usepackage{float} 
\usepackage{epsfig}
\usepackage{verbatim}
\usepackage{hyperref}
\usepackage{easyfig}
\usepackage{bbm}
\usepackage{amsmath}
\usepackage{amsfonts}
\usepackage[compact]{titlesec}
\usepackage{subdepth}
\usepackage{dsfont}
\usepackage{braket}
\usepackage{tikz}
\usepackage[utf8]{inputenc}
\usepackage[T1]{fontenc}
\usepackage{amsmath}
\usepackage[toc]{glossaries}
\usepackage{subfloat}
\usepackage{caption}
\usepackage{subcaption}
\usepackage{setspace}
\usepackage{physics}
\usepackage{amsfonts}
\usepackage{amssymb}
\usepackage{mathrsfs}
\usepackage{amsmath,amssymb}
\usepackage{bm}
\usepackage{caption}
\usepackage{amsmath}
\usepackage{graphicx}
\usepackage{cite}
\usepackage{pdfpages}
\usepackage{caption}
\usepackage{subcaption}
\usepackage{bbm}
\usepackage{braket}
\usepackage{slashed}
\usepackage{bm}

\graphicspath{{figures_new/}}

\renewenvironment{thebibliography}[1]
{\baselineskip=16pt plus 2pt minus 1pt
	\section*{\large\refname
		\@mkboth{\MakeUppercase\refname}{\MakeUppercase\refname}}%
	\list{\@biblabel{\@arabic\c@enumiv}}%
	{\settowidth\labelwidth{\@biblabel{#1}}%
		\leftmargin\labelwidth
		\advance\leftmargin\labelsep
		\@openbib@code
		\usecounter{enumiv}%
		\let\p@enumiv\@empty
		\renewcommand\theenumiv{\@arabic\c@enumiv}}%
	\sloppy
	\clubpenalty4000
	\@clubpenalty \clubpenalty
	\widowpenalty4000%
	\sfcode`\.\@m}

\let\fn\footnote
\renewcommand{\footnote}[1]{\linespread{1.1}\fn{#1}\linespread{1.29}}

\newcommand\scalemath[2]{\scalebox{#1}{\mbox{\ensuremath{\displaystyle #2}}}}

\def\tyng(#1){\hbox{\tiny$\yng(#1)$}}

\newcommand{\be}{\begin{equation}}
	\newcommand{\ee}{\end{equation}}
\newcommand{\bea}{\begin{array}}
	\newcommand{\ea}{\end{array}}
\newcommand{\beqa}{\begin{eqnarray}}
	\newcommand{\eeqa}{\end{eqnarray}}
\newcommand{\nn}{\nonumber}

\DeclareCaptionSubType*{figure}

\numberwithin{equation}{section}

\begin{document}
	\fontfamily{bch}\fontsize{11pt}{15pt}\selectfont
	\begin{titlepage}
		\begin{flushright}
		\end{flushright}
		\vskip 3 em
		\begin{center}
			{\Large Non-Abelian Magnetic Field and Curvature Effects on Pair Production}\\
			~\\
\centerline{$ \text{\large{\bf{S. K\"{u}rk\c{c}\"{u}o\u{g}lu}}}^{\dagger} \,, \, $ $ \text{\large{\bf{B. \"{O}zcan}}}{^\dagger} \,,\,$ $\text{\large{\bf{G. \"{U}nal}}}^{*}$}
\vskip 0.5cm
\centerline{\sl $^\dagger$ Middle East Technical University, Department of Physics,}
\centerline{\sl Dumlupınar Boulevard, 06800, Ankara, Turkey}
\centerline{\sl $^*$ Faculty of Engineering, Başkent University, 06790 Ankara, Turkey}

\centerline{\sl}
\centerline{\sl }
\begin{tabular}{r l}
	E-mails:
	&\!\!\!{\fontfamily{cmtt}\fontsize{11pt}{15pt}\selectfont kseckin@metu.edu.tr  }\\
	&\!\!\!{\fontfamily{cmtt}\fontsize{11pt}{15pt}\selectfont berk@metu.edu.tr} \\
	&\!\!\!{\fontfamily{cmtt}\fontsize{11pt}{15pt}\selectfont gonulunal23@gmail.com}
\end{tabular}
		\end{center}
\begin{quote}
  \begin{center}
     {\bf Abstract}
  \end{center}
We calculate the Schwinger pair production rates  in $\mathbb{R}^{3,1}$ as well as in the positively curved space $S^2 \times \mathbb{R}^{1,1}$ for both spin-$0$ and spin-$\frac{1}{2}$ particles under the influence of an external $SU(2) \times U(1)$ gauge field producing an additional uniform non-abelian magnetic field besides the usual uniform abelian electric field. To this end, we determine and subsequently make use of the spectrum of the gauged Laplace and Dirac operators on both the flat and the curved geometries. We find that there are regimes in which the purely non-abelian and the abelian parts of the gauge field strength have either a counterplaying or reinforcing role, whose overall effect may be to enhance or suppress the pair production rates. Positive curvature tends to enhance the latter for spin-$0$ and suppress it for spin-$\frac{1}{2}$ fields, while the details of the couplings to the purely abelian and the non-abelian parts of the magnetic field, which are extracted from the spectrum of the Laplace and Dirac operators on $S^2$, determine the cumulative effect on the pair production rates. These features are elaborated in detail.

\vskip 1em
\end{quote}
\end{titlepage}
\setcounter{footnote}{0}
\pagestyle{plain} \setcounter{page}{2}
\newpage

\section{Introduction and Summary of Results}
\label{sec:1}
Elucidating the non-perturbative effects in quantum field theories is an important area of research on which considerable contemporary efforts are focused. Problems in this context include a variety of phenomena some of which are both conceptually and computationally very difficult and usually escape a fully satisfactory theoretical description, such as the confinement of quarks in QCD. Nevertheless, there are also more tractable problems such as the Casimir effect \cite{Casimir1948} and the Schwinger pair production \cite{Schwinger}. Although, both of these phenomena are conceived as vacuum effects in QED, the former has already been verified in several different direct experiments over the past few decades, while it has not been possible to observe the latter as the amplitude for the production of massive particle and anti-particle pairs is exponentially suppressed and requires a very strong electric field $ \approx 10^{18} \, V \, m^{-1}$, while it is expected that the recent technological advances can open new avenues for exploring QED in intense background fields \cite{Ringwald2001, Fedotov:2022ely}.

There have been continual efforts ever since the original work of Schwinger \cite{Schwinger} (and the important earlier works due to Sauter \cite{Sauter1931} and Euler and Heisenberg \cite{HeisenbergEuler1936}) to look for alternative mechanisms as well as novel aspects and features that may enhance the effect and thereby reduce the electric field strength required for its observation. Time-dependent electric fields tend to enhance the pair production amplitudes at a significant rate \cite{Brezin, Dunne1998,Hebenstreit2008}, nevertheless experiments conducted using high intensity laser beams has not, so far, led to an observational confirmation (see, for instance, \cite{Fedotov:2022ely} and the references therein.). Effect of inhomogenous electric fields are also address in several papers \cite{Dunne2005, Gies2005, Schtzhold2008} (\cite{Hattori:2023egw} provides a comprehensive review). Very recently, it has been reported that a condensed matter analogue involving electrons and holes is observed at the Dirac point of graphene superlattices \cite{Superlattices} and ballistic graphene transistors \cite{Graphene}.

In two recent papers \cite{Kurkcuoglu, Kurkcuoglu2}, in which one of the present authors (S.K.) is a coauthor, possible influences of additional uniform abelian magnetic field as well as constant positive and negative curvature are considered by computing the pair production amplitudes in the Minkowski space and the product manifolds $S^2 \times \mathbb{R}^{1,1}$ and $H^2 \times \mathbb{R}^{1,1}$, and the results indicated a decrease for spin-$0$ and increase for spin-$\frac{1}{2}$ for the pair production amplitudes with the applied magnetic field, while positive curvature enhanced the effect for spin-$0$ and suppressed it for spin-$\frac{1}{2}$, and the opposite prevailed for the negative curvature. For, spin-$1$ particles, which were treated as a part of an $SU(2)$ gauge field, it was found that the pair production rate increases with the applied magnetic field, while the positive (negative) curvature acts to suppress (enhance) the effect \cite{Kurkcuoglu2}.

In the present work, we explore the effects of additional uniform non-abelian $SU(2) \times U(1)$ magnetic gauge field backgrounds\footnote{We may note that this configuration is distinct from that treated in \cite{Brown1979}, where a component of the uniform $SU(2)$ field strength could be of electric type and could cause the decay of the vacuum by pair production.}. on the pair production of spin-$0$ and spin-$\frac{1}{2}$ fields on flat as well as positively curved backgrounds, i.e. in the Minkowski space $\mathbb{R}^{3,1}$ and the product manifold $S^2 \times \mathbb{R}^{1,1}$, where $U(1)$ provides the usual uniform magnetic field in both cases. We assume that both the scalar and spinor fields are also charged under the $SU(2)$ part of the gauge field background and to distinguish this from the usual spin, we refer to it as the "isospin" degree of freedom alluding to the similar uses of the latter terminology in the literature \cite{Hasenfratz1976, Jackiw1976, Balachandran:2003ay}. To compute the pair production amplitudes, we make use of the spectrum of the gauged Laplacians on ${\mathbb R}^2$ and $S^2$, which are already available in \cite{Estienne_2011}, while we derive those for the square of the Dirac operators for both of the geometries. The latter are interesting problems in their own right with several intriguing and novel feature,which includes a detailed account of their zero-modes and we provide a comprehensive analysis. A brief summary of our results is as follows. Due to the isospin degree of freedom, the energy spectrum splits into two branches namely,$\Lambda_{n_1}^+(\beta^\prime)$ and $\Lambda_{n_1}^-(\beta^\prime)$ for the scalar and $\lambda_{n_1}^+(\beta^\prime)$ and $\lambda_{n_1}^-(\beta^\prime)$ spinor fields\footnote{Let us remark that the corresponding wave functions for $\Lambda_{n_1}^\pm(\beta^\prime)$ and $\lambda_{n_1}^\pm(\beta^\prime)$ are not the eigenfunctions of the isospin as explained in the appendices \ref{sec:A} and \ref{sec:B}.}. In the flat geometry, similar to the case of the pure uniform abelian magnetic field \cite{Kurkcuoglu}, pair production is in general suppressed for spin-$0$ and enhanced for spin-$\frac{1}{2}$ fields with increasing non-abelian magnetic field $\beta^\prime$ (This is the non-abelian field strength $\beta$ scaled by the square root of the abelian field $B_1$, i.e. $\beta^\prime = \frac{\beta}{\sqrt{B_1}}$ as will be defined in the next section.) These outcomes are mainly being due to the $\Lambda_{n_1}^+(\beta^\prime)$ monotonically increasing with $\beta^\prime$, which makes the corresponding states harder to fill by the produced particle anti-particle pairs due to higher energy cost in the former (spin-$0$), and due to the proliferation of zero energy states $\lambda_{0}^-(\beta^\prime)$ with increasing degeneracy (with increasing $y = \frac{B_1}{E}$) in the latter\footnote{Here, we should either conceive the uniform magnetic field to be over a finite portion of the space, or introduce an infrared cut-off via a mass term. We use the latter option in this paper, as will be discussed in the ensuing sections.} (spin-$\frac{1}{2}$ cases). However, there are also some very novel features. For spin-$0$ fields, for a certain range values of $\beta^\prime$, $\Lambda_{n_1}^-(\beta^\prime)$ becomes less than its value at $\beta^\prime = 0$ and therefore the corresponding levels are less energy costly to get filled.
This leads to the pair production rates, which exceed those at $\beta^\prime = 0$ at sufficiently large values of $y = \frac{B_1}{E}$. For $\beta^\prime < \beta^\prime_{c_1}$, these rates increase further with increasing $\beta$, while they decrease with it for $\beta^\prime_{c_1} < \beta^\prime < \beta^\prime_{c_2}$, with the estimates for the critical $\beta^\prime$ values provided in section $2$. For spin-$\frac{1}{2}$ particles, we find that, for sufficiently small values of $y = \frac{B_1}{E}$, pair production rates is further enhanced since $\lambda_{0}^-(\beta^\prime)$ monotonically decreases toward $\lambda_{0}^-(\beta^\prime = 0 ) = 0$ with $\beta^\prime$, making the corresponding eigenstates effectively degenerate with the zero energy states. Nevertheless, as $y = \frac{B_1}{E}$ increases, pair production rates converges back to that at $\beta^\prime = 0$, since $\lambda_{0}^+(\beta^\prime)$ increases monotonically with $\beta^\prime$ and states with corresponding energies become energetically costly to be filled, counterbalancing the effect of the former.  All of these features are elaborated in detail in section $2$.

In the curved geometry $S^2 \times \mathbb{R}^{1,1}$, very novel features are encountered as the continuous parameter $\alpha$ governing the non-abelian field strength vary, leading to either a competition with or support of the effect (which may enhance or suppress the RPP rates as noted above and discussed in \cite{Kurkcuoglu}) of the quantized abelian magnetic field. For scalar fields, two critical values of $\alpha$, which are determined in terms of the Dirac monopole charge $N$, govern the relative pair production rates (RPP), between the curved and flat backgrounds. The latter is measured via the function $\gamma_0(\omega, \alpha, N)$ which is defined in section 3. For $0 < \alpha < \alpha_{c_1}$,  $\gamma_0(\omega, \alpha, N) > 1$ and it indicates relatively larger pair production in the curved space, which tends to converge to the flat space results with increasing $N$.  For $ \alpha_{c}^{(1)} < \alpha < \alpha_{c}^{(2)}$, $\gamma_0(\omega,\alpha,N)$ is slightly above the value $1$ roughly within the interval $0 < \omega \lesssim 1$ (i.e. large electric field or small curvature) but eventually goes below it as $\omega$ is increased further, while we again find $\gamma_0(\omega,\alpha,N)  > 1$ for $\alpha > \alpha_{c}^{(2)}$ indicating an increasing RPP. Another critical value of $\alpha$, namely $\alpha_{c}^{(3)} = \frac{1}{2} (1 + \sqrt{2 N +1} )$, facilitates the comparison of the pair production amplitudes with and without the non-abelian field (i.e. $\alpha \neq 0$ and $\alpha =0$), which is measured using the function $R_0(\omega, \alpha, N)$, which will also be defined in section 3. We find that at large $\omega$, $R_0(\omega, \alpha, N) \rightarrow \infty \,, \frac{1}{2} \frac{N+2}{N+1}, 0 $, for $\alpha < , =, >\, \alpha_{c}^{(3)}$, indicating enhanced, saturated and suppressed pair production amplitudes, respectively. We also see that $R_0(\omega, \alpha^{(3)}_c, N)$ converging to $\frac{1}{2}$ at large $N$, and matching with the result we obtain independently in the flat background. For the spinor fields, the RPP is measured via the function $\gamma_{1/2}(\omega, \alpha, N)$ and in general it is less than $1$ and become more so with increasing $\alpha$ indicating a decrease in RPP as the non-abelian magnetic field becomes stronger. This is countered by the restoring effect of the zero modes, whose degeneracy grow with increasing Dirac monopole charge $N$ and drive the $\gamma_{1/2}(\omega, \alpha, N)$ back to $1$ and the pair production rates to those found in the flat background. Nevertheless, the overall effect is still larger for $\alpha \neq 0$ compared to $\alpha =0$, since the lowest lying energies after the zero modes remain below their value at $\alpha =0$ and the corresponding states are therefore more eligible to get filled by the produced pairs. Increasing $N$ or $\omega$ counter acts as in either case the dominance of the zero modes is elevated and the pair production rates converges to their vales at $\alpha =0$.

Finally, the case of vanishing abelian magnetic field (that is the presence of a pure uniform $SU(2)$ magnetic field) is also studied for the scalar and the spinor fields in both the flat and the curved backgrounds, and the results are compared and contrasted with those summarized above.

\section{Pair production rates for scalar fields fields on $\mathbb{R}^{3,1}$}
\label{sec:2}
In this section, we calculate the pair production rate for particles with spin-$0$ in the Minkowski space $\mathbb{R}^{3,1}$, under the influence of uniform abelian and non-abelian fields. We consider that there is a uniform electric field $\vec{E} = E \hat{x}_3 $ in the $x_3$-direction as usual and consider an additional $SU(2) \times U(1)$ gauge field generating a uniform magnetic field as we shall introduce shortly. In order to obtain the pair production rates, our strategy is to Wick rotate $\mathbb{R}^{3,1}$ to $\mathbb{R}^4 = \mathbb{R}^2\times \mathbb{R}^2$ and evaluate the Euclidean effective action due to appropriately constructed gauged Laplacian and Dirac operators for spin-$0$ and spin-$1/2$ particles respectively. On the first $\mathbb{R}^2$ copy, spanned by $(x_1,x_2)$-coordinates, we introduce the $SU(2) \times U(1)$ gauge field of the form
\begin{align}
 \label{eq:2:1}
\vec{A}_{(1)}= \frac{B_1}{2} (- x_2 \hat{x}_1 + x_1 \hat{x}_2) \mathbbm{1}_2 + \beta (- \sigma_2 \hat{x}_1 + \sigma_1 \hat{x}_2),
\end{align}
where $\vec{\sigma} := (\sigma_1, \sigma_2, \sigma_3)$ are the usual Pauli matrices. Clearly $\vec{A}$ is composed of a $U(1)$ gauge field $\vec{A}^{U(1)} := \frac{B_1}{2} (- x_2  \hat{x}_1 + x_1 \hat{x}_2)$, and a purely $SU(2)$ gauge field $A^{SU(2)}_i := -\beta \epsilon_{ij} \sigma_j$, where $i,j =1,2$. In what follows, we will call this $SU(2)$ field (as well as its analogue on $S^2$ to be introduced in section 3) as the "isospin" gauge field to distinguish it from the usual spin, which will naturally be present in the ensuing discussion for spin-$1/2$ particles. We can easily see that $\vec{A}$ generates a uniform magnetic field
\begin{align}
 \label{eq:2:5}
F_{12} =  B_1 \mathbbm{1}_2 + 2\beta^2 \sigma_3.
\end{align}
On the second copy of $\mathbb{R}^2$ spanned by $(x_3, x_4)$, we consider another $U(1)$ gauge field, say $\vec{A}_{(2)}$ generating a uniform abelian magnetic field $F_{34} := B_2$. We can take $\vec{A}_{(2)}$ in the Landau or in the symmetric gauge but this is going to be immaterial for our purposes. The magnetic field $B_2$ will be Wick rotated to the uniform electric field, ($F_{34} \rightarrow i F_{03} $, i.e. $B_2 \rightarrow i E$) at an appropriate stage in the calculation(to be given below), and the sole purpose to introduce it at this stage is to take advantage of the well-known solution of the Landau problem to write down the spectrum of the gauged Laplacian and Dirac operators to facilitate the evaluation of the Euclidean effective actions. Thus, the total gauged field on $\mathbb{R}^4$ is $A_\mu: = (\vec{A}_{(1)},\, \vec{A}_{(2)})$ and Wick rotating $\mathbb{R}^4$ to $\mathbb{R}^{3,1}$, i.e. $(x_1, x_2, x_3, x_4) \rightarrow (x_1, x_2, x_3, -i x_0)$, yields the electric and the magnetic fields $\vec{E} = E \mathbbm{1}_2 \hat{x}_3$, $\vec{B} = F_{12} \hat{x}_3$ (with the $U(1)$ accounting for the electromagnetic field.)

\subsection{Spectrum of the gauged Laplacian}
\label{sec:2:1:1}
Introducing the covariant derivative $D_\mu := \partial_\mu - i A_\mu$ on $\mathbb{R}^4$, $(\mu: 1,2,3,4)$, gauged Laplace operator on $\mathbb{R}^4$ may be written in a self-evident notation as
\beqa
\label{eq:2:6}
-D^2 & =& - ( D^2_{(1)} +  D^2_{(2)}) \nonumber \\
&=& - (\vec{\partial} - i \vec{A}_{(1)})^2 - (\vec{\partial} - i \vec{A}_{(2)})^2 \,.
\eeqa
$- D^2_{(1)}$ can be expressed in the form
\be
  \label{eq:2:8}
  - D^2_{(1)} = 2 B_1 \Big(a^{\dagger} a + \sqrt{2}\beta^\prime(a^{\dagger}\sigma_+ + a\sigma_-) + \frac{1}{2}(1+2\beta^{\prime 2}) \mathbbm{1}_2 \Big),
\ee
where $a$ and $a^{\dagger}$ are the annihilation and creation operators defined in the same manner as in the Landau problem (see appendix \ref{sec:A}) and $\sigma_{\pm} = \sigma_1 \pm i \sigma_2$ are the isospin ladder operators. Here we have also introduced the notation $\beta^\prime = \beta / \sqrt{B_1}$ as the dimensionless non-abelian magnetic field strength by scaling $\beta$ with respect to the square root of the abelian magnetic field. This operator is essentially very similar to the Hamiltonian of the Jaynes-Cummings model \cite{Jaynes} as it is already discussed in \cite{Estienne_2011} and it can easily be diagonalized. Its spectrum is
\be
  \label{eq:2:9}
  \Lambda^{\pm}_{n_1} = 2B_1 \Big(n_1 \pm \sqrt{2\beta^{\prime 2} n_1 +1/4} + \beta^{\prime 2}\Big) \,,
\ee
where $n_1=0,1,2\cdots$ for the upper and $n_1 =1,2,\cdots$ for the lower signs, respectively. Details of a straightforward calculation leading to \eqref{eq:2:9} is provided in the Appendix \ref{sec:A} for convenience. Note in particular that the lowest lying state is given by $\Lambda^+_0 = 2B_1( \beta^{\prime 2} + 1/2)$. 
 
On the other hand, spectrum of $- D^2_{(2)}$ is nothing but the solution of the Landau problem and the eigenvalues are
$B_2 (2 n_2 + 1)$, where $n_2=0,1,2,\cdots$. Putting these facts together we have
\beqa
  \label{eq:2:10}
  Spec (- D^2 + m^2) = \Lambda^{\pm}_{n_1}  + B_2 (2n_2 + 1) + m^2,  
\eeqa
with the density of states given as $\frac{B_1}{2 \pi} \times \frac{B_2}{2 \pi} $, since the presence of the non-abelian magnetic field does not alter the density of states corresponding to the spectrum $\Lambda^{\pm}_{n_1} $ of $-D^2_{(1)}$ as it can readily be inferred from the calculations provided in Appendix \ref{sec:A}.

\subsection{Pair production rates}
\label{sec:2:1:2}

We start with the computation of the Euclidean effective action $\Gamma_E\equiv Tr\,\log(- D^2+m^2)$, which will Wick rotate to Lorentzian signature at an appropriate stage. Following the approach in \cite{Kurkcuoglu}, we have
\begin{align}
  \label{eq:2:11}
  \Gamma_E &= -Tr\, \lim_{\epsilon\rightarrow 0} \int_\epsilon^\infty \frac{ds}{s} e^{-s(- D^2 + m^2)}, \nonumber \\
           &= -\lim_{\epsilon\rightarrow 0} \int_0^\infty d^4x \int_\epsilon^\infty \frac{ds}{s} \bra{x}e^{-s( - D^2 + m^2)}\ket{x} \,.
\end{align}
Expanding the position kets $\ket{x}$ on $\mathbb{R}^4$ with respect to the eigenkets $\ket{n_1,n_2,\alpha}$ (with the auxiliary index $\alpha$ labeling the degeneracy) of the gauged Laplacian, we may write

\begin{align}
  \label{eq:2:12}
  \Gamma_E = -\lim_{\epsilon\rightarrow 0}\int_0^\infty d^4x \int_\epsilon^\infty \frac{ds}{s} &\sum_{n_1,n_2,\alpha} \bra{x} e^{-s(- D^2 + m^2)} \ket{n_1,n_2, \alpha} \bra{n_1,n_2, \alpha} \ket{x} \nonumber \\
           = -\lim_{\epsilon\rightarrow 0}\int_0^\infty d^4x \int_\epsilon^\infty \frac{ds}{s} &\sum_{n_1,n_2,\alpha} \bra{x} \ket{n_1,n_2, \alpha} \big ( e^{-s(\Lambda_{n_1}^+ + B_2(2n_2+1) + m^2)} \nonumber \\
           & \quad \quad +e^{-s(\Lambda_{n_1}^- + B_2(2n_2+1) + m^2)} \big) \bra{n_1,n_2, \alpha} \ket {x}, \nonumber \\
            = -\lim_{\epsilon\rightarrow 0}\int_0^\infty d^4x \int_\epsilon^\infty \frac{ds}{s} &\sum_{n_1,n_2,\alpha} \psi^*_{n_1,n_2,\alpha}(x)  \psi_{n_1,n_2,\alpha}(x) \big( e^{-s(\Lambda_{n_1}^+ + B_2(2n_2+1) + m^2)}\nonumber\\
           & \quad \quad \quad  +e^{-s(\Lambda_{n_1}^- + B_2(2n_2+1) + m^2)} \big) \,,
\end{align}
where $\psi_{n_1,n_2,\alpha}(x)$ denote the eigenfunctions of the gauged Laplacian in the position basis. Using the normalization of  $\psi_{n_1,n_2,\alpha}(x)$ and the density of states $\frac{B_1 B_2}{(2\pi)^2}$, sum over these degenerate states can be performed\footnote{$\sum_{\alpha} \psi_{n_1,n_2,\alpha}^*(x) \, \psi_{n_1,n_2,\alpha}(x) = \sum_\alpha 1 = \int d^4 x \frac{B_1 B_2}{(2\pi)^2} \,. $} and we may write 
\begin{align}
  \label{eq:2:14}
  \Gamma_E &= -\lim_{\epsilon\rightarrow 0}\int_0^\infty d^4x \int_\epsilon^\infty \frac{ds}{s}\frac{B_1 B_2}{(2\pi)^2} \sum_{n_1,n_2} \left ( e^{-s(\Lambda_{n_1}^+ + B_2(2n_2+1) + m^2)} +e^{-s(\Lambda_{n_1}^- + B_2(2n_2+1) + m^2)} \right ), \nonumber \\
           &= -\lim_{\epsilon\rightarrow 0}\int_0^\infty d^4x\frac{B_1}{8\pi^2} \int_\epsilon^\infty \frac{ds}{s^2}\frac{s B_2}{sinh\, sB_2} \sum_{n_1} \left ( e^{-s(\Lambda_{n_1}^+ + m^2)} +e^{-s(\Lambda_{n_1}^- + m^2)} \right ) \,.
\end{align}
In the second line of \eqref{eq:2:14}, we have performed the summation over the index $n_2$. This is the appropriate stage to Wick rotate the Euclidean effective action via  $x_4 \rightarrow ix_0$ and $B_2 \rightarrow -iE$ and identify $\Gamma_E$ with $i S_{eff}$. Pair production rate is proportional to the real part of the latter, which we denote as $Re(iS_{eff})$. We have
\begin{align}
  \label{eq:2:15}
  iS_{eff} = \frac{i}{8\pi^2}\lim_{\epsilon\rightarrow 0}\int_0^\infty d^4xB_1 \int_\epsilon^\infty \frac{ds}{s^2}\frac{s B_2}{sinh\, sB_2} \sum_{n_1,n_2} \left ( e^{-s(\Lambda_{n_1}^+  + m^2)} +e^{-s(\Lambda_{n_1}^- + m^2)} \right ).
\end{align}
We can now perform the $s$ integration in the same manner as in \cite{Kurkcuoglu}. The integral has singularities at $s=n\pi/E$, $n=1,2,3, \cdots$. For the pair production rate we only need the real part of $iS_{eff}$ and the contributions to this comes from the integrals around small semicircles at each singularity. Therefore, taking our integration variables as $s = n\pi/E + z$ with $|z| \ll1$ and performing the integration over the parameter $z$, we get 
\begin{align}
  \label{eq:2:16}
  Re(iS_{eff}) = \frac{E B_1}{8\pi^2} \int d^4x \sum_{n=1}^\infty \frac{(-1)^n}{n} \sum_{n_1} \left ( e^{-(n\pi/E)(\Lambda^+_{n_1}+m^2)} + e^{-(n\pi/E)(\Lambda^+_{n_1}+m^2)} \right ) \,,
\end{align}
which, after performing the sum over $n$, can be cast into the form
\begin{multline}
  Re(iS_{eff}) = \frac{E B_1}{8\pi^2} \int d^4x \sum_{n_1} \ln (1+ e^{-(2 \pi B_1/E)(n_1 + \sqrt{2 \beta^{\prime 2} n_1 +1/4} + \beta^{\prime 2} + \frac{1}{2} m^2)}) \\
  + \ln (1+ e^{-( 2\pi B_1/E)(n_1 - \sqrt{2\beta^{\prime 2} n_1 +1/4} + \beta^{\prime 2} + \frac{1}{2} m^2)}) \,.
\end{multline}
Introducing the dimensionless parameter $y := B_1/E$ and taking the limit $m^2 \rightarrow 0$, we are able to write 
\begin{align}
  \label{eq:2:18}
  Re(iS_{eff}) = -\int d^4x \frac{E^2}{96 \pi }f_0(y, \beta^\prime),
\end{align}
where 
\beqa
\label{eq:2:19}
&& f_0(y, \beta^\prime) = \frac{12 y}{\pi}  \sum_{n=1}^\infty \frac{(-1)^{n+1}}{n} \sum_{n_1} \left( e^{- \pi n y \frac{\Lambda^+_{n_1}}{B_1}} + e^{- \pi n y \frac{\Lambda^-_{n_1}}{B_1}}  \right)  \\
&&=  \frac{12 y}{\pi}  \left (\sum_{n_1=0}^{\infty} \ln \,(1+ e^{-2\pi y (n_1 + \sqrt{2\beta^{\prime 2} n_1+1/4} + \beta^{\prime 2})}) + \sum_{n_1=1}^{\infty} \ln \,(1+ e^{-2\pi y (n_1 - \sqrt{2\beta^{\prime 2} n_1+1/4} + \beta^{\prime 2})}) \right) \nonumber \,.
\eeqa       
We see that, $f_0(y, \beta^\prime) \rightarrow 1$ as $y \rightarrow 0$ i.e. $B_1 \rightarrow 0$ and $\beta \rightarrow 0$ with $\beta^\prime = \frac{\beta}{\sqrt{B_1}}$ held fixed\footnote{This limit is essentially independent of the value of $\beta^{\prime}$ and easily evaluated at $\beta^{\prime} = 0$ and therefore holds the same at any value of $\beta^{\prime}$ by continuity. With $\beta^{\prime} = 0$, starting from the first line in \eqref{eq:2:19} and performing the sum over the index $n_1$ first, we have
\beqa
\lim_{y \rightarrow 0} \, f_0(y, \beta^\prime) &=& \lim_{y \rightarrow 0} \, \frac{12}{\pi}  \sum_{n=1}^\infty \frac{(-1)^{n+1}}{n}  \frac{y}{\sinh n \pi y}  \nonumber \\
&=&  \frac{12}{\pi^2} \sum_{n=1}^\infty \frac{(-1)^{n+1}}{n^2} =  \frac{12}{\pi^2} \,  \eta(2) = 1 \nonumber \,,
\eeqa	
using the value of  the Dirichlet eta function $\eta(2) = \frac{\pi^2}{12}$.
}.	 
Then, \eqref{eq:2:18} yields nothing but twice the usual Schwinger result in this limit, the overall factor of 2 being due to equal contributions from isospin up and down degrees of freedom in this limit.  

In order to assess the pair production rates in this setting, we may inspect the behavior of the function $f_0(y, \beta^\prime)$ as we vary $y$ at various values of $\beta^\prime$. 
\begin{figure}[h!]
    \centering
    \begin{subfigure}[b]{0.49\textwidth}
    	\includegraphics[width=\textwidth]{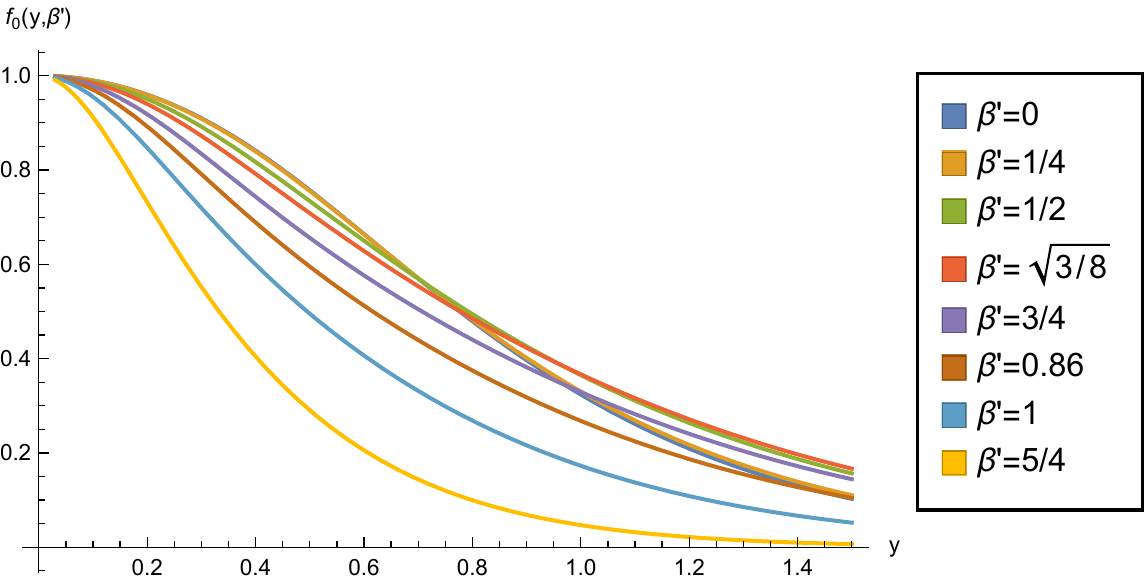}
       \caption{\label{fig:2:1} $f_{0}(y,\beta^\prime)$}
    \end{subfigure}
    \hfill
    \begin{subfigure}[b]{0.49\textwidth}
    	\includegraphics[width=\textwidth]{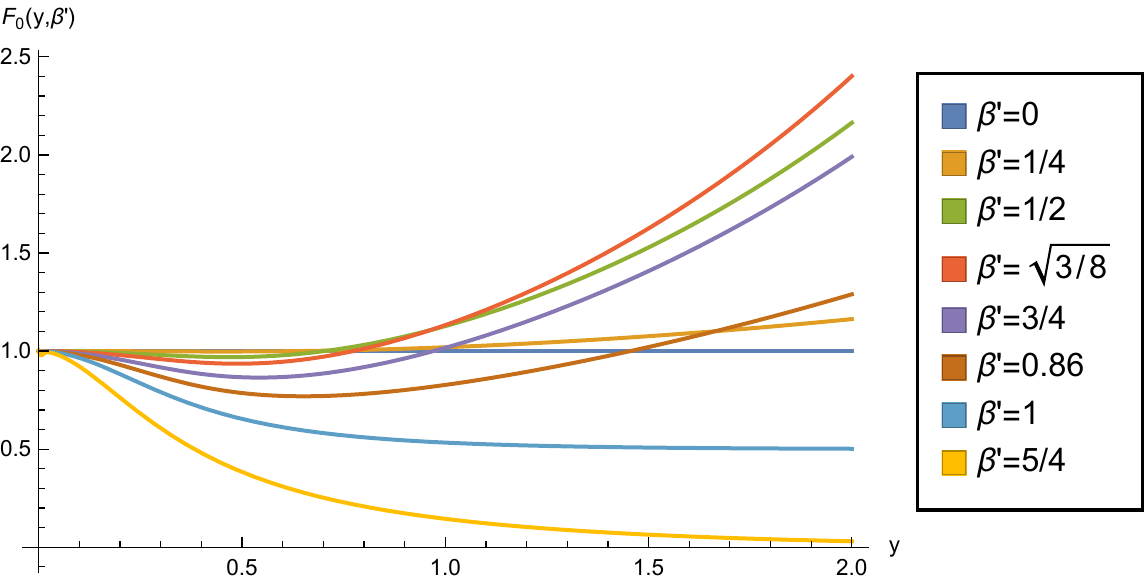}
    		\caption{\label{fig:2:2} $F_{0}(y,\beta^\prime)$}
    \end{subfigure}
    \hfill
	\centering
    \begin{subfigure}[b]{0.49\textwidth}
		\includegraphics[width=\textwidth]{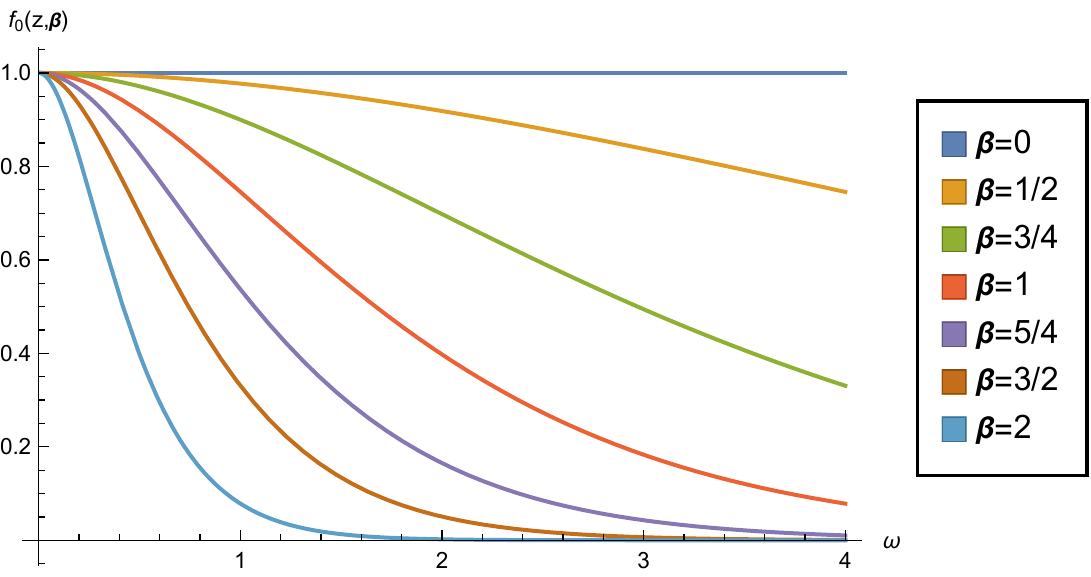}
		 \caption{\label{fig:fl0} $f_{0}(z,\beta)$}
    \end{subfigure}
    \caption{\label{fig:main1} Profiles of $f_{0}(y,\beta^\prime)$, $F_{0}(y,\beta^\prime)$ and $f_{0}(z,\beta)$.}
\end{figure}
In Figure~\ref{fig:2:1} the profile of this function is plotted for $\beta^\prime = 0, 1/4, 1/2, \sqrt{\frac{3}{8}}, 7/8, 1, 3/2$. We immediately observe that $f_0(y,\beta^\prime)$ decreases with increasing $y$. This means that with the increasing abelian field strength, $B_1$, ($y = B_1/E$) leads to a decrease in the pair production rates. This feature of the function $f_0(y, \beta^\prime)$ is expected and in accord with the case studied in \cite{Kurkcuoglu} with purely abelian magnetic field. In fact, we observe that, since the eigenvalues $\Lambda_{n_1}^+(\beta^\prime)$ are always larger than $\Lambda_{n_1}^+(\beta^\prime = 0) = 2 B_1 (n_1 + \frac{1}{2})$, (i.e. those obtained for purely abelian magnetic field) and become larger with increasing $\beta^\prime$ too; We conclude that these quantum states become increasingly harder to be filled by particle, anti-particle pairs as $\beta^\prime$ increases and we observe a significant decrease in the value of the function $f_0(y,\beta^\prime)$ for increasing $\beta^\prime$ values. However, for a given level $n_1$, $\Lambda_{n_1}^-(\beta^\prime)$ starts with the value $\Lambda_{n_1}^-(\beta^\prime = 0)= 2 B_1 (n_1 - \frac{1}{2})$ at $\beta^\prime = 0$, decreases to the minimum $2 B_1 \frac{n_1^2- 1/4}{2 n_1}$ at $\beta^\prime_{c_1} = \sqrt{\frac{n_1^2- 1/4}{2 n_1}}$ and monotonically increases starting from this point and attains $2 B_1(n_1 - \frac{1}{2})$ value again at $\sqrt{2n_1-1}$. Thus, compared the to $\beta^\prime =0$ configuration, between $0 < \beta^\prime \leq \beta^\prime_{c_1} $, with increasing $\beta^\prime$, states with $\Lambda_{n_1}^-(\beta^\prime)$ are energetically more favorable to be filled by produced pairs, while between $\beta^\prime_{c_1} \leq \beta^\prime \leq \sqrt{2n_1-1}$, they still are more favorable to be filled, but become less so with increasing $\beta^\prime$. These features are reflected in the profile $f_0(y,\beta^\prime)$ as readily observed from the Figure~\ref{fig:2:1}; At around\footnote{At sufficiently large values of $y$, contribution of both the sums in \eqref{eq:2:19} become sufficiently small, allowing us to distinguish the effects of the energies $\Lambda_{n_1}^- (\beta^\prime)$ of the available quantum states on the pair production rates.} $y \approx 1$, we see that $f_0(y,\beta^\prime)$ takes larger values with increasing $\beta^\prime \leq \beta^\prime_{c_1}$, with $\beta^\prime_{c_1} \approx \sqrt{3/8}$ well approximated by that associated to the lowest energy level $\Lambda_{1}^-$, while for $\beta^\prime_{c_1} \leq \beta^\prime \lesssim \beta^\prime_{c_2}$, we have $f_0(y,\beta^\prime)$ decreasing with increasing $\beta^\prime$ but exceeding its value at $\beta^\prime =0$ at sufficiently large values of $y$. Our numerical estimates give $f_0(y=1.5, \beta^\prime = 0.86) \approx 0.1046 $, $f_0(y = 1.5, \beta^\prime = 0) \approx 0.1024 $, $f_0(y=1.5, \beta^\prime = 0.87) \approx 0.1006$, placing $\beta^\prime_{c_2} \approx 0.86$ at $y = 1.5$ with $\beta^\prime_{c_2} \lesssim 1$ for $y \rightarrow \infty$. For $\beta^\prime > 1 $,  we have $\Lambda_{1}^-(\beta^\prime) > \Lambda_{1}^-(\beta^\prime = 0)$ and any further increase in the value of $\beta^\prime$ results in a sharp decrease of the function $f_0(y,\beta^\prime)$ and hence the pair production rates.

We may also define the function $F_0(y,\beta^\prime) := \frac{f_0(y,\beta^\prime)}{f_0(y,0)}$, which is not only a good measure for the relative pair production rates, but also allows us to further elaborate on the significance of the values $\beta^\prime_{c_1} = \sqrt{3/8}$ and $\beta^\prime = 1$. Inspecting the profiles of $F_0(y,\beta^\prime)$ in Figure~\ref{fig:2:2}, we see that, $F_0(y,\beta^\prime)$ exceed the value $1$ at $y \approx 1$ and becomes larger with increasing $\beta^\prime$ within the interval $0 < \beta^\prime \lesssim \sqrt{3/8}$. For, $\sqrt{3/8} \leq \beta^\prime < 1$, $F_0(y,\beta^\prime)$ is above the value $1$ only for sufficiently large $y$, and increases with $\beta^\prime$, approaching the value $1$, while for $\beta^\prime > 1$, it quickly decreases and converges to zero. In addition to these features, which are in complete accord with conclusions we have reached by inspecting the profile of $f_0(y,\beta^\prime)$, we further see that at $\beta^\prime = 1$, $F_0(y,\beta^\prime)$ approaches to the value $1/2$ at large $y$, with $F_0(y = 2,\beta^\prime =1) \approx 0.5020 $. This means that at the critical value $\beta^\prime = 1$, pair production rate quickly converges to half of what was found for the purely abelian case ($\beta^\prime =0$) in \cite{Kurkcuoglu}. This profile of $F_0(y,\beta^\prime)$ at $\beta^\prime = 1$ can be predicted by comparing the energies and degeneracy of the lowest lying quantum states. Indeed, we have $\Lambda_0^+(\beta^\prime=0) = \Lambda_1^-(\beta^\prime=0) = B_1$ and $\Lambda_1^-(\beta^\prime = 1) = B_1$\footnote{Here, it is sufficient to focus on the lowest lying energy eigenvalues since they are the ones, which are most easily filled by the produced pairs.}, thus we have double the number of states at this energy in the purely abelian case leading to $F_0(y,\beta^\prime)$ converging to the value of $\frac{1}{2}$, which is corroborated from its profile given in Figure~\ref{fig:2:2}.
  
The spectrum of $-D_{(1)}^2$ with purely non-abelian magnetic field $\beta$ is obtained from $\Lambda_{n_1}^\pm$ by taking the limit $B_1 \rightarrow 0$, $n_1 \rightarrow \infty$ such that $2 B_1 n_1 \rightarrow k^2$ and $\Lambda_n^\pm \rightarrow = k^2 + 2\beta^2 \pm 2\beta k$, where $k = \sqrt{k_x^2 +k_y^2}$ and $k_x, k_y$ are the eigenvalues of $\vec{p}$ in the transverse directions. This spectrum is also directly worked out from first principles in Appendix \ref{sec:A} for completeness.  Computation of the Euclidean effective action $\Gamma_E$ proceeds straightforwardly, and the short cut is that in \eqref{eq:2:19}, $B_1 \sum_{n_1}$ gets replaced with the integral $\int k \, dk$ and this yields
\be
Re(iS_{eff}) = -\int d^4x \,\frac{E^2}{96 \pi} \, f_0 (\pi/E, \beta) \,,
\ee
where
\be
\label{eq:beta0flat}
f_0(z, \beta) = \frac{12}{\pi^2 } \, z \int \, k \, dk \, \Big ( \ln \,(1+e^{- z [(k+\beta)^2 + \beta^2 + m^2]}) + \ln \, (1+e^{- z [(k - \beta)^2 + \beta^2 + m^2]}) \Big )\,.
\ee
This is a decreasing function of $\beta$, as observed from Figure~\ref{fig:fl0},leading to a decrease in the pair production rates with increasing $\beta$. In the limit $\beta \rightarrow 0$, we obtain twice the Schwinger result, since the dimension of the eigenstate space is trivially doubled due to the isospin degree of freedom. 

\section{Pair production rates for spinor fields fields on $\mathbb{R}^{3,1}$}
\label{sec:2:2}

We now proceed to consider the pair production for spin-$\frac{1}{2}$ particles on $\mathbb{R}^{3,1}$ under the influence of the same additional $SU(2) \times U(1)$ magnetic field. For this purpose, we again consider the Wick rotated configuration with the magnetic fields $F_{12}$ and $F_{34}$ on $\mathbb{R}^2 \times \mathbb{R}^2 \equiv \mathbb{R}^4$.  To our knowledge, the spectrum of the Dirac operator in such a background gauge field is not considered in the literature before. Therefore, we proceed to handle this task first.  We may note that the solution of this problem is interesting in its own right as it leads to zero modes in a manner similar to the spectrum of the Dirac operator with purely abelian uniform magnetic field. 

\subsection{Spectum of the gauged Dirac operator}
\label{sec:2:2:1}

We may launch the discussion by writing out the gauged Dirac operator on the first copy of $\mathbb{R}^2$. This is given as
    \begin{equation}
        \label{eq:2:20}
        \slashed{D}_{(1)} = \gamma_i D_i = \gamma_i (\partial^i - i A^i_{(1)})
    \end{equation}
where $\gamma_i$ are the $2 \times 2$ span the Clifford algebra on $\mathbb{R}^2$. We take them as $\gamma_1 = \tau_1$ and $\gamma_2 = \tau_2$, where $\tau_1$, $\tau_2$ are the $2\times 2$ Pauli matrices. The gauge field $A^i_{(1)}$ is as given already in \eqref{eq:2:1}. After some straightforward algebra which is relegated to the appendix \ref{sec:B}, we may write the operator $- \slashed{D}^2_{(1)}$ as
\begin{align}
    \label{eq:2:22}
    - \slashed{D}^2_{(1)} = 2 B_1
    \begin{pmatrix}
        a^{\dagger}a & \sqrt{2} \beta^\prime a^{\dagger} & 0 & 0 \\
        \sqrt{2} \beta^\prime a & a^{\dagger}a + 2\beta^{\prime 2} & 0 & 0 \\
        0 & 0 & a a^{\dagger} + 2\beta^{\prime 2}  & \sqrt{2} \beta^\prime a^{\dagger} \\
        0 & 0 & \sqrt{2} \beta^\prime a & a a^{\dagger}
    \end{pmatrix}.
\end{align}
where $a\,, a^\dagger$ are the usual annihilation and creation operator (see appendix A). Clearly $ - \slashed{D}^2_{(1)}$ is acting on the Hilbert space $\mathcal{H} = \mathbb{C}^4 \otimes \mathcal{F}$, where the Fock space, $\mathcal{F}$, is spanned by the eigenstates of the number operator $N = a^\dagger a$ as usual. Diagonalizing $- \slashed{D}^2_{(1)}$ in an appropriate subspace $\mathcal{H}_n \subset \mathcal{H}$ leads to the spectrum (see appendix \ref{sec:B})
\begin{align}
    \label{eq:2:23}
    \lambda_{n_1}^\pm = B_1 \left (1+2n_1+2\beta^{\prime 2} \pm \sqrt{1+4\beta^{\prime 2}(1+2n_1+\beta^{\prime 2})} \right)\,,
\end{align}
where $n_1 = 0,1, \cdots$. We may remark that the $\pm$ signs in the spectrum $\lambda_{n_1}^\pm $ are neither correlated with the spin nor with the isospin of the system. Eigenstates of $ - \slashed{D}^2_{(1)}$ are simultaneous eigenstates of the spin but not those of the isopin, since $ - \slashed{D}^2_{(1)}$ commutes with $\tau_3$ (for that matter with all $\tau_i$), but not with $\sigma_3$. Thus, each eigenvalue occurs twice (once for spin up and once for spin down) and the density of states is $2 \times \frac{B_1}{2 \pi}$. Details leading to these facts are provided in appendix \ref{sec:B}. In particular, we may note that the $n_1=0$ level with the lower sign gives the zero modes and $\lambda_{n_1}^+$ ($\lambda_{n_1}^-$) is a monotonically increasing (decreasing) function of $\beta^\prime$. Thus all the eigenvalues $\lambda_{n_1}^-$ tend to zero as $\beta^\prime \rightarrow \infty$.  

We may note that the spectrum of $-\slashed{D}^2_{(2)}$ on the second $\mathbb{R}^2$ copy is that of the Dirac-Landau problem and given by $2B_2 ( n_2 + 1)$ and $2 B_2 n_2 $, for $n_2=0,1,\cdots$, for spin up and down respectively. Thus, the spectrum of $- \slashed{D}^2 + m^2 =  - (\slashed{D}^2_{(1)} + \slashed{D}^2_{(2)}) + m^2$ on $\mathbb{R}^2 \times \mathbb{R}^2 = \mathbb{R}^4$ is easily written as 
\begin{align}
	\label{eq:2:24}
	Spec(- \slashed{D}^2 + m^2) =
	\begin{cases}
		 \lambda_{n_1}^\pm + 2B_2(n_2 + 1) + m^2, \\
		 \lambda_{n_1}^\pm + 2B_2n_2 + m^2 \,,
	\end{cases}
\end{align}
where, $n_1\,, n_2 = 0,1, \cdots$ and the density of states is  $2 \times B_1 B_2/(2\pi)^2$ in each branch.

\subsection{Pair production rates}
\label{sec:2:2:2}

\begin{figure}[h!]
    \centering
    \begin{subfigure}[b]{0.49\textwidth}
    	\includegraphics[width=\textwidth]{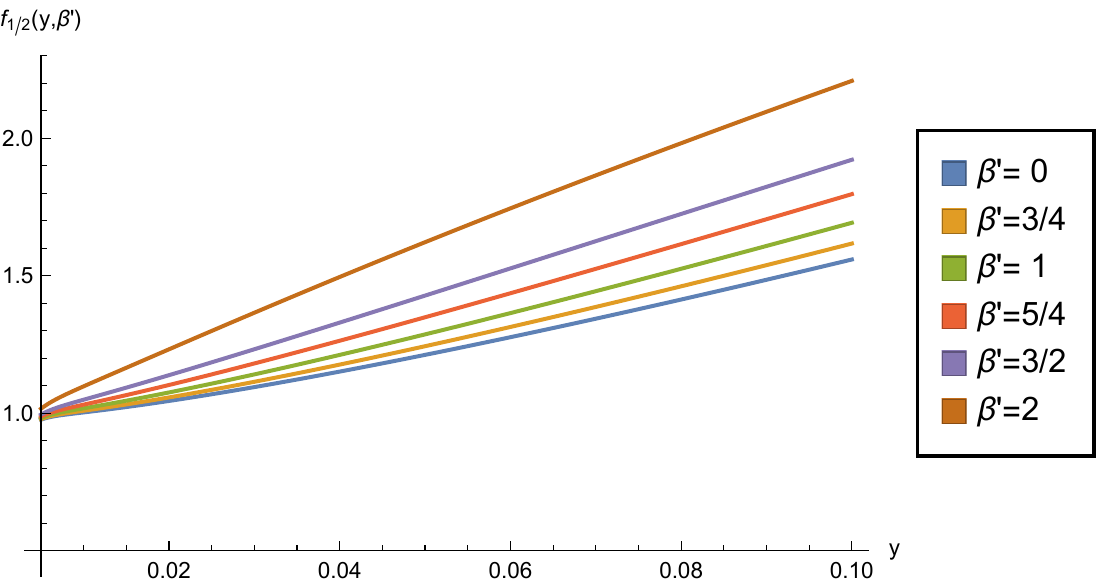}
        \caption{\label{fig:2:3} $f_{1/2}(y,\beta^\prime)$, $\frac{m^2 \pi}{E} =  2 \times 10^{-2}$.}
    \end{subfigure}
    \hfill
    \begin{subfigure}[b]{0.49\textwidth}
    	\includegraphics[width=\textwidth]{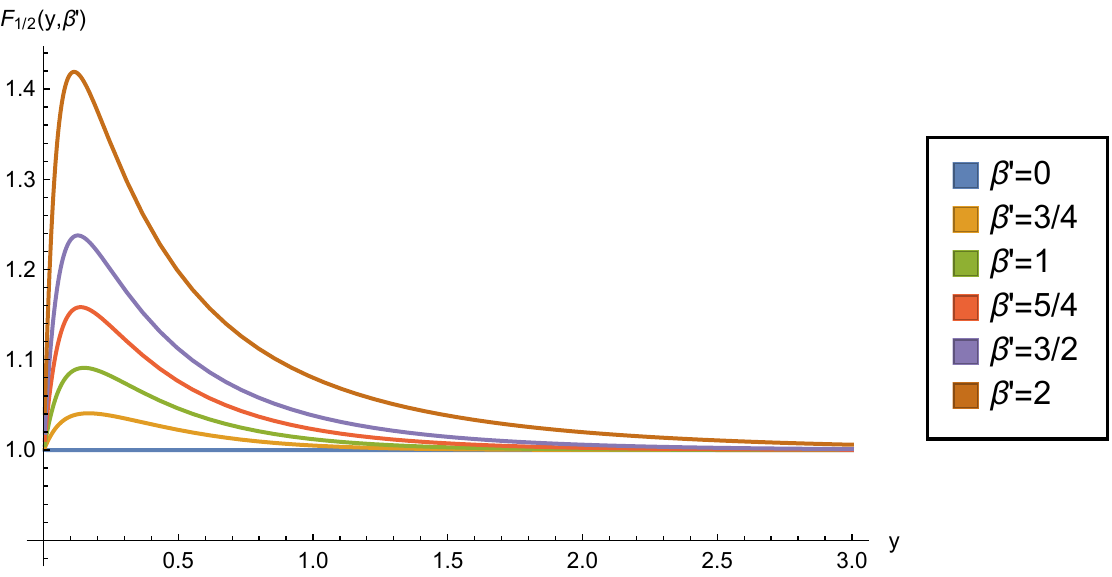}
    	\caption{\label{fig:2:4} $F_{1/2}(y,\beta^\prime)$}
    \end{subfigure}
    \hfill
	\centering
    \begin{subfigure}[b]{0.49\textwidth}
		\includegraphics[width=\textwidth]{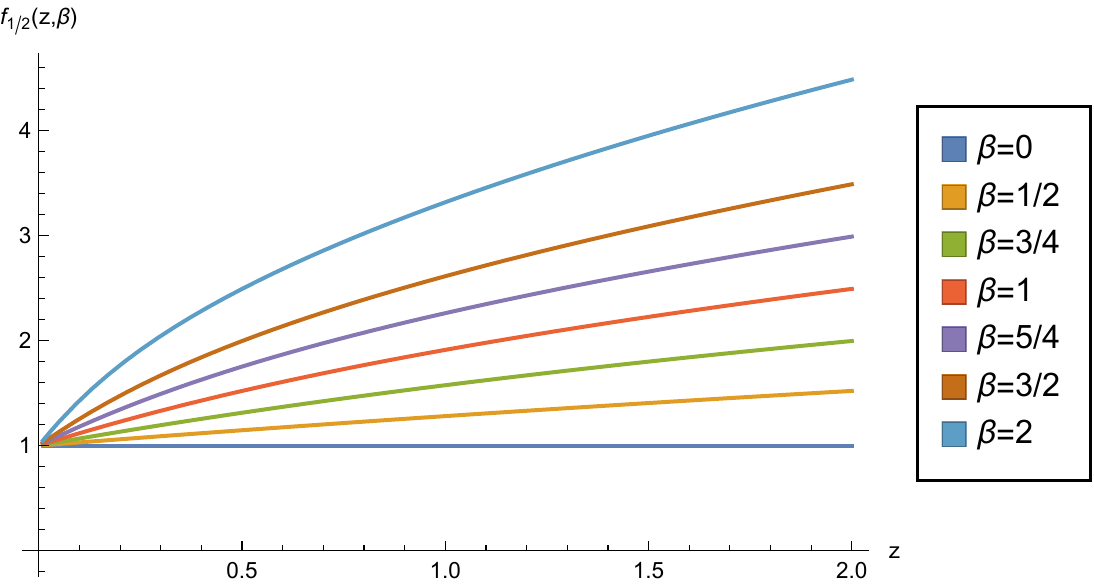}
		\caption{\label{fig:fl05} $f_{1/2}(z,\beta)$, $\frac{m^2 \pi}{E} =  2 \times 10^{-2}$.}
    \end{subfigure}
    \caption{\label{fig:main2} Profiles of $f_{1/2}(y,\beta^\prime)$, $F_{1/2}(y,\beta^\prime)$, and $f_{1/2}(z,\beta)$.}
\end{figure}

In the Euclidean signature the one-loop effective action is given as 
    \begin{align}
        \label{eq:2:25}
        \Gamma_E & =  - Tr\, \ln \, (\slashed{D} + m) \,, \nonumber \\
        & = -\frac{1}{2} \, Tr\, \ln \,(\slashed{D}^\dagger + m) (\slashed{D} + m) \,, \nonumber \\ 
        & = -\frac{1}{2} \, Tr\, \ln \,(-\slashed{D}^2 + m^2) \,.
    \end{align}
Separating out the zero mode contribution explicitly and performing the summation over the index $n_2$, we get
    \begin{multline}
        \label{eq:2:26}
        \Gamma_E = \frac{B_1B_2}{4\pi^2} \int d^4x \lim_{\epsilon \rightarrow 0} \frac{ds}{s} coth\, sB_2 \Biggl[ \sum_{n_1=0}^{\infty} e^{-s \left(1 + 2n_1 + 2\beta^{\prime 2} + \sqrt{1+4\beta^{\prime 2}(1 + 2n_1 + \beta^{\prime 2})}  +m^2 \right )} \\
          + \sum_{n_1=1}^{\infty} e^{-s \left ( 1 + 2n_1 + 2\beta^{\prime 2} - \sqrt{1+4\beta^{\prime 2}(1 + 2n_1 + \beta^{\prime 2})} +m^2\right ) } + e^{-sm^2}  \Biggr].
    \end{multline}
\eqref{eq:2:26} includes an overall of factor of $2$ since each of the eigenvalues $\lambda_{n_1}^\pm$ occurs with multiplicity $2$. Performing the integral over $s$, evaluating the sum due to the ensuing residue integration and Wick rotating the $\Gamma_E$ to Minkowski time allows us to write $Re(iS_{eff})$ as
        \begin{multline}
            \label{eq:2:27}
            Re(iS_{eff}) =  - \int d^4x \, \frac{E^2}{2\pi^2} \, y \, \Biggl[ \ln \,(1-e^{-\pi m^2 /E}) \\ 
            + \sum_{n_1= 0} \ln \,(1-e^{-\pi m^2 /E - y\pi\bigl(1+2n_1+2\beta^{\prime 2}+\sqrt{1+4\beta^{\prime 2}(1+2n_1+\beta^{\prime 2})}\bigr)}) \\
            + \sum_{n_1= 1} \ln \,(1-e^{-\pi m^2 /E -y\pi\bigl(1+2n_1+2\beta^{\prime 2}-\sqrt{1+4\beta^{\prime 2}(1+2n_1+\beta^{\prime 2})}\bigr)})
            \Biggr] \,,
        \end{multline}
where, we have, once again, used $y \equiv B_1/E$. We may write our final result as
    \be
    \label{eq:2:28}
        Re(iS_{eff}) = - \int d^4x \frac{E^2}{24\pi} f_{1/2} (y,\beta^\prime),
    \ee
where
    \begin{multline}
        \label{eq:2:29}
        f_{1/2}(y,\beta^\prime) = - \frac{6 y}{\pi} \Biggl[ \ln \, (1 - e^{-\pi m^2 /E}) 
            + \sum_{n_1= 0} \ln \,(1-e^{-\pi m^2 /E -y\pi\bigl(1+2n_1+2\beta^{\prime 2}+\sqrt{1+4\beta^{\prime 2}(1+2n_1+\beta^{\prime 2})}\bigr)}) \\
            + \sum_{n_1= 1}  \ln \,(1-e^{- \pi m^2 /E -y\pi\bigl(1+2n_1+2\beta^{\prime 2}-\sqrt{1+4\beta^{\prime 2}(1+2n_1+\beta^{\prime 2})}\bigr)}) \Biggr] \,.
    \end{multline}
We may first note that, the first term in this expression is the contribution of the zero modes, which will make $f_{1/2}(y,\beta^\prime)$ diverge to infinity unless it is regulated. The mass term is kept as an infrared cut-off to avoid this divergent behavior. In the absence of the non-abelian field, i.e. $\beta^\prime = 0$, we already know that \cite{Kurkcuoglu}, $f_{1/2}(y,\beta^\prime) \rightarrow 1$ as $y \rightarrow 0$ and it increases with $y$. In the present case, since the degeneracy of the zero modes still increases with increasing values of $B_1$ (hence increasing $y$), we expect an increase in the pair production rates as $y$ becomes larger as there is no energy cost for the produced pairs to fill these zero energy states. The function $f_{1/2}(y,\beta^\prime)$ is rather sensitive to the choice of the infrared cut-off and its profile is given Figure~\ref{fig:2:3}, where we have picked $\frac{m^2 \pi}{E} =  2 \times 10^{-2}$ and evaluated the sums up to $n_1=125$. Figure~\ref{fig:2:3} clearly depicts these expectations and we further observe that the function $f_{1/2}(y,\beta^\prime)$ and hence the pair production goes up as we raise the value of $\beta^\prime$ at any fixed value of $y$. To better understand this feature, we note from \eqref{eq:2:29} that at any given value of $y$, the argument of the first sum tends to $\approx \ln 1 \rightarrow 0$, since $\lambda^+_{n_1}$ in the exponential is monotonically increasing with $\beta^\prime$, while the argument of the second sum tends to $\ln \, (1 - e^{-\pi m^2 /E})$ since $\lambda^-_{n_1}$ in the exponential is monotonically decreasing with $\beta^\prime$ starting with a maximum value of $2 B_1 n_1$ at $\beta^\prime = 0$. Thus, as $\beta^\prime$ becomes larger, all the terms in the second sum tend to contribute almost the same as with that of the zero modes resulting in a sharp increase in the pair production rates. Let us note in passing that, the hierarchy in the plots with respect to $\beta^\prime$ in Figure~\ref{fig:2:3} is preserved at other physically sensible values of the cut-off $\frac{m^2 \pi}{E}$.
A good measure for the relative production rate is provided by the function $F_{1/2}(y,\beta^\prime) = \frac{f_{1/2}(y,\beta^\prime)}{f_{1/2}(y,0)}$, which is essentially independent of the infrared cut-off\footnote{We have verified this numerically using Mathematica.} and it is plotted in Figure~\ref{fig:2:4} at several values of $\beta^\prime$. We immediately see the quickly formed peaks in the profile of this function at low values of $y$ and hence the sharp increase in the pair production rates compared to the purely Abelian case with $\beta^\prime = 0$ as we keep on increasing $\beta^\prime$. As $y$ increases further only the zero mode term $\ln \, (1 - e^{-\pi m^2 /E})$ contributes significantly both to the numerator and denominator of $F_{1/2}(y,\beta^\prime)$ and the pair production rates tend back to its value at $\beta^\prime = 0$.

The spectrum of $-\slashed{D}_{(1)}^2$ with purely non-abelian magnetic field $\beta$ is obtained from $\lambda_{n_1}^\pm$ by taking the limit $B_1 \rightarrow 0$, $n_1 \rightarrow \infty$ such that $2 B_1 n_1 \rightarrow k^2$. This  gives $\lambda_n^\pm \rightarrow  k^2 + 2\beta^2 \pm 2\beta \sqrt{k^2 +\beta^2}$, where $k = \sqrt{k_x^2 +k_y^2}$ as already defined in the previous section. For completeness, we provide a complete derivation of this spectrum from first principles in Appendix B. Absorbing the the zero mode term in 
\eqref{eq:2:27} back into the last sum in that expression and subsequently replacing $B_1 \sum_{n_1}$ with the integral $\int k \, dk$, we obtain
\be
Re(iS_{eff}) = -\int d^4x \,\frac{E^2}{24 \pi} \, f_{1/2}(\pi/E, \beta) \,,
\ee
where
\begin{multline}
  \label{eq:f1/2}
	f_{1/2}(z, \beta) = - \frac{6}{\pi^2} \, z \int \, k \, dk\, \Big\{\ln \,(1-e^{- z [(k^2 + 2\beta^2 + 2 \beta \sqrt{k^2 + \beta^2} + m^2]}) \\ + \ln \,(1-e^{- z [k^2 + 2\beta^2 - 2 \beta \sqrt{k^2 + \beta^2} + m^2]})\Big\} \,.
\end{multline}
This is an increasing function of $\beta$, leading to a significant increase in the pair production rates as $\beta$ becomes larger, as readily observed from the plots given in Figure~\ref{fig:fl05}. In the $\beta \rightarrow 0$ limit, we recover twice the Schwinger result, due to equal contributions coming from the isospin up and isospin down degrees of freedom in this limit.

\section{Pair production rates for scalar fields fields on $S^2 \times \mathbb{R}^{1,1}$}

In this section, we compute the pair production rates for spin-$0$ and spin-$\frac{1}{2}$ particles subject to a radial $SU(2) \times U(1)$ magnetic field on the product manifold $S^2 \times \mathbb{R}^{1,1}$. To proceed, we follow a similar line of development as in the previous section and first evaluate the one-loop effective action on $S^2 \times \mathbb{R}^2$ with a radial $SU(2) \times U(1)$ magnetic field on $S^2$ and the usual uniform magnetic field on $\mathbb{R}^2$. The latter will then be Wick rotated to the uniform electric field as before.

\subsection{Spectrum of the gauged Laplacian on $S^2 \times \mathbb{R}^{2}$}

To proceed, we need the spectrum of the gauged Laplacian 
\be
D^2 = \frac{\Vec{\Lambda}^2}{a^2}\,, \quad \Vec{\Lambda} \equiv \Vec{r} \times (\Vec{p} - \Vec{A}) \,,
\ee
on $S^2$, where 
\be
\label{eq:3:1}
\Vec{A} = \Vec{A}_{abelian} + \alpha \frac{\Vec{r}\times \Vec{\sigma}}{a^2} \,,
\ee
$a$ being the radius of $S^2$. Here $\Vec{A}_{abelian}$ is the gauge potential for the Dirac monopole. The associated field strength is 
\be
\label{eq:4:2}
B = \frac{N}{2a^2} + \left (2 \left ( \alpha -\frac{1}{2}\right )^2 -\frac{1}{2} \right) \frac{\Vec{\sigma} \cdot \hat{r}}{a^2} \,,
\ee
where $N \in {\mathbb Z}$ stands for the Dirac monopole charge. 

$D^2$ can be brought into the form 
\be
D^2  = \frac{1}{a^2} \left (\Vec{J}^2 + \frac{1}{4} -  \frac{N^2}{4} + 2 \left (\alpha -\frac{1}{2} \right)^2 -\frac{1}{2} + 2 \left(\alpha - \frac{1}{2}\right)( \Vec{J}\cdot \Vec{\sigma} -\frac{1}{2} + \frac{N}{2} \Vec{\sigma} \cdot \hat{r} )+ \frac{N}{2} \Vec{\sigma} \cdot \hat{r} \right ),
\ee
where $\Vec{J}$ is the total angular momentum, which is given as 
    \begin{align}
        \label{eq:3:3}
        \Vec{J} = \Vec{r} \times (\Vec{p} - \Vec{A}_{abelian}) - \frac{N}{2}\hat{r}  - \frac{\Vec{\sigma}}{2} = \Vec{\Lambda}_{abelian}  + \frac{N}{2}\hat{r} + \frac{\Vec{\sigma}}{2}.
    \end{align}
$\Vec{J}$ involves the contribution of the orbital angular momentum of the charged particle, that of the electromagnetic field generated by the 
particle-Dirac monopole pair as well as the contribution of the $SU(2)$ isospin. Spectrum of $D^2$ is already obtained in \cite{Estienne_2011} and given by
\be
\label{eq:4:6}
\Lambda^\pm_{n_1}(\alpha) = \frac{1}{a^2} \left ( n_1 (N + n_1) + 2 \left ( \alpha -\frac{1}{2} \right )^2 -\frac{1}{2} \pm \sqrt{4 \left (\alpha- \frac{1}{2} \right )^2 (n_1 + N)n_1 + \frac{N^2}{4}} \right ),
\ee
where we have $n_1 = 0,1,2,\dots$ for $\Lambda^+_{n_1}$ and $n_1 = 1,2,\dots$ for $\Lambda^-_{n_1}$ and $N =1,2, \dots$. For convenience, we reproduce the calculation leading to this spectrum in Appendix \ref{sec:C} and also provide the group theoretical details that leads us to the $(2n_1 + N)$-fold degeneracy at each level and branch. Let us also note that $\Lambda^\pm_{n_1}(\alpha) = \Lambda^\pm_{n_1}(1-\alpha)$, which is a consequence of the fact that $B$ in \eqref{eq:4:2} is symmetric under the interchange $\alpha \leftrightarrow 1- \alpha$.
Spectrum of $D^2_{S^2} +D^2_{\mathbb{R}^2} + m^2$ on  $S^2 \times \mathbb{R}^2$ is given as
    \be
        \label{eq:3:5}
        Spec(D^2_{S^2} +D^2_{\mathbb{R}^2} + m^2) = \Lambda_{n_1}^\pm(\alpha) + B_2(2n_2 + 1) + m^2 \,,
    \ee
where $n_1$ is as given above and $n_2 =0,1,2,\dots$.

\subsection{Pair production rates}

The Euclidean one-loop effective action takes the form,
    \begin{multline}
        \label{eq:3:6}
        \Gamma_E = - \frac{1}{16\pi^2a^2}\int d^2x \int d\Omega_2 \int \frac{ds}{s} \frac{B_2}{sinh\,sB_2} \\
        \times \Bigg\lbrack \sum_{n_1 = 0}^{\infty} (2n_1 + N) \,e^{- \frac{s}{a^2} \left [ n_1 (N + n_1) + 2 \left (\alpha -\frac{1}{2} \right)^2 -\frac{1}{2}  + \sqrt{4 \left  (\alpha- \frac{1}{2} \right)^2 (n_1 + N)\, n_1 + \frac{N^2}{4} } \right ]} \\
        + \sum_{n_1 = 1}^{\infty} (2n_1 + N)\, e^{- \frac{s}{a^2} \left \lbrack n_1 (N + n_1) + 2 \left (\alpha -\frac{1}{2} \right)^2 -\frac{1}{2} - \sqrt{4 \left (\alpha- \frac{1}{2} \right)^2 (n_1 + N)\, n_1 + \frac{N^2}{4} } \right \rbrack} \Bigg\rbrack \,.
    \end{multline}
Evaluating the integral over $s$ and performing the Wick rotation $\mathbb{R}^2 \rightarrow \mathbb{R}^{1,1}$, $B_2 \rightarrow -iE$, we find 
    \begin{align}
        \label{eq:3:7}
        Re(iS_{eff}) = -\int d^2x \int d\Omega_2\, \frac{E^2}{16\pi^3} \beta_0(\omega, \alpha, N),
    \end{align}
where $\beta_0(\omega, \alpha, N)$ is given as
    \begin{multline}
        \beta_0(\omega, \alpha, N) = \omega \Bigg \lbrack \sum_{n_1 = 0}^{\infty} (2n_1 + N)\, \ln \,(1+e^{-\omega \left \lbrack n_1 (N + n_1) + 2 \left (\alpha -\frac{1}{2} \right)^2 -\frac{1}{2}  + \sqrt{4 \left  (\alpha- \frac{1}{2} \right)^2 n_1 (n_1 + N) + \frac{N^2}{4}} \right \rbrack}) \nonumber \\
        +\sum_{n_1 = 1}^{\infty} (2n_1 + N)\, \ln \,(1+e^{-\omega \left \lbrack n_1 (N + n_1) + 2 \left (\alpha -\frac{1}{2} \right)^2 -\frac{1}{2}  - \sqrt{4 \left  (\alpha - \frac{1}{2} \right)^2 n_1(n_1 + N) + \frac{N^2}{4}} \right \rbrack}) \Bigg \rbrack \,.
    \end{multline}
In above, we have defined $\omega := \pi /Ea^2$. Let us note immediately that $\beta_0(\omega, \alpha, N) = \beta_0(\omega, 1-\alpha, N)$, which is clearly a consequence of the same symmetry that we have already noted for the spectrum in \eqref{eq:4:6}.     

In order to compare the pair production rates on this geometry to that on $\mathbb{R}^{3,1}$, we first evaluate the limit $S^2 \rightarrow \mathbb{R}^2$. To compute this, we take $a \rightarrow \infty$ and also $N, \alpha  \rightarrow \infty$, while keeping both $\frac{N}{2 a^2}$ and $\frac{\alpha^2}{a^2}$ constant. Since the definition of $\omega$ already contains the term $1/a^2$, we can keep $\omega N$ (or similar combinations) as is. We find
\begin{multline}
\label{eq:3:10}
\beta_0^{flat}(\omega, \alpha, N) = \omega N \Bigg\lbrack \sum_{n_1 = 0}^{\infty}\, \ln \,(1+e^{-\omega \lbrack n_1 N + 2 \alpha^2 + \sqrt{4 \alpha^2 N n_1 + \frac{N^2}{4}} \rbrack } ) \\ 
+ \sum_{n_1 = 1}^{\infty}\, \ln \,(1+e^{-\omega \lbrack n_1 N + 2 \alpha^2 - \sqrt{4 \alpha^2 N n_1 + \frac{N^2}{4}} \rbrack }) \Bigg \rbrack \,.
\end{multline}
In what follows, for notational ease, we write $\Lambda_{n_1}^{\pm \, flat} := \frac{1}{a^2} ( n_1 N + 2 \alpha^2 \pm \sqrt{4 \alpha^2 N n_1 + N^2/4})$. We observe that, the latter identifies with $\eqref{eq:2:9}$ and $\beta_0^{flat}(\omega) \rightarrow \frac{\pi^2}{6} f_0(y, \beta^\prime)$ once we set $B_1 = \frac{N}{2a^2}$ and $\beta^{2} = \frac{\alpha^2}{a^2}$ with $\beta^{\prime 2} = \frac{\beta^2}{B_1}$.

\begin{figure}[h!]
    \centering
    \begin{subfigure}[b]{0.49\textwidth}
    	\includegraphics[width=\textwidth]{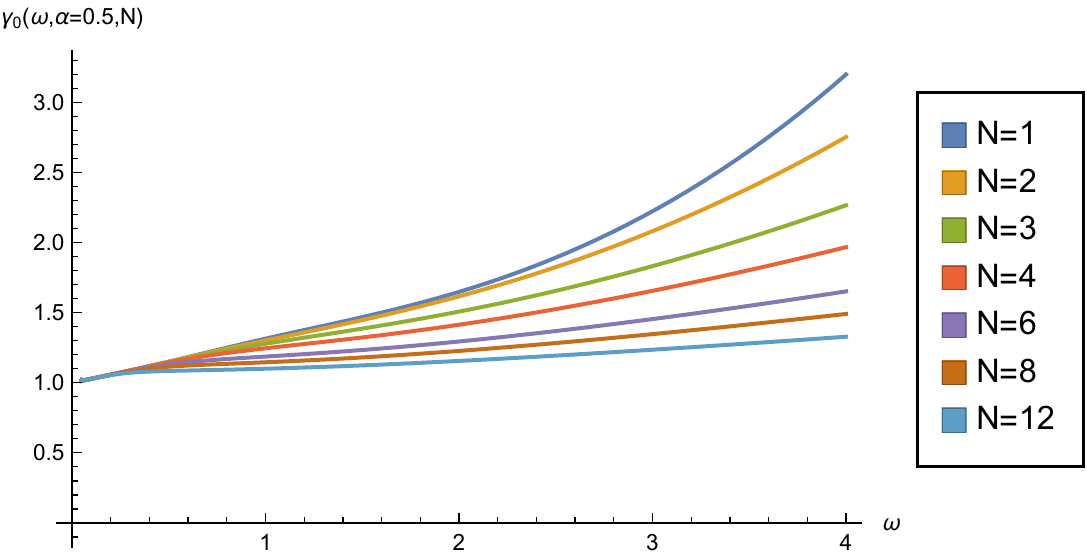}
        \caption{\label{fig:3:1} $\alpha = 1/2$}
    \end{subfigure}
    \hfill
    \begin{subfigure}[b]{0.49\textwidth}
    	\includegraphics[width=\textwidth]{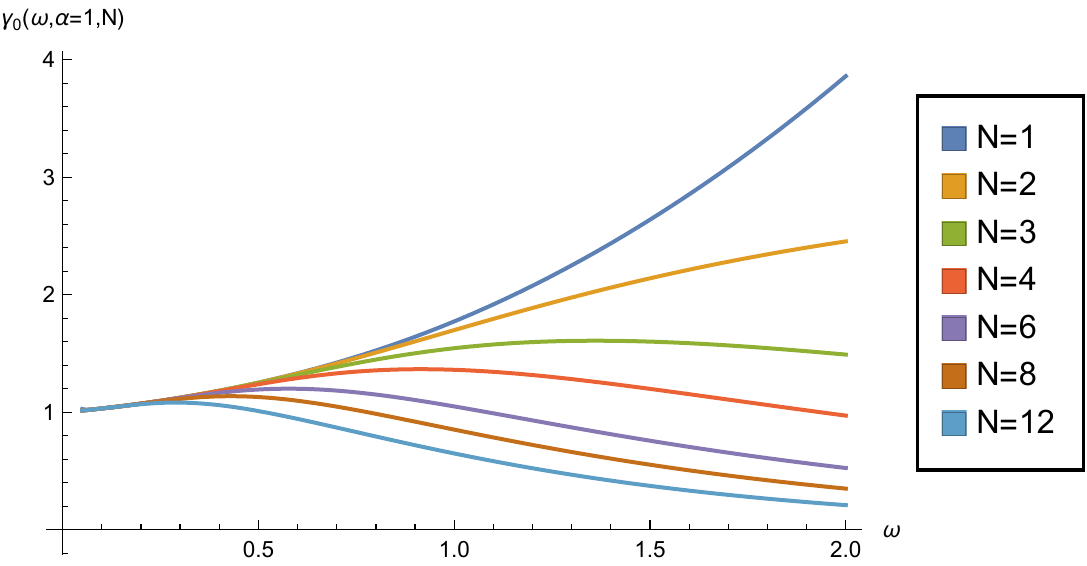}
    	\caption{\label{fig:3:2} $\alpha = 1$}
    \end{subfigure}
    \hfill
	\centering
    \begin{subfigure}[b]{0.49\textwidth}
		\includegraphics[width=\textwidth]{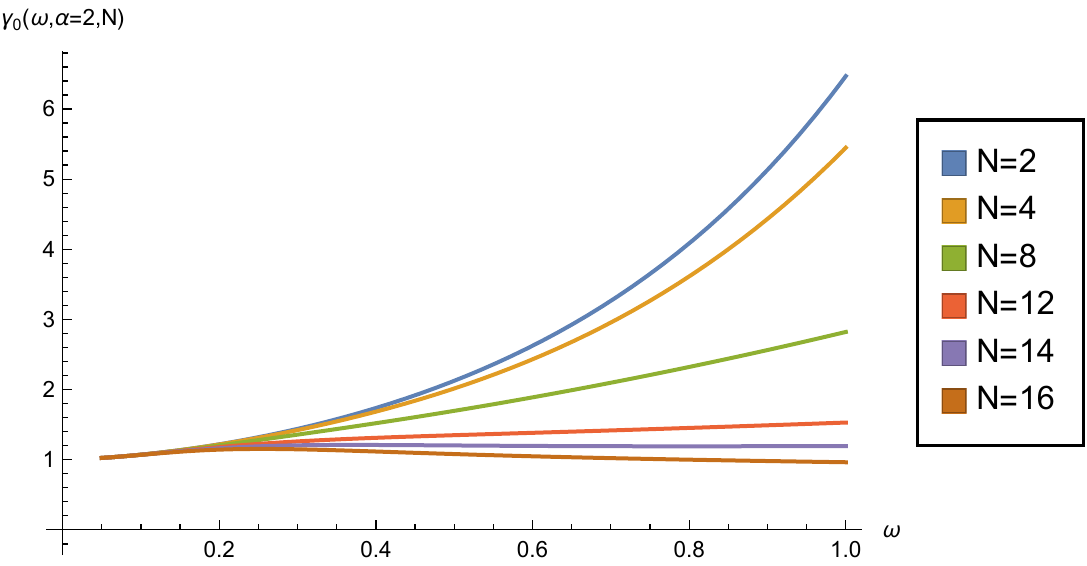}
		\caption{\label{fig:3:3} $\alpha = 2$}
    \end{subfigure} \\
    \hfill
	\centering
	\begin{subfigure}[b]{0.49\textwidth}
		\includegraphics[width=\textwidth]{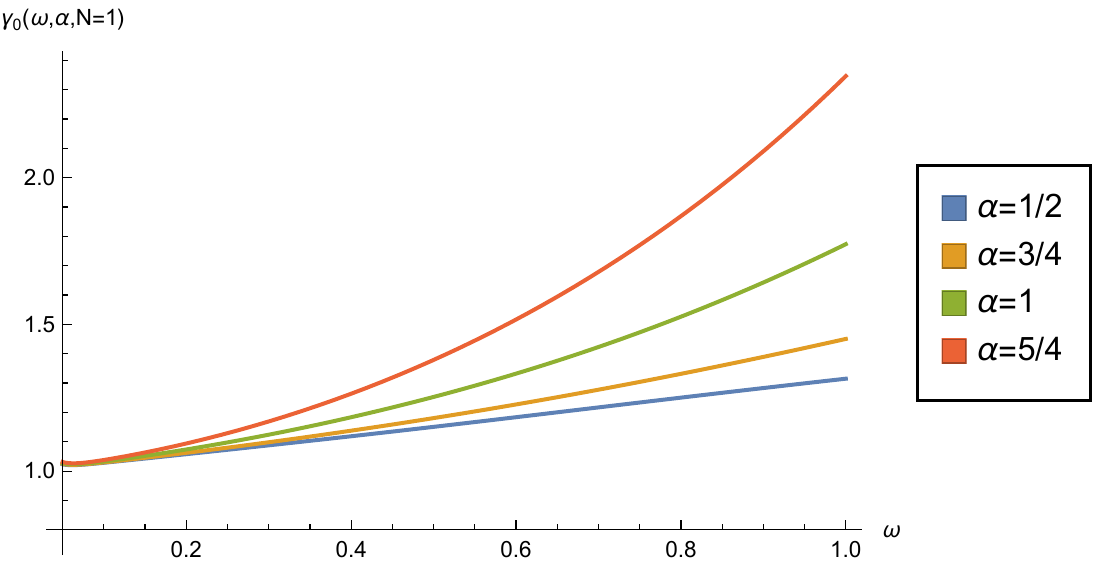}
	\caption{\label{fig:3:5a} $N = 1$}
	\end{subfigure}
	\hfill
	\begin{subfigure}[b]{0.49\textwidth}
		\includegraphics[width=\textwidth]{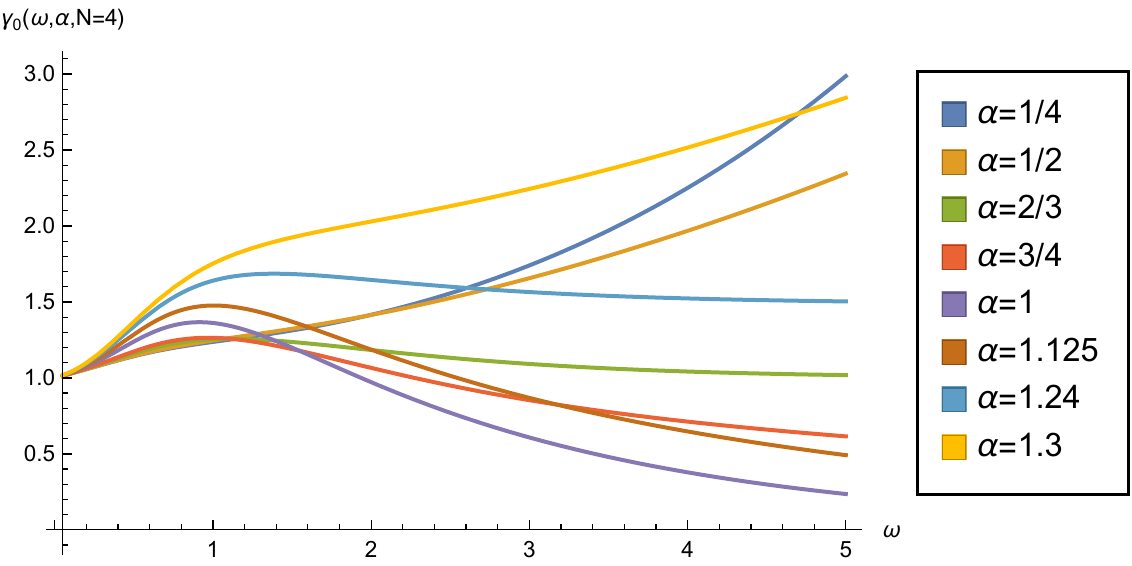}
	\caption{\label{fig:3:5b} $N = 4$}
	\end{subfigure}
  \caption{\label{fig:main3} Relative pair production rates (RPP) $\gamma_0(\omega, \alpha, N)$ as a function of $\omega$.}
\end{figure}

The ratio $\gamma_0(\omega, \alpha, N) \equiv \frac{\beta_0(\omega , \alpha, N)}{\beta_0^{flat}(\omega,\alpha, N)}$ is useful to compare the pair production rates on $\mathbb{S}^2 \times \mathbb{R}^{1,1}$ to that on $\mathbb{R}^{3,1}$\footnote{Let us note that $\gamma_0(\omega, \alpha, N) \neq \gamma_0(\omega, 1-\alpha, N)$, since the denominator $\gamma_0(\omega, \alpha, N)$ is not symmetric under $\alpha \leftrightarrow 1-\alpha$. In particular, we may examine the relative pair production rates for $\alpha < 0$, using only $\alpha >0$ by forming the ratio $\tilde{\gamma}_0(\omega, \alpha, N) :\equiv \frac{\beta_0(\omega , 1- \alpha, N)}{\beta_0^{flat}(\omega , \alpha, N)}$.}. We first inspect the profile of $\gamma_0(\omega, \alpha, N)$ from the Figures~\ref{fig:3:1}, \ref{fig:3:2} and  \ref{fig:3:3} as we vary $\omega$ at  $\alpha=1/2, 1, 2$ at several $N$ values.
To elaborate on the physical meaning of the profiles of $\gamma_0(\omega, \alpha, N)$ and hence the relative pair production rates, we essentially need to consider the hierarchy of only the lowest lying energy levels for the spherical and flat geometries, which are $\Lambda_{0}^+$, $\Lambda_{1}^-$ and $\Lambda_{0}^{+ \, flat}$, $\Lambda_{1}^{- \, flat}$, respectively. In addition, we observe that the ratio of degeneracies of the curved to flat case goes as $\frac{N+2n_1}{N} = 1 + \frac{2 n_1}{N}$, which may be understood as being due to the contribution of the curvature to the density of states, which is $\frac{B_1}{2 \pi} + \frac{2 n_1}{4 \pi a^2}$ as opposed to only $\frac{B_1}{2 \pi}$ for the flat case \cite{Kurkcuoglu}, indicating improved availability of quantum states at any energy level except at $n_1 = 0$. Under the circumstances, that will be described shortly, the contribution of the second term in $\beta_0(\omega, \alpha, N)$ becomes important,  $\gamma_0(\omega, \alpha, N)$ includes the factor $1 + \frac{2}{N}$, whose enhancing effect tends to decrease with increasing $N$. These features completely govern the behavior of $\gamma_0(\omega, \alpha, N)$, with $\gamma_0(\omega, \alpha, N) > 1$ indicating relatively higher and $\gamma_0(\omega, \alpha, N) < 1$ lower pair production amplitudes.

\begin{figure}[h!]
	\centering
	\begin{subfigure}[b]{0.49\textwidth}
		\includegraphics[width=\textwidth]{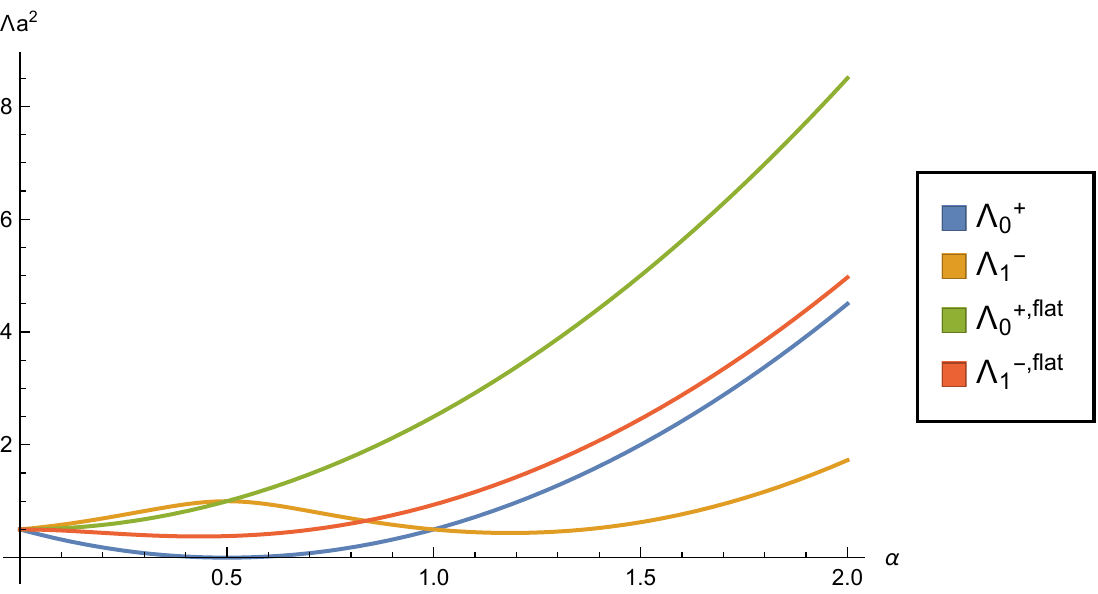}
	\caption{\label{fig:3:4a} $N = 1$}
	\end{subfigure}
	\hfill
	\begin{subfigure}[b]{0.49\textwidth}
		\includegraphics[width=\textwidth]{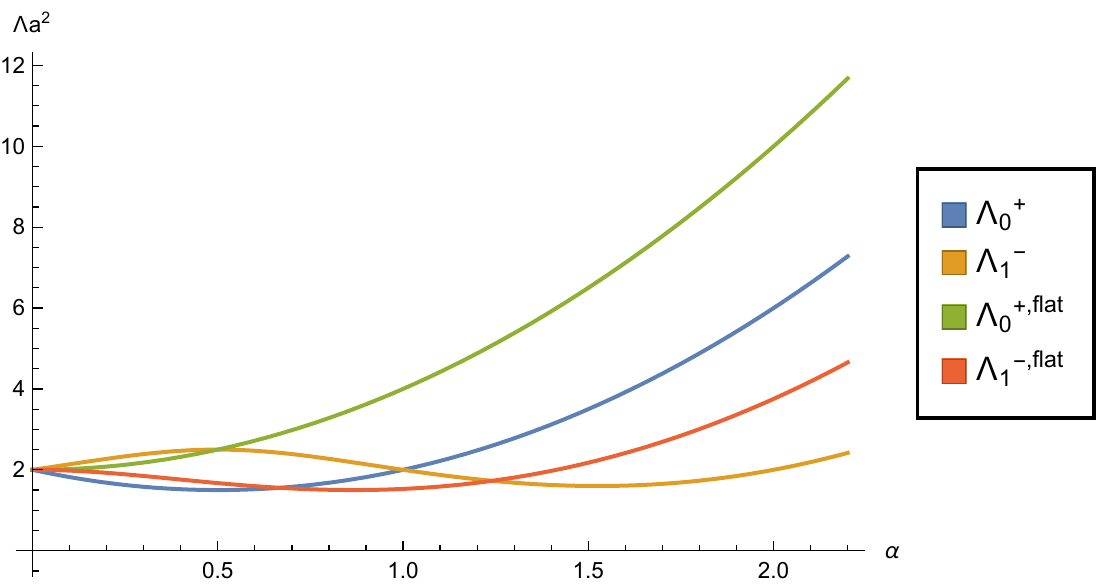}
	\caption{\label{fig:3:4b} $N = 4$}
	\end{subfigure}
  \caption{\label{fig:main4} $\Lambda_{0}^+$, $\Lambda_{1}^-$, $\Lambda_{0}^{+ \, flat}$ and $\Lambda_{1}^{- \, flat}$ as a function of $\alpha$.}
\end{figure}

We note that, $\Lambda_{0}^+ < \Lambda_{0}^{+ \, flat}$ (in fact, $\Lambda_{n_1}^+ < \Lambda_{n_1}^{+ \, flat}$ for all $n_1 =0,1,2, \dots$) for all values of $\alpha$ and $\Lambda_{0}^+$ and $\Lambda_{1}^-$ intersect at $\alpha = 0$ and $\alpha = 1$ with $\Lambda_{0}^+ < \Lambda_{1}^-$ for $0 < \alpha < 1$ and $\Lambda_{0}^+ > \Lambda_{1}^-$ for $\alpha > 0$. Furthermore, we have $\Lambda_{0}^+$ intersecting $\Lambda_{1}^{- \, flat}$ at $\alpha_c^{(1)} = \frac{1}{2} \frac{N}{N-1}$ for $N \geq 2$, with 
$\Lambda_{0}^+ < \Lambda_{1}^{- \, flat}$ for $\alpha < \alpha_c^{(1)}$ and $\Lambda_{0}^+ > \Lambda_{1}^{- \, flat}$ for $\alpha > \alpha_c^{(1)}$, while for $N=1$,  $\alpha_c^{(1)} =0$, since $\Lambda_{0}^+ = \frac{1}{2} = \Lambda_{1}^{- \, flat}$ at $\alpha =0$ and $\Lambda_{0}^+$ remains always less than $\Lambda_{1}^{- \, flat}$ for $\alpha >0$.  
Finally, $\Lambda_{1}^-$ and $\Lambda_{1}^{- \, flat}$ intersect at $\alpha = 0$ and at $\alpha = \alpha_c^{(2)} > 1$, with $\Lambda_{1}^- > \Lambda_{1}^{- \, flat}$ for $\alpha < \alpha_c^{(2)}$ and $\Lambda_{1}^- < \Lambda_{1}^{- \, flat}$ for $\alpha > \alpha_c^{(2)}$, and where $\alpha_c^{(2)}$ is determined by the unique solution of $\frac{1}{N} = \frac{1}{16} \frac{1}{(\alpha- \frac{1}{2})^2} +  \frac{1}{8} \frac{1}{(\alpha- \frac{1}{2}) \alpha}$ for $\alpha  \geq 1$ at a given value of $N = 2,3\dots$, except at $N=1$, for which $\alpha_c^{(2)} = 0.836$. At $N=2$ we have $\alpha_c^{(2)} = 1$ and it gradually increases from this value with increasing $N$. For instance $\alpha_c^{(2)} = 1.24$ at $N=4$. We also note that at $N=2$, $\alpha_c^{(1)} = \alpha_c^{(2)} = 1$. Plots of these energy eigenvalues as a function of $\alpha$ at $N=1$ and $N=4$ are provided in Figures~\ref{fig:3:4a} and \ref{fig:3:4b} and clearly illustrate all the features which we have explained above.
Putting these facts together, for $0 < \alpha < \alpha_c^{(1)}$, we have $\Lambda_{0}^+$ as the lowest energy eigenvalue, so these states are energetically favored to be filled by the produced pairs in comparison with $\Lambda_{1}^-$ as well as $\Lambda_{0}^{+ \, flat}$, $\Lambda_{1}^{- \, flat}$ and we therefore see from Figures~\ref{fig:3:1}, \ref{fig:3:2} and  \ref{fig:3:3} that $\gamma_0(\omega, \alpha, N)$ increases above its starting value $1$ and hence we conclude that there is an increase in the relative pair production probabilities. As the Dirac monopole charge $N$ is increased, $\Lambda_{0}^+$ (as well as $\Lambda_{1}^-$, $\Lambda_{0}^{+ \, flat}$, $\Lambda_{1}^{- \, flat}$) also increases and therefore the rate of increase in $\gamma_0(\omega, \alpha, N)$ slows down and it gradually tends to $1$ as $N \rightarrow \infty$, indicating that the pair production rates tend to converge toward what is found for the flat case. For $\alpha_c^{(1)} < \alpha < \alpha_c^{(2)}$, we have $\Lambda_{1}^{- \, flat}$ as the lowest among these four energies and we eventually see that $\gamma_0(\omega, \alpha, N)$ goes below the value $1$ (in fact tend to zero at large $\omega$) as expected. Nevertheless, within a narrow range of values of $\omega$, approximately $0 < \omega \lesssim 1$, we also observe that $\gamma_0(\omega, \alpha, N) \geq 1$. The latter can be explained being due to approximately equal contributions from $\Lambda_{0}^+$ and $\Lambda_{1}^-$ to the numerator beating those of $\Lambda_{0}^{+ \, flat}$ and $\Lambda_{1}^{- \, flat}$ to the denominator\footnote{For instance, we may estimate $\gamma_0(\omega = 1/2, \alpha, N)$ at $N= 4$ and $\alpha = 1$. We have, in units of $\frac{1}{a^2}$, $\Lambda_{0}^+ = \Lambda_{1}^- = 2$, while $\Lambda_{0}^{+ \, flat} = 4$, $\Lambda_{1}^{- \, flat} = 1.528$ and 
\be
\gamma_0(1/2, \alpha, N) \approx \frac{5}{4}  \frac{2 \ln (1 + e^{-1})}{\ln (1 + e^{-2}) + \ln (1 + e^{-0.85})} \approx 1.537 \nn
\ee
while we have $\gamma_0(2,, \alpha, N) \approx \frac{5}{4} \frac{2 \ln (1 + e^{-4})}{\ln (1 + e^{-8}) + \ln (1 + e^{-3.056})} \approx 0.979$.}. Finally, for $\alpha > \alpha_c^{(2)}$, we have $\Lambda_{1}^-$ become the lowest among these four energies leading to $\gamma_0(\omega, \alpha, N) > 1$ and a significant increase in the relative pair production amplitudes. We may note that at $\alpha = \alpha_c^{(2)}$, $\gamma_0(\omega, \alpha, N)$ converges to the ratio of the degeneracies $1 + \frac{2}{N}$, except at $N=1$. In the latter case, two facts: $\Lambda_{0}^{+}$ always remaining less than $\Lambda_{1}^{-, flat}$ and $\Lambda_{0}^{+, flat}$ being almost always the largest among the lowest lying energies are sufficient to ensure that $\gamma_0(\omega, \alpha, N=1) > 1$ with a positive rate of change with increasing $\alpha$.  We may illustrate all of these conclusions by inspecting the profile of $\gamma_0(\omega, \alpha, N)$ at a fixed value of $N$ and several distinct values of $\alpha$. In Figures~\ref{fig:3:5a} and \ref{fig:3:5b}, we give the plots of $\gamma_0(\omega, \alpha, N)$ at $N=1$ and $N=4$, for which $(\alpha_c^{(1)}, \alpha_c^{(2)}) = (0, 0.836)$ and $(\alpha_c^{(1)}, \alpha_c^{(2)}) = (1, 1.24)$, respectively. We see that for $\omega \gtrsim 1$, $\gamma_0(\omega, \alpha, N=4)$ is above the value $1$ for all $0 < \alpha < \frac{2}{3}$ albeit decreasing with increasing $\alpha$ and it goes below $1$ for $\frac{2}{3} < \alpha < 1.24$\footnote{Even within this interval there is a sub-hierarchy: $\gamma_0(\omega, \alpha, N=4)$ decreasing with increasing $\alpha$ up to $\alpha \approx 1$ and increasing for $\alpha > 1$ while it remains below the value $1$. This is also understood as being due the counterplay among $\Lambda_{0}^+$, $\Lambda_{1}^-$ and $\Lambda_{1}^{- \, flat}$ all together.}. At $\alpha = 1.24$, $\gamma_0(\omega, \alpha, N=4) \rightarrow \frac{3}{2}$ for large $\omega$, while it increases for $\alpha > 1.24$.  

\begin{figure}[h!]
	 	\centering
 	\begin{subfigure}[b]{0.49\textwidth}
 		\includegraphics[width=\textwidth]{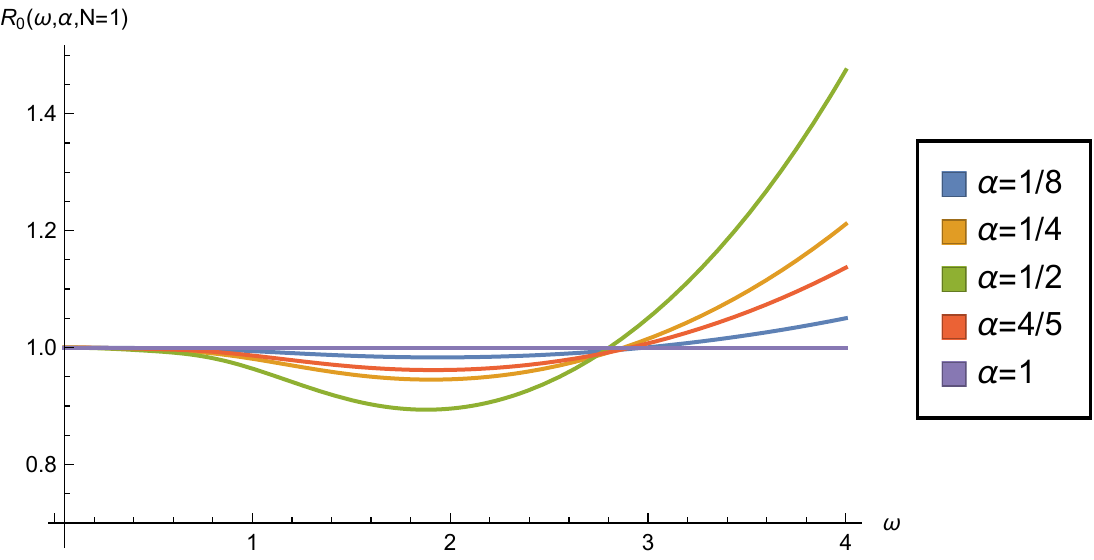}
 		\caption{\label{fig:3:6a} $N = 1$}
 	\end{subfigure}
 	\hfill
 	\begin{subfigure}[b]{0.49\textwidth}
 		\includegraphics[width=\textwidth]{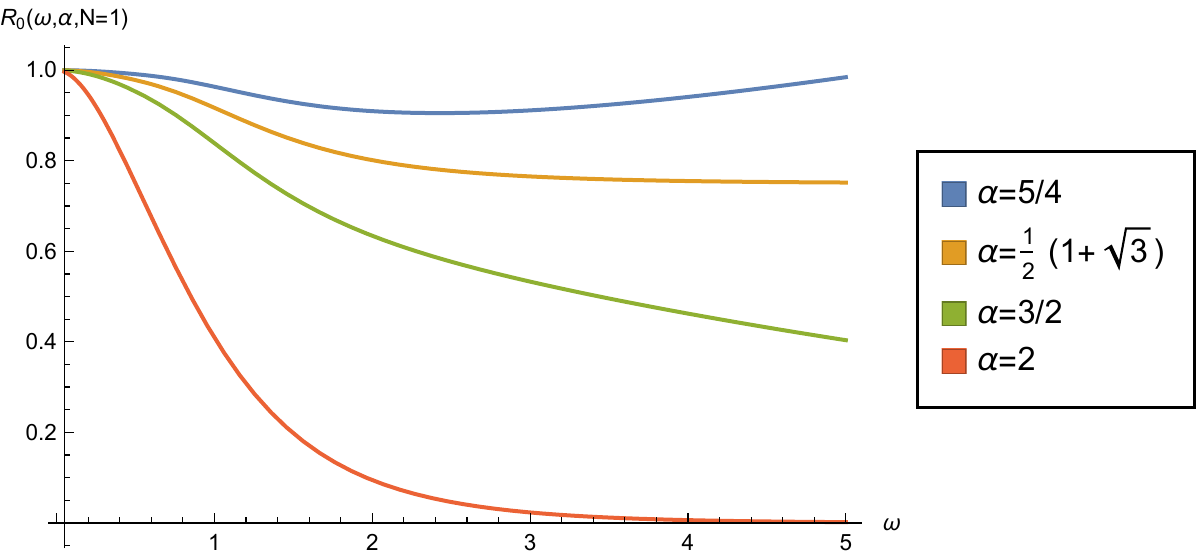}
 		\caption{\label{fig:3:6b} $N = 1$}
 	\end{subfigure}
	\centering
	\begin{subfigure}[b]{0.49\textwidth}
		\includegraphics[width=\textwidth]{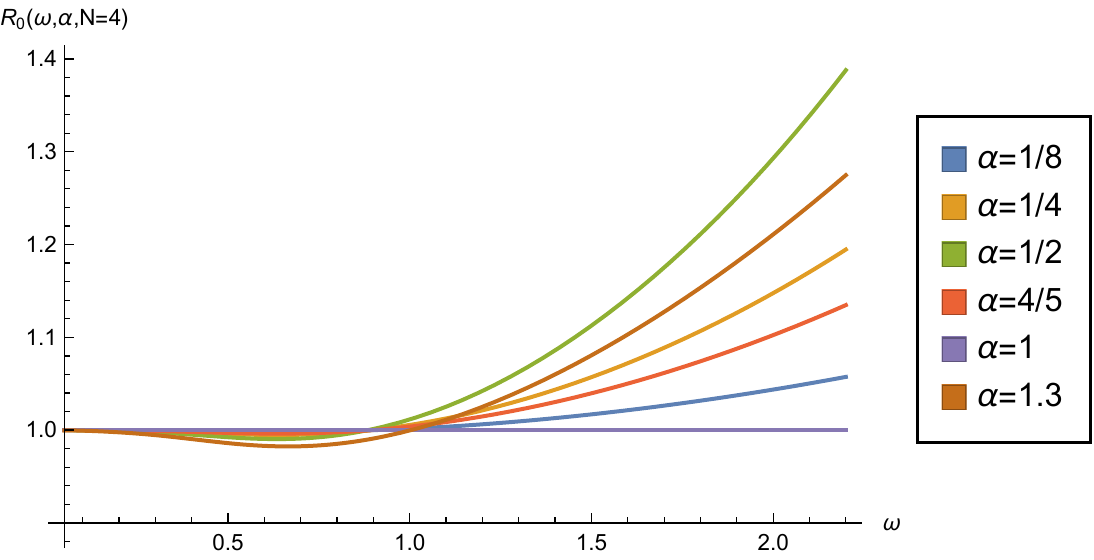}
		\caption{\label{fig:3:7a} $N = 4$}
	\end{subfigure}
	\hfill
	\begin{subfigure}[b]{0.49\textwidth}
		\includegraphics[width=\textwidth]{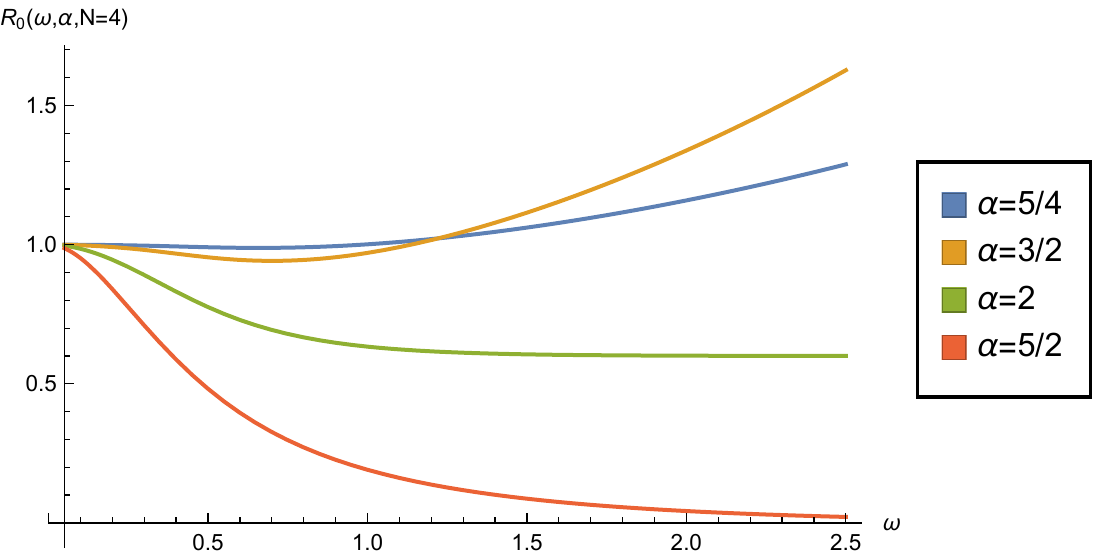}
		\caption{\label{fig:3:7b} $N = 4$}
	\end{subfigure}
  \caption{\label{fig:main5} $R_0(\omega, \alpha, N)$ as a function of $\omega$.}
\end{figure}

We may also compare the pair production rates with and without the non-abelian magnetic fields. For this purpose, we define $R_0(\omega, \alpha, N) \equiv \frac{\beta_0(\omega, \alpha, N)}{\beta_0(\omega, 0, N)}$. Plots of $R_0(\omega, \alpha, N)$ for $N=1$ and $N=4$ are provided in the figures \ref{fig:3:6a}, \ref{fig:3:6b} and \ref{fig:3:7a}, \ref{fig:3:7b}, respectively. To explain the physics underneath, we first note that in units of $\frac{1}{a^2}$, $\Lambda_{n_1}^\pm (\alpha = 0) = \Lambda_{n_1}^\pm (\alpha = 1)$ and $\Lambda_{0}^+ (\alpha = 0,1) = \frac{N}{2} = \Lambda_{1}^- (\alpha = 0,1)$. We also have $\Lambda_{0}^+ \leq \frac{N}{2}$, $\Lambda_{1}^- \geq \frac{N}{2}$ for $0 \leq \alpha \leq 1$. At $\alpha = \frac{1}{2}$, $\Lambda_{0}^+$ has a global minimum (taking the value $\frac{N-1}{2}$) and $\Lambda_{1}^-$ has a local maximum (taking the value $\frac{N+1}{2}$). For $\alpha >1$, $\Lambda_{0}^+ > \frac{N}{2}$ and it increases monotonically with $\alpha$, whereas $\Lambda_{1}^-$ continues to decrease further, makes a minimum\footnote{This is at $\alpha = \frac{1}{2} + \frac{1}{4} \sqrt{\frac{(N+2)(3N+2)}{N+1}}$, although it is not relevant for our purposes.} and monotonically increases from thereon attaining the value $\frac{N}{2}$ again at $\alpha_c^{(3)} : = \frac{1}{2} (1 + \sqrt{2N+1})$. All of these features can be visibly recognized from Figure~\ref{fig:3:4b}. Keeping these facts in mind, we first note that $R_0(\omega, \alpha, N) = R_0(\omega, 1- \alpha, N)$. Next, we observe that the large $\omega$ behavior of $R_0(\omega, \gamma, N)$ is described by
\begin{align}
\label{eq:3:12}
R_0(\omega, \alpha, N) \xrightarrow{\omega \rightarrow \infty}
\begin{cases}
\infty & \alpha < \alpha_c^{(3)} \,, \quad \alpha \neq 0 \,, 1 \,, \\ 
\frac{1}{2} \frac{N+2}{N+1} & \alpha = \alpha_c^{(3)} \,, \\
0 & \alpha > \alpha_c^{(3)} \,.
\end{cases}
\end{align}
where the limiting value $\frac{1}{2} \frac{N+2}{N+1}$ at $\alpha_c^{(3)}$ is readily seen to be simply the ratio of the degeneracy $N+2$ of the energy $\Lambda_1^-(\alpha_c^{(3)}) = \frac{N}{2}$ to the sum of the degeneracies $N$ and $N+2$ of $\Lambda_0^+(\alpha = 0) = \frac{N}{2} = \Lambda_1^-(\alpha = 0)$, since the single former and the two latter energy levels yield an equivalent as well as the most dominant contribution $\ln (1 + e^{- \omega \frac{N}{2}} )$ to the numerator and denominator of $R_0(\omega, \alpha, N)$, respectively. At large $N$, we have $R_0(\omega, \alpha_c^{(3)}, N) \rightarrow \frac{1}{2} + \frac{1}{2 N}$, thus it converges to $F_0(y,\beta^\prime = 1) \approx \frac{1}{2}$ that we have found for the flat geometry while $\frac{ \alpha_c^{(3)} }{ a \sqrt{\frac{N}{2 a^2}}} \approx 1 \rightarrow \beta^\prime_c = 1$ as $a, N \rightarrow \infty$ (cf. Figure \ref{fig:2:2} and the ensuing discussion). More generally, we have $R_0(\omega, \alpha, N) \rightarrow F_0(y,\beta^\prime)$ as $a, \alpha, N \rightarrow \infty$ with $B_1 = \frac{N}{2a^2}$, $\beta^{2} = \frac{\alpha^2}{a^2}$ and $\beta^{\prime 2} = \frac{\beta^2}{B_1}$. We also see that within $0 < \alpha < 1 $, $R_0(\omega, \alpha, N) < 1$ around $\omega \approx \frac{2}{N}$ and increases above $1$ afterwards. This is a consequence of the counterplay between the contributions due to $\Lambda_{0}^+ \leq \frac{N}{2}$ and $\Lambda_{1}^- \geq \frac{N}{2}$ on one side and those of $\Lambda_{0}^+ = \frac{N}{2} = \Lambda_{1}^-$ at $\alpha = 0$ on the other. Since the degeneracies are $N$ for $\Lambda_{0}^+$ and $N+2$ for $\Lambda_{1}^-$, the latter being slightly larger, tilts the $R_0(\omega, \alpha, N)$ value below $1$. Let us also that, the rate of change of $R_0(\omega, \alpha, N)$ for both of these regimes (i.e. $R_0(\omega, \alpha, N) < 1$ and $R_0(\omega, \alpha, N) > 1$)  is increasing for $0 < \alpha < \frac{1}{2}$ and decreasing for $\frac{1}{2} < \alpha  < 1$ with a maximum at $\alpha = \frac{1}{2}$, while the rate of change of $R_0(\omega, \alpha, N)$ with respect to $\omega$, becoming slower for $R_0(\omega, \alpha, N) < 1$ and faster for $R_0(\omega, \alpha, N) > 1$ with increasing $N$. For $1 < \alpha < \alpha_c^{(3)}$, although $R_0(\omega, \alpha, N) \rightarrow \infty$ as $\omega \rightarrow \infty$, interval over which $R_0(\omega, \alpha, N) < 1$ stretches longer with increasing $\alpha$ and eventually making $R_0(\omega, \alpha, N)$ converge to $\frac{1}{2} \frac{N+2}{N+1}$ at $\alpha = \alpha_c^{(3)}$. Finally, for $\alpha > \alpha_c^{(3)}$ and $R_0(\omega, \alpha, N)$ rapidly approaches to zero since both $\Lambda_{0}^+$ and $\Lambda_{1}^-$ are larger than the $\frac{N}{2}$ value they take at $\alpha = 0$. 

\begin{figure}[h!]
	\centering
	\begin{subfigure}[b]{0.49\textwidth}
		\includegraphics[width=\textwidth]{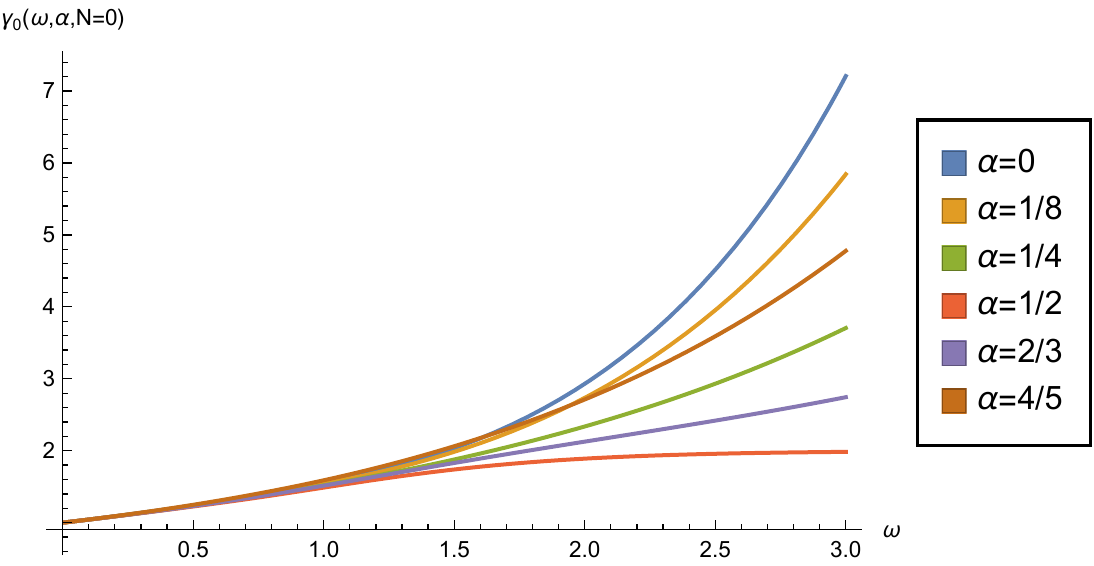}
		\caption{\label{fig:3:8a} $\gamma_0(\omega, \alpha)\,, N = 0$.}
	\end{subfigure}
	\hfill
	\begin{subfigure}[b]{0.49\textwidth}
		\includegraphics[width=\textwidth]{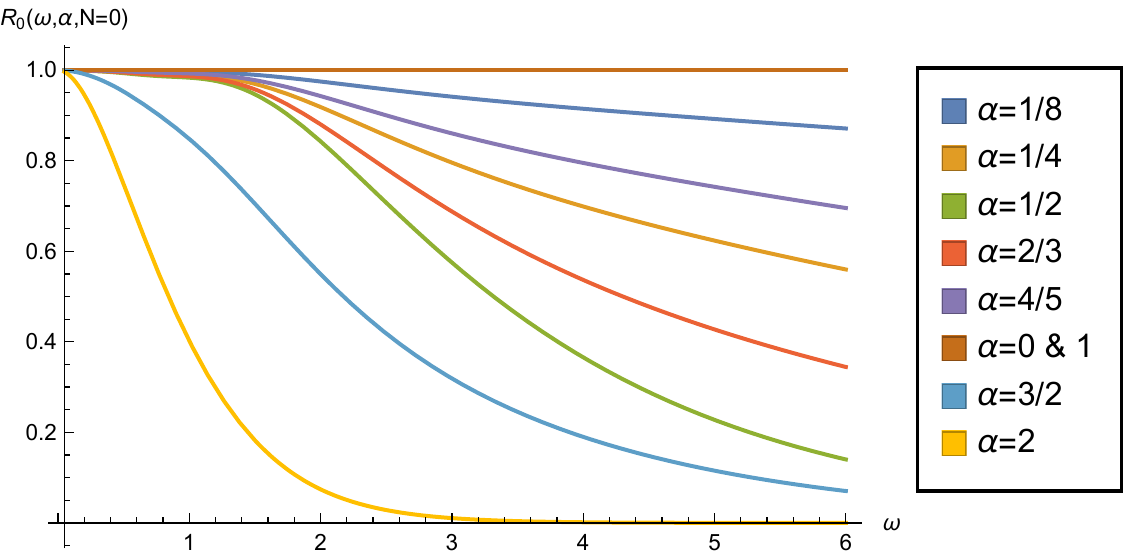}
		\caption{\label{fig:3:8b} $R_0(\omega, \alpha)\,, N = 0$.}
	\end{subfigure}
	\centering
	\begin{subfigure}[b]{0.49\textwidth}
		\includegraphics[width=\textwidth]{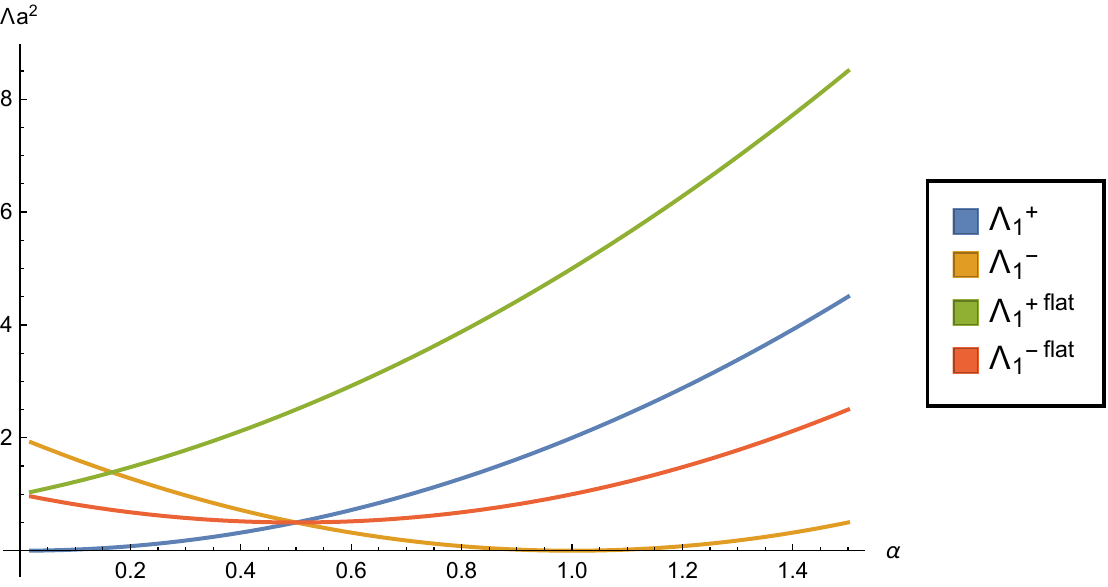}
		\caption{\label{fig:3:8c} Lowest lying energy eigenvalues as a function of $\alpha$.}
	\end{subfigure}
  \caption{\label{fig:main6}, $\gamma_0(\omega, \alpha)$, $R_0(\omega, \alpha)$ at $N=0$. }
\end{figure}

For the case of vanishing abelian monopole charge the spectrum of $D^2$ on $S^2$ is given in \eqref{eq:C:16} in Appendix C. Using the latter, we find that
\be
Re(iS_{eff}) = - \int d^2x \int d\Omega_2 \, \frac{E^2}{8\pi^3} \, \beta_0(\omega, \alpha) \,,
\ee
where
\be
\label{eq:beta0}
\beta_0(\omega, \alpha) = \omega \sum_{l = 1}^\infty l \,\Big\{ \ln (1+e^{-\omega[l^2 + 2 \left ( \alpha -\frac{1}{2} \right )^2 -\frac{1}{2} + 2 \left ( \alpha - \frac{1}{2} \right) \, l]}) +\ln (1+e^{-\omega[l^2 + 2 \left ( \alpha -\frac{1}{2} \right )^2 -\frac{1}{2} - 2 \left ( \alpha - \frac{1}{2} \right ) \, l ]})\Big\}.
\ee
We may note that for $\alpha = 0$, the sum over the second term in \eqref{eq:beta0} starts from $l=0$, since the spectrum of $\Lambda_{l, N=0}^-$ starts with $l = 0$ as remarked in Appendix \ref{sec:C}. Nevertheless, the expression for $\beta_0(\omega, \alpha)$ holds the same in this case too, due to the factor $l$ in the summand which ensures that the $l=0$ term of the sum yields a vanishing contribution. Let us note that, we also have $\beta_0(\omega, \alpha) = \beta_0(\omega, 1-\alpha)$, implying in particular that $\beta_0(\omega, 0) = \beta_0(\omega, 1)$. In the limit $a \rightarrow \infty$, keeping $\frac{\alpha}{a}$ fixed we may write $\beta_0^{flat}(\omega, \alpha) = \omega \sum_{l = 1}^\infty l \,\big\{ \ln (1+e^{-\omega[(l +\alpha)^2 + \alpha^2]}) +\ln (1+e^{-\omega[ (l - \alpha)^2 + \alpha^2] })\big\}$. In this limit, setting $\frac{l}{a} \rightarrow k$, $\frac{\alpha}{a} \rightarrow \beta$ and replacing $ \frac{1}{a} \sum_{l} \frac{l}{a}$ with $\int k \, dk$ we find that $\beta_0^{flat}(\omega) \rightarrow \frac{\pi^2}{12} f_0(\frac{\pi}{E},\beta, m=0)$, where the last factor on the r.h.s. is given in \eqref{eq:beta0flat}. Below, we plot the profiles of $\gamma_0(\omega, \alpha) \equiv \frac{\beta_0(\omega,\alpha)}{\beta_0^{flat}(\omega, \alpha)}$ and $R_0(\omega, \alpha) \equiv \frac{\beta_0(\omega, \alpha)}{\beta_0(\omega, 0)}$.  From the Figures~\ref{fig:3:8a} and \ref{fig:3:8b}, we see that the pair production amplitude is larger compared to the flat case at all values of $\alpha$ and approaches to twice the latter as $\alpha \rightarrow \frac{1}{2}$ either from below or above. $\gamma_0(\omega, \alpha)$ decreases toward this value with increasing $\alpha$ for $0 < \alpha \leq \frac{1}{2}$ and decreasing $\alpha$ for $\alpha > 1$, as is readily inferred from the hierarchy among the lowest lying energy states for the curved and the flat case. From the Figure~\ref{fig:3:8c}, we see that $ \Lambda^+_{l=1}(\alpha)$ is the lowest energy at $\alpha =0$, and it monotonically increases with $\alpha$ passing through the value $\frac{1}{2}$ at $\alpha = \frac{1}{2}$ while both $\Lambda^-_{l=1}(\alpha)$ and $ \Lambda^{- flat}_{l=1}(\alpha)$ decrease to their minimum value of $\frac{1}{2}$ at $\alpha = \frac{1}{2}$. Thus, we have $\Lambda^+_{l=1}(\frac{1}{2}) =  \Lambda^-_{l=1}(\frac{1}{2})= \Lambda^{- flat}_{l=1}(\frac{1}{2}) = \frac{1}{2}$, implying immediately that the limit $\gamma_0(\omega, \alpha = \frac{1}{2}) \rightarrow 2$ at large $\omega$, since the denominator is dominated by the same eigenvalue with twice the multiplicity. 

\section{Pair production rates for spinor fields fields on $ S^2 \times\mathbb{R}^{1,1}$}
\label{sec:3:2}

\subsection{Spectrum of the gauged Dirac operator}
\label{sec:3:2:1}

In this section, we will calculate the pair production rate for spinor fields on $S^2 \times\mathbb{R}^{1,1}$ under the influence of the same background fields introduced in the previous section. For this purpose, we need the spectrum of the square of the gauged Dirac operator $\slashed{D}^2 = \frac{1}{a^2} (\vec{\tau} \cdot \vec{\Lambda} +1)^2 $ on $S^2$, where $\tau_i$ are the usual Pauli matrices spanning the spin space. This is an interesting problem in its own right, which we fully solve in Appendix~\ref{sec:D}. In terms of the total angular momentum $\vec{K} =\vec{L} + \frac{\vec{\sigma}}{2} + \frac{\vec{\tau}}{2}$, we have 
\be
a^2 \slashed{D}^2 = K^2 - \left(\frac{N^2}{4} - 2 \left (\alpha - \frac{1}{2} \right)^2 \right) + \chi \,,
\ee
where $\chi : = 2 \left(\alpha - \frac{1}{2}\right ) (\vec{K} \cdot \vec{\sigma} - \frac{1}{2}) + N\, \alpha \, \vec{\sigma} \cdot \hat{r} -2 \, \alpha  \left(\alpha - \frac{1}{2}\right) (\vec{\sigma} \cdot \hat{r})(\vec{\tau} \cdot \hat{r})$ and squares to the diagonal operator 
\begin{align}
\chi^2 = 4 \left(\alpha - \frac{1}{2}\right)^2 \left( K^2 + \frac{1}{4} \right) - \left ( \left (\alpha -\frac{1}{2} \right)^2 -\frac{1}{4} \right) \left (N^2 - 4\left(\alpha - \frac{1}{2}\right )^2 \right ) \,,
\end{align}
in the $\vec{K}$ basis. Relegating the details of the calculations to Appendix~\ref{sec:D}, we write the spectrum of $\slashed{D}^2_{S^2} +\slashed{D}^2_{\mathbb{R}^2} + m^2$ on $S^2 \times \mathbb{R}^2$ is given in the table below:
\begin{table}[h!]
	\centering
	\begin{tabular}{|c|c|c|c|} 
		\hline
		$Spec(D^2_{S^2} +D^2_{\mathbb{R}^2} + m^2)$ & Density of states\\
		\hline
		$\lambda_{n_1-1} + 2n_2 B_2 $ & $\frac{B_2}{2\pi} \frac{(2n_1 + N - 1)}{4\pi a^2}$\\ 
		$\lambda_{n_1}^\pm + 2n_2 B_2 $ & $\frac{B_2}{2\pi} \frac{(2n_1 + N + 1)}{4\pi a^2}$\\ 
		$\lambda_{n_1 + 1} + 2n_2 B_2 $ & $\frac{B_2}{2\pi} \frac{(2n_1 + N + 3)}{4\pi a^2}$ \\
		$\lambda_{n_1-1} + 2n_2B_2 + 2 $  & $\frac{B_2}{2\pi} \frac{(2n_1 + N - 1)}{4\pi a^2}$ \\ 
		$\lambda_{n_1}^\pm + 2n_2B_2 + 2 $ & $\frac{B_2}{2\pi} \frac{(2n_1 + N + 1)}{4\pi a^2}$ \\ 
		$\lambda_{n_1-1} + 2n_2B_2 + 2 $ & $\frac{B_2}{2\pi} \frac{(2n_1 + N + 3)}{4\pi a^2}$ \\
		\hline
	\end{tabular}
	\caption{Spectrum of $\slashed{D}^2_{S^2}$ and the corresponding density of states.}
	\label{table:4:2}
\end{table}

\noindent where 
\be
\lambda_{n_1}^\pm(\alpha) = \frac{1}{a^2} \left (  \xi_{n_1} + \left(\alpha - \frac{1}{2}\right)^2 \pm \sqrt{4 \left(\alpha - \frac{1}{2}\right)^2 \xi_{n_1} + \frac{N^2}{4}} \right) \,,
\ee
$\xi_{n_1} = \left ( n_1 + \frac{1}{2} \right)^2 + N  \left( n_1 + \frac{1}{2} \right) + \left (\left(\alpha - \frac{1}{2}\right)^2 - \frac{1}{4} \right)$. In the table, we have used the notation $\lambda_{n_1 + 1} := \lambda_{n_1 + 1}^-$, $\lambda_{n_1-1} := \lambda_{n_1-1}^+$, and in the latter for $n_1 =0$ and $N \geq 2$, $\sqrt{4 \left(\alpha - \frac{1}{2}\right)^2 \xi_{n_1-1} + N^2/4}$ is replaced with $\frac{N}{2} - 2 (\alpha - \frac{1}{2})^2$ as explained in Appendix D. We have $n_1 =0,1,2,\cdots \,, n_2 = 0,1,2,\cdots$ except for $\lambda_{n_1 - 1}$ at $N=1$ for which $n_1 =1,2,\cdots$. From detailed considerations given in Appendix~\ref{sec:D}, we see that, for $N \geq 2$ spectrum of $\slashed{D}^2_{S^2}$ has the zero modes $\lambda_{0}^- = 0 $ and $\lambda_{-1} = 0$ at $n_1 = 0$. However, for $N=1$, we have from \eqref{eq:D:12:c} that the set of eigenvalues $\lambda_{n_1-1}$ start with $n_1 = 1$, meaning that this branch does not include a zero mode, while the zero mode from the branch $\lambda_{n_1}^-$ is retained. Let also note that the entire spectrum is symmetric under $\alpha \leftrightarrow 1-\alpha$.

\subsection{Pair production rates}
\label{sec:3:2:2}

We first evaluate the effective action on $\mathbb{R}^{2} \times S^2$ and then Wick rotate to $S^2 \times \mathbb{R}^{1,1}$. We have,
    \begin{align}
    \label{eq:3:14}
      \Gamma_E = \frac{1}{2} \int \frac{ds}{s} Tr\, \Big[e^{-s(\slashed{D}^2 + m^2)}\Big]\,.
    \end{align}
Taking the integral over $s$ and Wick rotating by letting $x_4 \rightarrow ix_0$, $B_2 \rightarrow -iE$, real part of $iS_{eff}$ on $S^2 \times \mathbb{R}^{1,1}$ can be written (for $N \geq 2$) as 
    \begin{align}
        \label{eq:3:15}
        Re(iS_{eff}) = &- \frac{1}{16\pi^2 a^2} \int d^2x \, \int d \Omega_2 \sum_{n=1}^{\infty}  e^{-n\pi m^2/E}\,\frac{E}{n}  \Biggl\{  \\
        &   \sum_{n_1 = 0 }(2n_1+N-1) \Big[e^{-n\pi(\xi_{n_1-1} +  \left (\alpha -\frac{1}{2} \right)^2 + \sqrt{4 \left (\alpha -\frac{1}{2} \right)^2 \xi_{n_1-1} + \frac{N^2}{4}} + m^2a^2)/(E a^2)} \Big]\nonumber \\
        + & \sum_{n_1=0}(2n_1+N+1) \Big[ e^{-n\pi(\xi_{n_1} +  \left (\alpha -\frac{1}{2} \right)^2 + \sqrt{4 \left (\alpha -\frac{1}{2} \right)^2\xi_{n_1} + \frac{N^2}{4}} + m^2a^2)/(E a^2)} \Big]\nonumber \\  
        + & \sum_{n_1= 0}(2n_1+N+1) \Big[ e^{-n\pi(\xi_{n_1} + \left (\alpha -\frac{1}{2} \right)^2 - \sqrt{4 \left (\alpha -\frac{1}{2} \right)^2 \xi_{n_1} + \frac{N^2}{4}} + m^2a^2)/(E a^2)} \Big]\nonumber \\
        + & \sum_{n_1=0}(2n_1+N+3)\Big[ e^{-n\pi (\xi_{n_1+1} +  \left (\alpha -\frac{1}{2} \right)^2 - \sqrt{4 \left (\alpha -\frac{1}{2} \right)^2 \xi_{n_1+1} + \frac{N^2}{4}} + m^2a^2 )/(E a^2)} \Big]\Biggr\}.\nonumber
    \end{align}
Using $\omega = \frac{\pi}{Ea^2}$, we can write $Re(iS_{eff})$ as
    \begin{align}
        \label{eq:3:16}
        Re(iS_{eff}) &= - \int d^2x \, \int d \Omega_2 \, \frac{E^2}{8\pi^3} \, \beta_{1/2}(\omega, \alpha, N),
    \end{align}
where
    \begin{align}
        \label{eq:3:17}
        \beta_{1/2}(\omega, \alpha, N) = - &\frac{\omega}{2} \Biggl\{ 2 N \,  \ln \, (1-e^{-\omega \, m^2 a^2}) \\
        +&\sum_{n_1=1}^{\infty} (2n_1+N-1) \ln \, (1-e^{-\omega(\xi_{n_1-1} +  \left (\alpha -\frac{1}{2} \right)^2 +  \sqrt{4  \left (\alpha -\frac{1}{2} \right)^2 \xi_{n_1-1} + \frac{N^2}{4}})+m^2a^2)})\nonumber \\
        +&\sum_{n_1=0}^{\infty} (2n_1+N+1) \ln \, (1-e^{-\omega(\xi_{n_1} + \left (\alpha -\frac{1}{2} \right)^2 +  \sqrt{ 4 \left (\alpha -\frac{1}{2} \right)^2 \xi_{n_1} + \frac{N^2}{4}})+m^2a^2)}) \nonumber \\
        +&\sum_{n_1=1}^{\infty} (2n_1+N+1) \ln \, (1-e^{-\omega(\xi_{n_1} + \left (\alpha -\frac{1}{2} \right)^2 -  \sqrt{ 4 \left (\alpha -\frac{1}{2} \right)^2 \xi_{n_1} + \frac{N^2}{4}})+m^2a^2)}) \nonumber \\
        +&\sum_{n_1=0}^{\infty} (2n_1+N+3) \ln \, (1-e^{-\omega(\xi_{n_1+1} + \left (\alpha -\frac{1}{2} \right)^2 -  \sqrt{ 4 \left (\alpha -\frac{1}{2} \right)^2\xi_{n_1+1} + \frac{N^2}{4}})+m^2a^2)}) 
     \Biggr\}.\nonumber	 
    \end{align}
In \eqref{eq:3:17}, we have written the contribution of the zero modes explicitly. For $N =1$, we have the first sum in \eqref{eq:3:15} starting from $n_1 = 1$ and hence $\beta_{1/2}(\omega, \alpha, N)$ has the same form except that there is no factor of $2$ in front of the first term in \eqref{eq:3:17}. We note that  $\beta_{1/2}(\omega, \alpha, N) =  \beta_{1/2}(\omega, 1-\alpha, N)$.

We can calculate the form $\beta_{1/2}(\omega, \alpha, N)$ takes in the limit $S^2 \times \mathbb{R}^{1,1} \rightarrow \mathbb{R}^{3,1}$. We let $a \rightarrow \infty$, $N\rightarrow \infty$ and $\alpha \rightarrow \infty$, while keeping $N/ 2 a^2$ and $\alpha^2/a^2$ constant. Since $\omega$ is proportional to $1/a^2$, this practically means we can keep $\omega \, \alpha^2$ and $\omega \, N$ as such, while the terms which vanish as $a \rightarrow \infty$ are dropped. We find, for $N \geq 2 $, $\lambda_{n_1}^\pm \rightarrow \lambda_{n_1}^{\pm \, flat} = N  \left (n_1 + \frac{1}{2} \right )  + 2 \alpha^2 +\pm \sqrt{4 \alpha^2 \left (N  \left (n_1 +\frac{1}{2} \right )   + \alpha^2 \right ) + \frac{N^2}{4}}$ and $\lambda_{n_1 \pm 1} \rightarrow \lambda_{n_1 \pm 1}^{flat}$ and hence
\begin{align}
        \label{eq:3:18}
        \beta_{1/2}^{flat}(\omega, \alpha, N) =& -\frac{\omega N}{2}\Biggl\{2 \, \ln \, (1-e^{-\omega \, m^2 a^2}) \nn \\
        +&\sum_{n_1 = 1}^{\infty}  \ln\, (1 - e^{-\omega ( N \left (n_1 -\frac{1}{2} \right ) + 2 \alpha^2 + \sqrt{4 \alpha^2 \left ( N  \left (n_1 -\frac{1}{2} \right )  + \alpha^2 \right ) + \frac{N^2}{4} } + m^2 a^2)})  \nonumber \\
        +&\sum_{n_1 = 0}^{\infty}  \ln \, (1-e^{-\omega( N  \left (n_1 + \frac{1}{2} \right )  + 2 \alpha^2 + \sqrt{4 \alpha^2 \left (N  \left (n_1 +\frac{1}{2} \right )   + \alpha^2 \right ) + \frac{N^2}{4}}+m^2a^2)})  \nonumber \\
        +&\sum_{n_1 = 1}^{\infty}  \ln \, (1-e^{-\omega( N  \left (n_1 + \frac{1}{2} \right )  + 2 \alpha^2 - \sqrt{4 \alpha^2\left ( N  \left (n_1 + \frac{1}{2} \right )  + \alpha^2 \right ) + \frac{N^2}{4}}+m^2a^2)})  \nonumber \\
        +&\sum_{n_1 = 0}^{\infty}  \ln \, (1-e^{-\omega(N  \left (n_1 + \frac{3}{2} \right )  + 2 \alpha^2 - \sqrt{4 \alpha^2 \left ( N  \left (n_1 + \frac{3}{2} \right )  + \alpha^2 \right ) + \frac{N^2}{4}}+m^2a^2)})
        \Biggr\} \,.
    \end{align}
Shifting the index of first sum in \eqref{eq:3:18}, it is seen  to be equal to the second sum in \eqref{eq:3:18}, similarly shifting the index of the last sum to $1$, it is seen to be equivalent to the third sum in \eqref{eq:3:18}. Thus, we obtain
 \begin{align}
        \label{eq:3:19}
        \beta_{1/2}^{flat}(\omega, \alpha, N) = -\omega N\Biggl\{ &  \ln \, (1 - e^{- \omega \, m^2 a^2}) \nn \\
        +\sum_{n_1 = 0}^{\infty} & \ln\, (1-e^{-\omega (N  \left (n_1 +\frac{1}{2} \right ) + 2 \alpha^2 + \sqrt{4 \alpha^2 \left ( N  \left (n_1 +\frac{1}{2} \right ) + \alpha^2 \right ) + \frac{N^2}{4}}+m^2a^2)}) \nonumber \\
        +\sum_{n_1 = 1}^{\infty} & \ln\, (1-e^{-\omega (N  \left (n_1 +\frac{1}{2} \right ) + 2 \alpha^2 - \sqrt{4 \alpha^2 \left ( N \left (n_1 +\frac{1}{2} \right ) +  \alpha^2 \right ) + \frac{N^2}{4}} + m^2a^2)}) 
        \Biggr\} \,.
    \end{align}
 Using $B_1 = \frac{N}{2 a^2}$ and $\beta = \frac{\alpha^2}{a^2}$, we find that $\beta_{1/2}^{flat}(\omega, \alpha, N) = \frac{\pi^2}{3} f_{1/2}(y, \beta^\prime)$,(cf. \eqref{eq:2:29}). We note that, for $N=1$, there is no factor of $2$ in front of the first term in the right hand side of \eqref{eq:3:18} and consequently there is factor of $\frac{1}{2}$ in front of the first term in the r.h.s. of \eqref{eq:3:19}.

\begin{figure}[h!]
\centering
	\begin{subfigure}[b]{0.49\textwidth}
		\includegraphics[width=\textwidth]{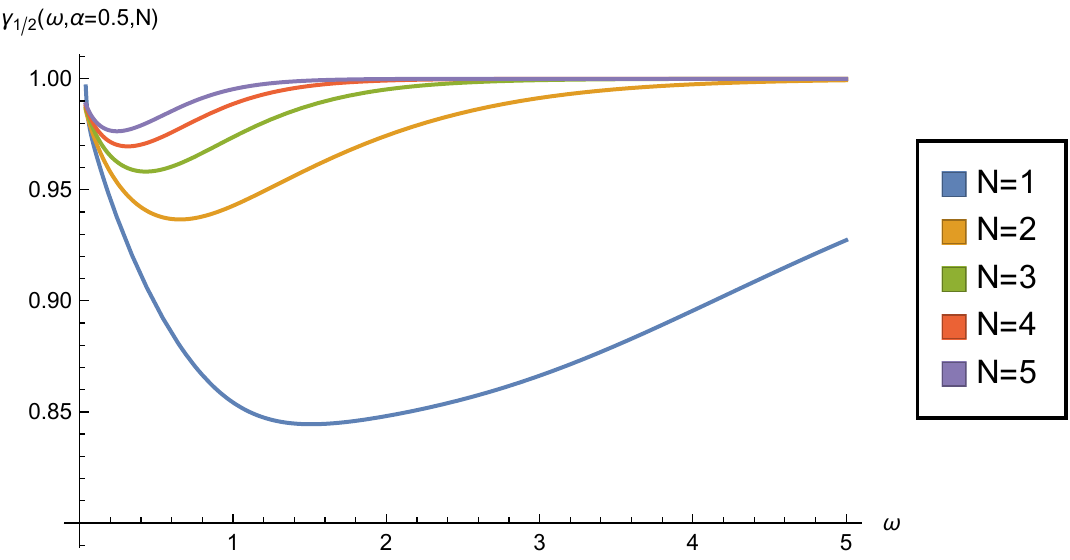}
		\caption{\label{fig:5:1} $\alpha = \frac{1}{2}$}
	\end{subfigure}
	\hfill
	\begin{subfigure}[b]{0.49\textwidth}
		\includegraphics[width=\textwidth]{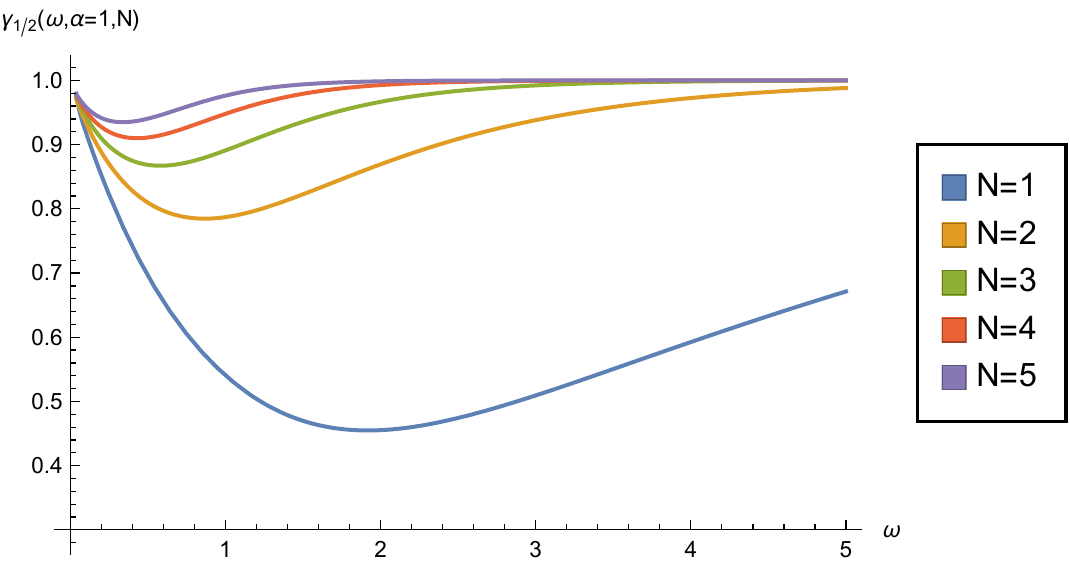}
		\caption{\label{fig:5:2} $\alpha = 1$}
	\end{subfigure}
	\centering
	\begin{subfigure}[b]{0.49\textwidth}
		\includegraphics[width=\textwidth]{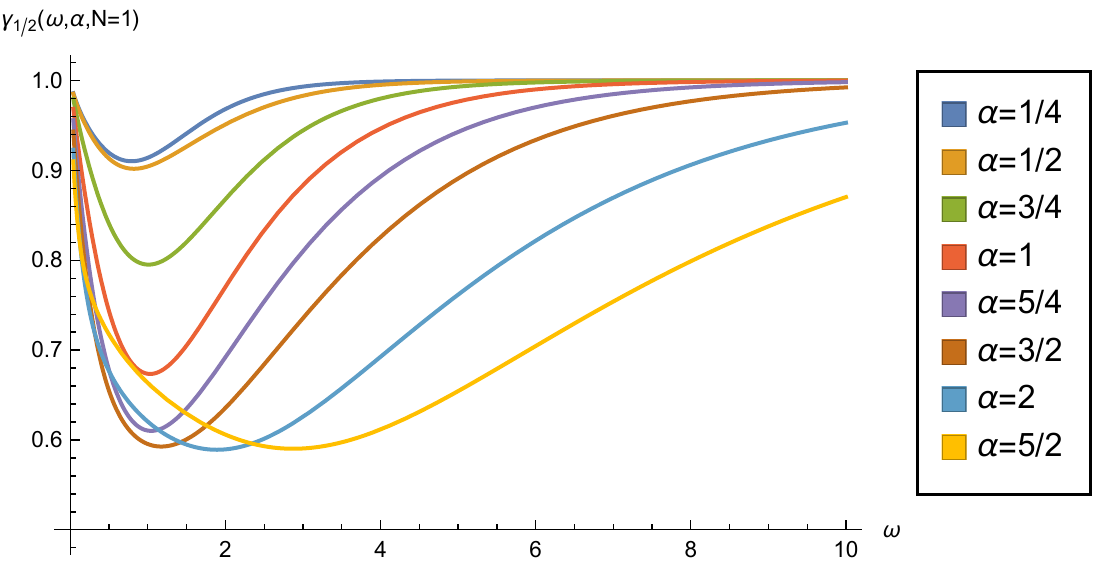}
		\caption{\label{fig:5:3} $N=1$}
	\end{subfigure}
	\hfill
	\begin{subfigure}[b]{0.49\textwidth}
		\includegraphics[width=\textwidth]{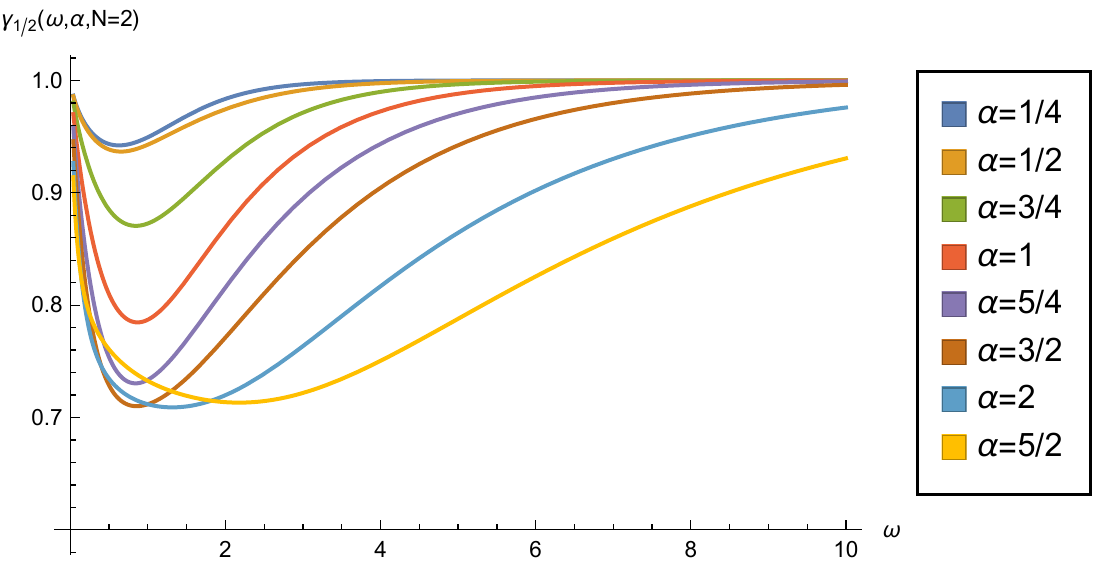}
		\caption{\label{fig:5:4} $N = 2$}
	\end{subfigure}
  \caption{\label{fig:main7} $\gamma_{1/2}(\omega, \alpha, N)$ as a function of $\omega$.}
\end{figure}

\begin{figure}[h!]
	\centering
	\includegraphics[width=0.49\textwidth]{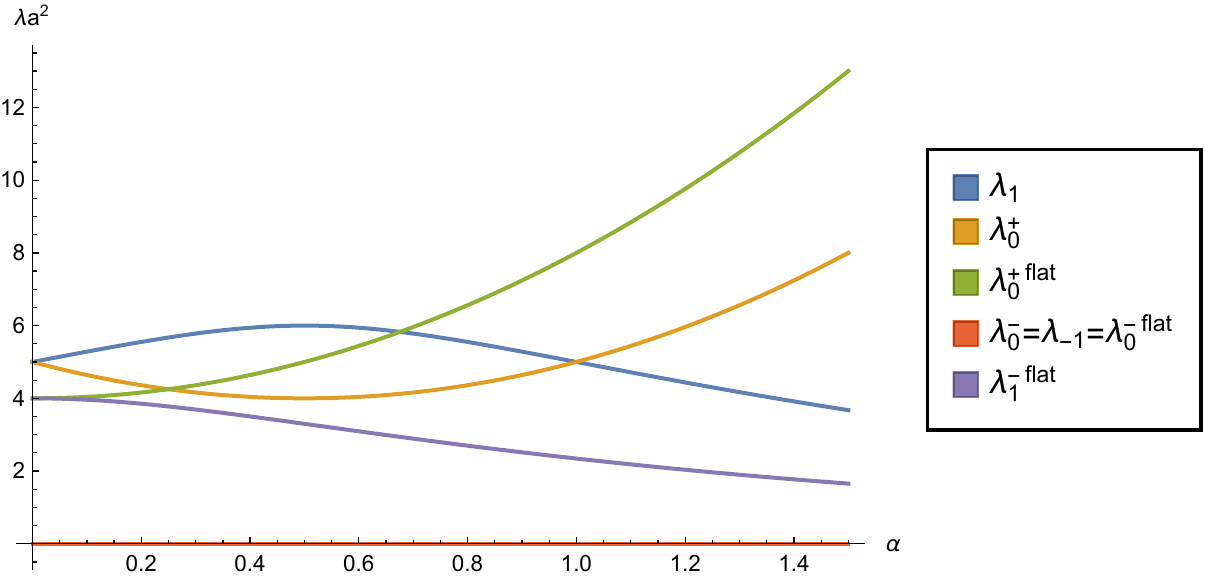}
	\caption{\label{fig:5:x} Lowest lying eigenvalues as a function of $\alpha$.}
\end{figure}

In order to compare the effects of curvature and the magnetic fields on the pair production rates, we define the ratio $\gamma_{1/2}(\omega, \alpha, N):= \frac{\beta_{1/2}(\omega, \alpha, N)}{\beta_{1/2}^{flat}(\omega, \alpha, N)}$. In Figure \ref{fig:main7}, we plot the profile of $\gamma_{1/2}(\omega, \alpha, N)$ both at a fixed value of the non-abelian charge $\alpha$ at different monopole strength $N$ as well as at several values of $\alpha$ at a fixed $N$. We observe that the pair production rate remains less than that on the flat space at given values of $N$ and $\alpha$. This is mainly due to the $n_1^2$ dependence of the eigenvalues (which is due to the curvature effects) in the numerator causing it to be less than the denominator of $\gamma_{1/2}(\omega, \alpha, N)$. More concretely, the energies $\lambda_{n_1+1}$, $\lambda_{m_1-1}^+ = \lambda_{m_1-1}$ are larger than $\lambda_{n_1+1}^{- \, flat} = \lambda_{n_1+1}^{flat}$, where $n_1 =0,1,2,\cdots$ and $m_1 =1,2,\cdots$ and hence the former are relatively harder to get filled by the produced pairs. Increase in $\alpha$ triggers a decrease in $\gamma_{1/2}(\omega, \alpha, N)$ since $\lambda_{n_1+1}^{- \, flat}$ decreases with $\alpha$ and the associated flat levels become relatively easier to get filled compared to those in the curved background. These facts are readily observed from the plots given in Figure \ref{fig:5:x}. With increasing $N$, terms due to the zero modes become the dominant contribution and they drive $\gamma_{1/2}(\omega, \alpha, N)$ back to the value $1$ at large $\omega$ and hence the pair production rate to that of the flat case. Our results are essentially independent  of the infrared cut-off value for any physically reasonable choice of the latter. In our calculations we have used $m^2 a^2 = \frac{1}{2}$.  

\begin{figure}[h!]
    	\centering
    	\begin{subfigure}[b]{0.49\textwidth}
    		\includegraphics[width=\textwidth]{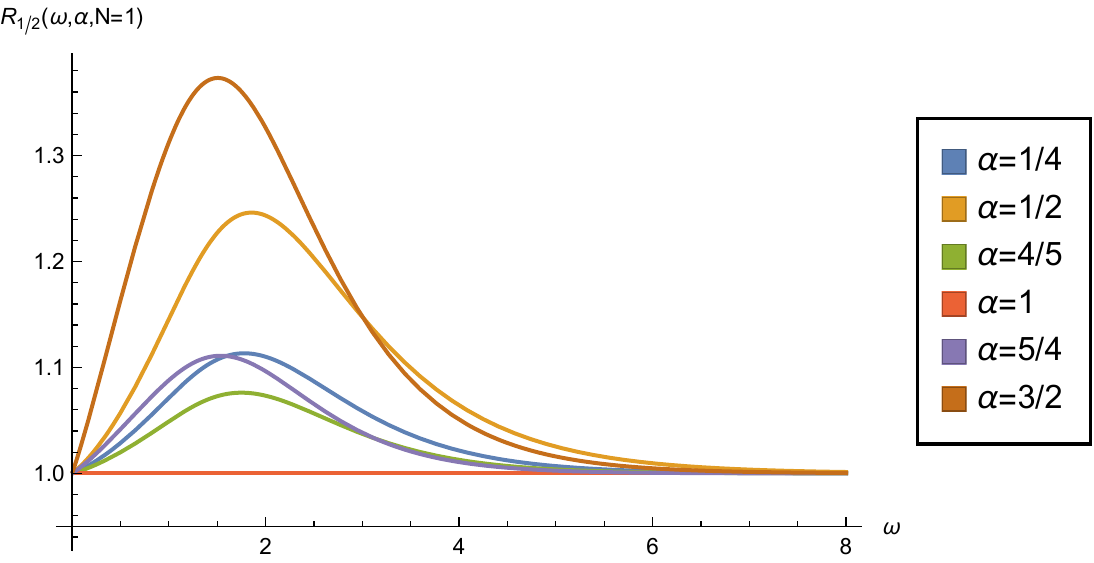}
    		\caption{\label{fig:5:5} $N = 1$}
    	\end{subfigure}
    	\hfill
    	\begin{subfigure}[b]{0.49\textwidth}
    		\includegraphics[width=\textwidth]{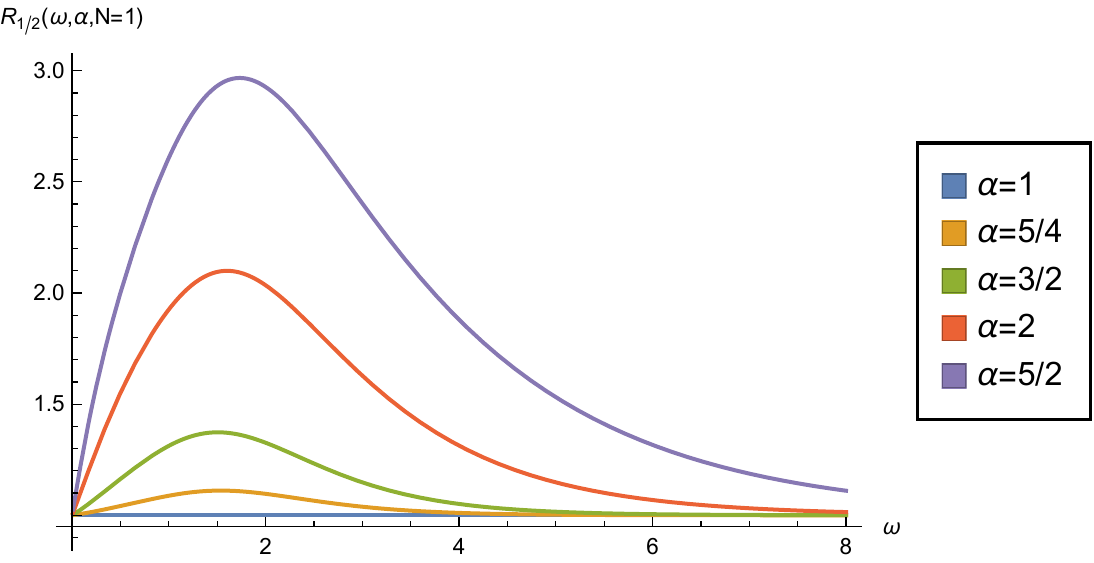}
    		\caption{\label{fig:5:6} $N = 1$}
    	\end{subfigure}
	\centering
	\begin{subfigure}[b]{0.49\textwidth}
		\includegraphics[width=\textwidth]{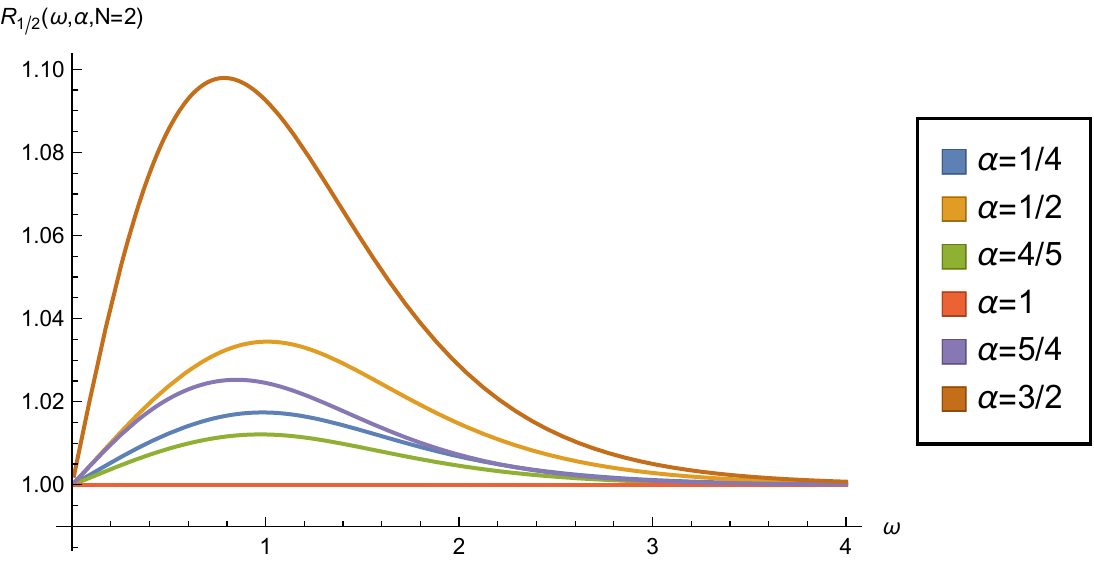}
		\caption{\label{fig:5:7} $N = 2$}
	\end{subfigure}
	\hfill
	\begin{subfigure}[b]{0.49\textwidth}
		\includegraphics[width=\textwidth]{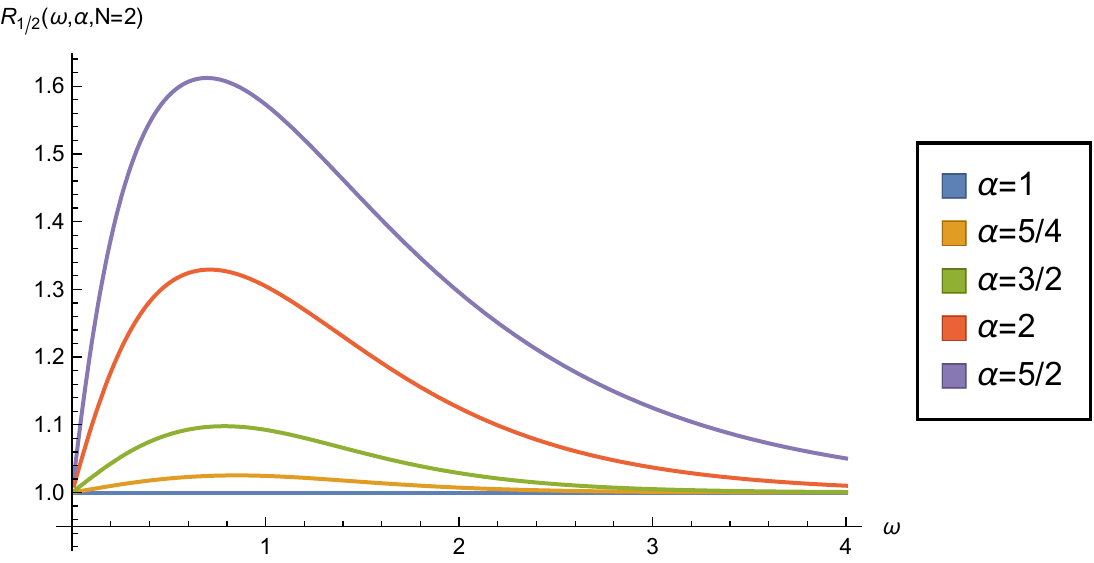}
		\caption{\label{fig:5:8} $N = 2$}
	\end{subfigure}
	\centering
	\begin{subfigure}[b]{0.49\textwidth}
		\includegraphics[width=\textwidth]{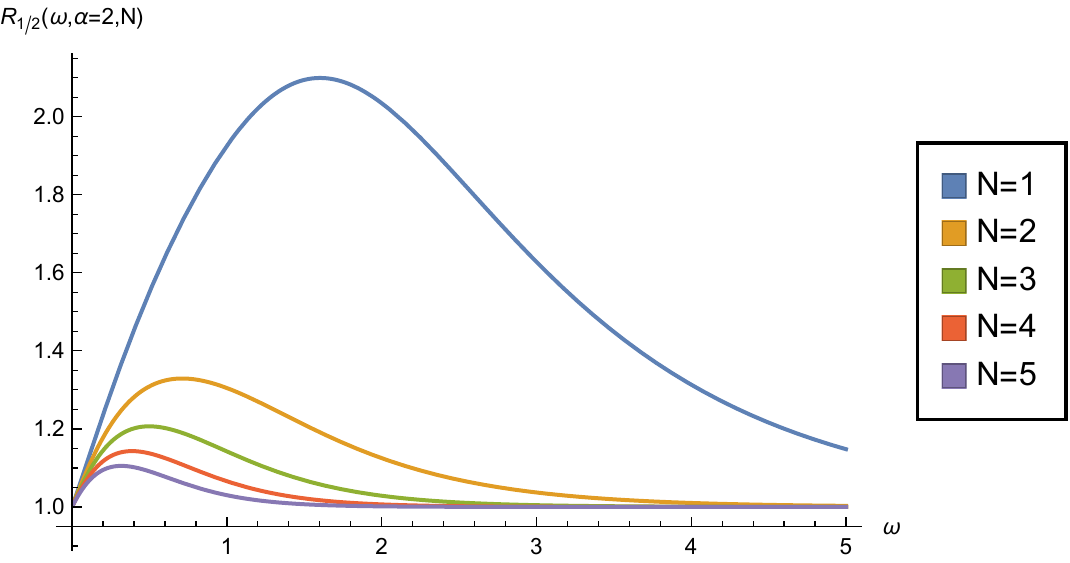}
		\caption{\label{fig:5:9} $\alpha = 2$}
	\end{subfigure}
  \caption{\label{fig:main8} $R_{1/2}(\omega, \alpha, N)$ as a function of $\omega$.}
\end{figure}

It is also noteworthy to emphasize that, the overall effect for $\alpha \neq 0$ is greater than that without the non-abelian field (i.e. $\alpha = 0$). To see this, we can readily inspect the profiles of $R_{1/2}(\omega, N, \alpha) \equiv \frac{\beta_{1/2}(\omega, N, \alpha)}{\beta_{1/2}(\omega, N, 0)}$. Plots at $N=1,2$ at several values of $\alpha$ are given in the Figures~\ref{fig:5:5}-\ref{fig:5:8} and for $\alpha =2$ at several values of $N$ in Figure~\ref{fig:5:9}. We see that there is an increase in the pair production for $\alpha \neq 0$ compared to $\alpha = 0$, while the rate of change in $R_{1/2}(\omega, N, \alpha)$ decreases as $N$ assumes larger values (see Figure ~\ref{fig:5:9}). We also observe that as $\omega$ becomes large, the pair production rates converge back to what they are at $\alpha =0$.  The latter two results are due to the dominating effect of the zero modes at larger values of $N$ and $\omega$. For $\alpha > 1$, at fixed $N$, $R_{1/2}(\omega, N, \alpha)$ becomes larger with increasing $\alpha$ before it converges to $1$ at large $\omega$, while for $0 \leq \alpha < \frac{1}{2}$, we see that $R_{1/2}(\omega, N, \alpha)$ increases with $\alpha$ and decreases toward the value $1$ in the interval $\frac{1}{2} < \alpha \leq 1$ and   $R_{1/2}(\omega, N, \alpha =1) =1$.  All of these features are readily explained by noting that, for $0 \leq \alpha \leq \frac{1}{2}$, $\lambda_{n_1}^+ < \lambda_{n_1}^+\big|_{\alpha = 0} = \lambda_{n_1+1}\big|_{\alpha = 0} = \lambda_{(n_1+1)-1}\big|_{\alpha = 0} = (n_1+1)(N+ n_1+ 1)$, $n_1=0,1,2,\cdots $ with $\lambda_{n_1}^+$  becoming smaller with increasing $\alpha$ and reaching a local minimum at $\alpha = \frac{1}{2}$, while for $\frac{1}{2} \leq \alpha \leq 1$ same inequality continues to hold with $\lambda_{n_1}^+$ increasing with $\alpha$ and becoming $(n_1+1)(N+ n_1+ 1)$ at $\alpha=1$. For $\alpha > 1$, we have $\lambda_{n_1+1} < (n_1+1)(N+ n_1+ 1)$ and $\lambda_{n_1}^+ > (n_1+1)(N+ n_1+ 1)$, with the former being energetically more favorable to get filled and hence leading to an increased amplitude for the pair production. Taking, $a, \alpha, N \rightarrow \infty$ with $B_1 = \frac{N}{2a^2}$, $\beta^{2} = \frac{\alpha^2}{a^2}$ and $\beta^{\prime 2} = \frac{\beta^2}{B_1}$, we have $R_{1/2}(\omega, \alpha, N) \rightarrow F_{1/2}(y,\beta^\prime)$ as readily inferred from the definition of $R_{1/2}(\omega, N, \alpha)$.  

For the case of vanishing abelian monopole charge the spectrum of $\slashed{D}^2_{S^2}$ is given in \eqref{alpha0short}. Using the latter, we find that
\begin{align}
Re(iS_{eff}) = - \int d^2 x \, \int d \Omega_2 \, \frac{E^2}{8\pi^3} \, \beta_{1/2}(\omega, \alpha) \,,
\end{align}
where
\begin{multline}
	\label{eq:4:44}
	\beta_{1/2}(\omega, \alpha) = - \omega \Biggl\{ \sum_{l=0}^{\infty} (2l+1)\ln \, (1-e^{-\omega(a^2 \lambda_l^+ + m^2a^2)}) \\
	+ \sum_{l = 1 }^{\infty} (2l+1) \ln \, (1-e^{-\omega(a^2 \lambda_l^- + m^2 a^2)}) \Biggl\} \,,
\end{multline}
and $\lambda_l^\pm = \frac{1}{a^2} \left(l^2+ l +  2 \left(\alpha - \frac{1}{2}\right)^2 \pm 2 \left | \alpha - \frac{1}{2}\right | \sqrt{(l+1/2)^2+\left ( \left (\alpha -\frac{1}{2} \right)^2 -\frac{1}{4} \right ) } \right ) $ are given in \eqref{alpha0short} with $l =0,1,2,\cdots$ for the upper and $l =1,2,\cdots$ for the lower sign. Let us also note that $\lambda_0^+ = \frac{4}{a^2} \left(\alpha - \frac{1}{2}\right)^2$, giving a zero mode at $\alpha = \frac{1}{2}$ and also that $\beta_{1/2}(\omega, \alpha) = 	\beta_{1/2}(\omega, 1-\alpha)$.

Taking the limit $S^2 \rightarrow {\mathbb R}^2$ we find
\begin{multline}
\beta_{1/2}^{flat}(\omega, \alpha) = - 2 \, \omega \, \Biggl\{\sum_{l=0}^{\infty}l \ln \, (1-e^{-\omega(l^2 +  2 \alpha^2 + 2  |\alpha | \sqrt{l^2+ \alpha^2} + m^2 a^2)}) \\ 
+ \sum_{l = 1}^{\infty} l \ln \, (1 - e^{-\omega(l^2 +  2 \alpha^2 - 2 \left | \alpha \right | \sqrt{l^2+\alpha^2} + m^2 a^2)} \Biggl\} \,.
\end{multline}
Setting $\frac{l}{a} \rightarrow k$, $\frac{\alpha}{a} \rightarrow \beta$ and replacing $ \frac{1}{a} \sum_{l} \frac{l}{a}$ with $\int k \, dk$, we find that $\beta_0^{flat}(\omega, \alpha) \rightarrow \frac{\pi^2}{6} f_{1/2}(\frac{\pi}{E},\beta)$, where the last factor on the r.h.s. is given in \eqref{eq:f1/2}.

\begin{figure}[h!]
	\centering
	\begin{subfigure}[b]{0.49\textwidth}
		\includegraphics[width=\textwidth]{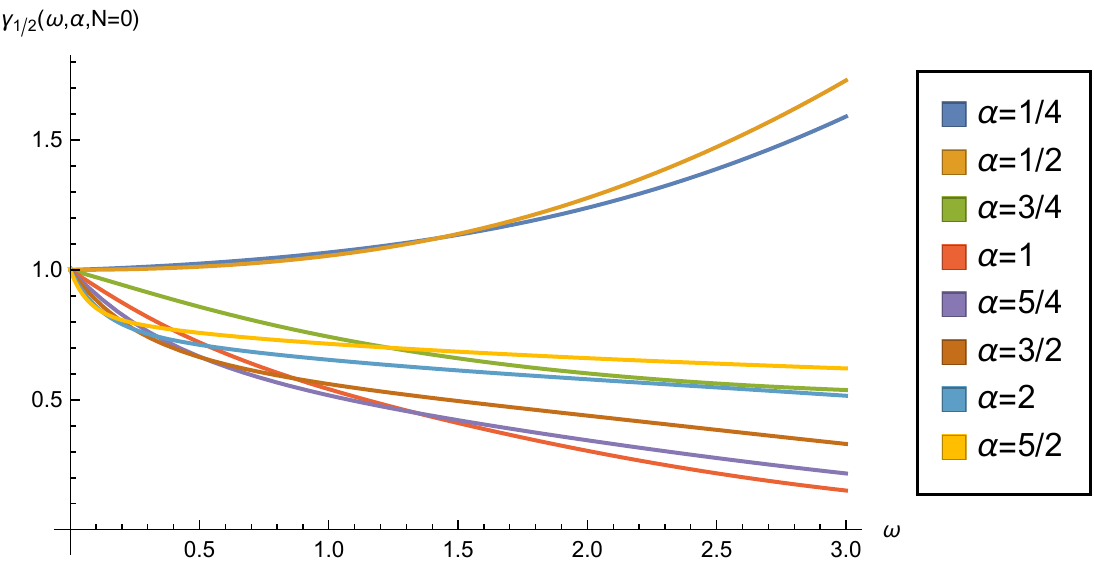}
		\caption{\label{fig:5:10} $\gamma_{1/2}(\omega, \alpha)\,, N = 0$}
	\end{subfigure}
	\hfill
	\begin{subfigure}[b]{0.49\textwidth}
		\includegraphics[width=\textwidth]{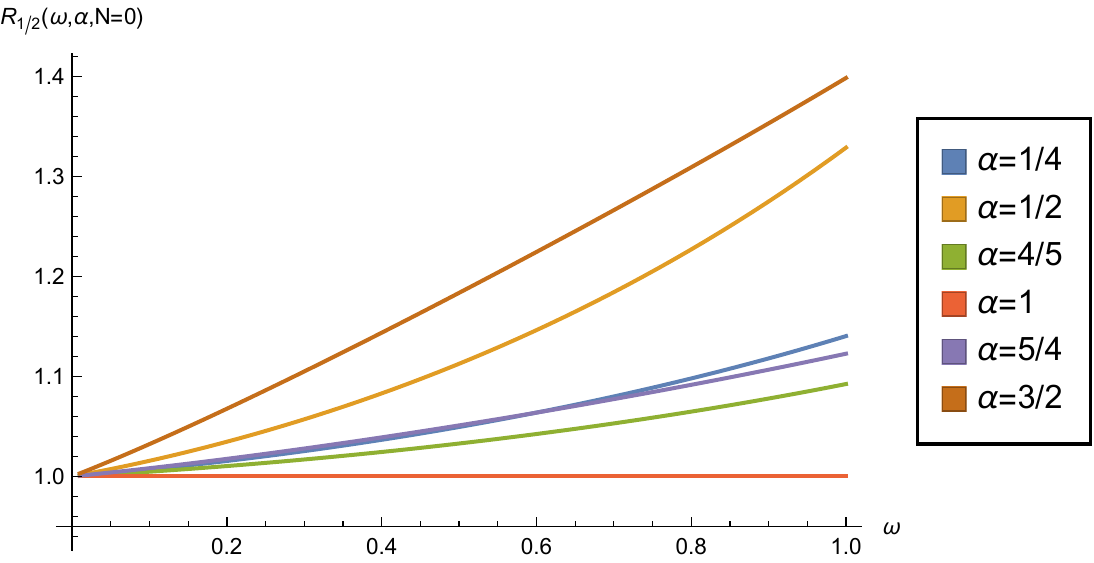}
		\caption{\label{fig:5:11} $R_{1/2}(\omega, \alpha)\,, N = 0$}
	\end{subfigure}
	\centering
	\begin{subfigure}[b]{0.49\textwidth}
	    \includegraphics[width=\textwidth]{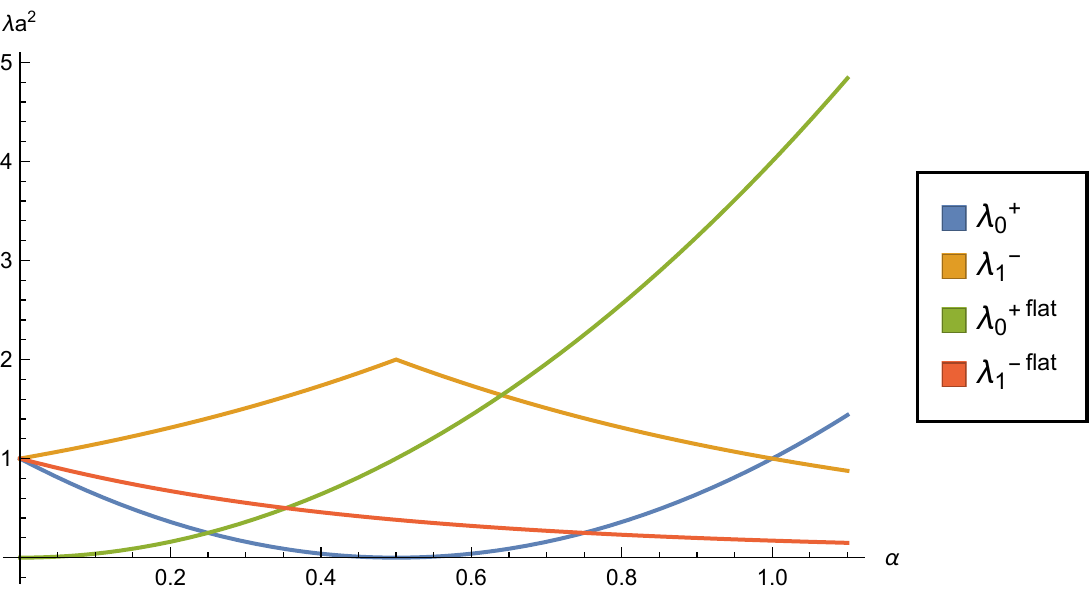}
	    \caption{\label{fig:5:12} Lowest lying energy eigenvalues as a function of $\alpha$. }
    \end{subfigure}
    \caption{\label{fig:main9} $\gamma_{1/2}(\omega, \alpha)$, $R_{1/2}(\omega, \alpha)$ at $N=0$.}
\end{figure}

Below, we plot the profiles of $\gamma_{1/2}(\omega, \alpha) \equiv \frac{\beta_{1/2}(\omega,\alpha)}{\beta_{1/2}^{flat}(\omega, \alpha)}$ and $R_{1/2}(\omega, \alpha) \equiv \frac{\beta_{1/2}(\omega, \alpha)}{\beta_{1/2}(\omega, 0)}$. From Figure~\ref{fig:5:10}, we see that for $0 \leq \alpha \leq \frac{1}{2}$ the pair production amplitude is larger than that of the flat case. We also see an increasing rate of change with $\alpha$ which is maximized at $\alpha = \frac{1}{2}$. The latter is essentially due  the zero modes in the spectrum at $\alpha = \frac{1}{2}$, which are filled without any energy cost. For $\frac{1}{2} < \alpha \leq 1$, relative pair production rates decrease with increasing $\alpha$, with the maximal rate reached at $\alpha = 1$. For $\alpha > 1$, $\gamma_{1/2}(\omega, \alpha)$ remains less than the value $1$, but slowly converges to it at large $\omega$. We can also note that the overall pair production amplitudes for $\alpha \neq 0$ are always greater compared to that at $\alpha =0$. From the plots in Figure~\ref{fig:5:11}, we observe that rate of change in $R_{1/2}(\omega, \alpha)$ increases with $\alpha$ for $0 \leq \alpha \leq \frac{1}{2}$, decreases for $\frac{1}{2} < \alpha \leq 1$, with $R_{1/2}(\omega, \alpha = 1) = 1$ as expected due to the the symmetry $\alpha \leftrightarrow 1 - \alpha$ and increasing further above the value $1$ for $\alpha >1$. These features can be readily attributed to the hierarchy among the lowest lying energy levels $\lambda_l^\pm$ and $\lambda_l^{\pm \,, flat}$ as seen from Figure~\ref{fig:5:12}.

\section{Conclusion}

In this paper, we have calculated the Schwinger pair production rates  in $\mathbb{R}^{3,1}$ as well as in the positively curved space $S^2 \times \mathbb{R}^{1,1}$ for spin-$0$ and spin-$\frac{1}{2}$ particles under the influence of an external $SU(2) \times U(1)$ gauge field producing an additional uniform non-abelian magnetic field besides the usual uniform abelian electric field. For this purpose, we have obtained the spectrum of the gauged Laplace and Dirac operators on both the flat and the curved geometries and used them to evaluate the Euclidean effective actions, which are Wick rotated at an appropriate stage of the calculation to determine the pair production amplitudes. From our results, we have seen that, depending on their relative strength the purely non-abelian and the abelian parts of the magnetic field have either a counterplaying or reinforcing role, whose overall effect may be to enhance or suppress the pair production rates. Positive curvature acts to enhance the latter for spin-$0$ and suppress it for spin-$\frac{1}{2}$ fields, while the details of the couplings to the purely abelian and the non-abelian parts of the magnetic field, which were studied through the dependence of the energy spectrum to the strength of these fields, played a critical role to determine the cumulative effect on the pair production rates. 

\appendix
\section{Spectrum of the gauged Laplacian on $\mathbb{R}^2$}
\label{sec:A}

We have the operator $- D^2_{(1)} = - (\vec{\partial} - i \vec{A}_{(1)})^2$ where $\vec{A}_{(1)}$ is valued in the Lie algebra of the gauge group $SU(2) \times U(1)$ and already given in \eqref{eq:2:1}. The latter yields the field strength
\begin{align}
\label{eq:A:3}
F_{12} & =  \partial_1 A_2 - \partial_2 A_1 + i [A_1,A_2], \nonumber \\
	&= \frac{B_1}{2} (\partial_1 x_1 + \partial_2 x_2) \mathbbm{1}_2 - i \beta^2 [\sigma_2, \sigma_1], \nonumber \\
	&=  B_1 \mathbbm{1}_2 + 2 \beta^2 \sigma_3.
\end{align}
Here, the first term represents an abelian magnetic field, while the second term is a non-abelian uniform magnetic field due to a pure $SU(2)$ gauge field. Spectrum of the operator $D^2_{(1)}$ is determined in \cite{Estienne_2011}. Here we reproduce the result of this paper for convenience. We drop the subscript $(1)$, for ease in notation in what follows. We have 
\begin{align}
 \label{eq:A:4}
        - D^2 
            &= - (\partial_1 - i A_1)^ - (\partial_2 - i A_2)^2, \nonumber \\
            &= - (\partial_1^2 - 2 i A_1 \partial_1 - A_1^2) - (\partial_2^2 - 2 i A_2 \partial_2 - A_2^2), \nonumber \\
            &= - (\partial_1^2 + \partial_1^2) - i (- B_1 x_2 \partial_1 + B_1 x_1 \partial_2 - 2 \beta \sigma_2 \partial_1 + 2 \beta \sigma_1 \partial_2)  +\frac{B_1^2}{4} (x_1^2 + x_2^2) \nonumber \\
            & \quad \quad + B_1 \beta (x_1 \sigma_1 + x_2 \sigma_2) + 2\beta^2, \nonumber \\
            &= - 4 \partial \bar{\partial} + B_1(\bar{z}\bar{\partial} - z \partial) + 4 \beta ( \sigma_-\bar{\partial}  - \sigma_+ \partial ) + \frac{B_1^2}{4} |z|^2 + B_1 \beta (z \sigma_-  +\bar{z} \sigma_+) + 2\beta^2, \nonumber \\
            &= 2 B_1\Big(- \frac{2\partial \bar{\partial}}{B_1} + \frac{1}{2}(\bar{z}\bar{\partial} - z \partial) + \frac{2\beta}{B_1} ( \sigma_-  \bar{\partial} - \sigma_+ \partial ) + \frac{B_1}{8} |z|^2 +\frac{\beta}{2} (z \sigma_-  +\bar{z} \sigma_+) + \frac{\beta^2}{B_1} \Big), \nonumber \\
            &= 2B_1\Big( a^\dagger a +\sqrt{2} \frac{\beta}{\sqrt{B}} (a^\dagger \sigma_+ + a \sigma_-) + \frac{1}{2} + \frac{\beta^2}{B_1}  \Big), \nonumber \\
            &= 2 B_1  \Big(a^{\dagger}a + \sqrt{2}\beta^\prime(a^{\dagger}\sigma_+ + a\sigma_-) + \frac{1}{2}(1+2\beta^{\prime 2})\Big).
    \end{align}
where we have introduced $z= x_1 + i x_2$, $\bar{z} = x_1 - i x_2$, $\partial = \frac{1}{2}(\partial_1 - i \partial_2)$,  $\bar{\partial} = \frac{1}{2}(\partial_1 +i \partial_2)$, $\sigma_+ = \frac{1}{2}(\sigma_1 + i \sigma_2)$, $\sigma_- = \frac{1}{2}(\sigma_1 - i \sigma_2)$ on the fourth and the creation and annihilation operators in the penultimate line of  \eqref{eq:A:4} via
\begin{align}
\label{eq:A:6}
a = \frac{1}{\sqrt{2B_1}} \Big(\frac{B_1}{2} z + 2 \bar{\partial} \Big) \,, \quad 
a^\dagger = \frac{1}{\sqrt{2B_1}} \Big(\frac{B_1}{2} \bar{z} - 2 \partial \Big) \,.
\end{align}
It can be readily checked that $[a, a^\dagger] = 1$. In the last line of \eqref{eq:A:4}, we have also defined and used the dimensionless non-abelian magnetic field, which is scaled with respect to the abelian magnetic field $\beta^\prime\equiv \beta/\sqrt{B_1}$ assuming $B_1 >0$. In fact, if we change the direction of the abelian magnetic field, i.e. $B_1 \rightarrow - B_1$, we have $-D^2$ retain the form in the last line of \eqref{eq:A:4} with $B_1$ replaced with $|B_1|$. Thus, in general we may write $\beta^\prime\equiv \beta/\sqrt{|B_1|}$. 

In matrix form, operator $-D^2$ can be written as 
\begin{align}
	\label{eq:A:7}
	- D^2 = 2B_1\begin{pmatrix}
		a^\dagger a + \frac{1}{2}(1+2\beta^{\prime 2}) & \sqrt{2}\beta^\prime a^\dagger \\
		\sqrt{2}\beta^\prime a & a^\dagger a + \frac{1}{2}(1+2\beta^{\prime 2})
	\end{pmatrix} \,.
\end{align}
Clearly, $D^2$ acts on the Hilbert space $\mathcal{H} = \mathbb{C}^2 \otimes \mathcal{F}$, where $\mathcal{F}$ is the usual Fock space spanned by the eigenstates of the number operator $N = a^\dagger a $. Spectrum of $D^2$ is obtained in a manner similar to that of the Jaynes-Cummings Hamiltonian \cite{Jaynes}. This means that it can be diagonalized in the subspace of $\mathcal{H}_2 \subset \mathcal{H}$ spanned by the states, $\ket{n+1,+}$ and $\ket{n,-}$, where $n$ is the eigenvalue of the number operator $N$ and $\pm$ denotes the isospin up and isospin down, respectively. In this subspace, we can write the matrix elements of $D^2_{(n)}$ as 
\begin{align}
    \label{eq:A:8}
        - D^2_{(n)} := 2B_1\begin{pmatrix}
        n + \frac{1}{2}(1+2\beta^{\prime 2}) & \sqrt{2(n+1)}\beta^\prime\\
        \sqrt{2(n+1)}\beta^\prime & n + \frac{1}{2}(1+2\beta^{\prime 2}) \,.
        \end{pmatrix}
\end{align}
Diagonalizing $D^2_{(n)}$, yields the eigenvalues
\begin{align}
        \label{eq:A:9}
        \Lambda^{\pm}_{n} = 
       2B_1 \left ( n  + \beta^{\prime 2} \pm \sqrt{2\beta^{\prime 2} n +1/4} \right) \,,
       \end{align}
with $n=0,1,2\cdots$, for the upper and $n=1,2,\cdots$ for the lower sign. We note that the ground state is given by $\Lambda^+_0 = 2B_1( \beta^{\prime 2} + 1/2)$, using the upper sign in \eqref{eq:A:9}. Let us remark that the corresponding eigenkets of energy are not simultaneous eigenkets of the isospin operator. This is expected, since $-D^2$ does not commute with the third component of the isospin operator $\sigma_3$ (nor it does with any component of $\vec{\sigma}$ for that matter). As in the Jaynes-Cummings model \cite{Jaynes}, a generalized number operator commuting with $-D^2$ can be constructed. Since we do not need these operators, we will not pursue their construction here.

Let us note the two distinct limiting cases. We may take $B_1 \rightarrow 0$ (hence $y =B_1/E \rightarrow 0$) and $\beta \rightarrow 0$ such that $\beta^\prime$ held fixed. This is the  limit in which $f_0(y, \beta^\prime) \rightarrow 1$ as discussed in section 2.  We may also consider $B_1 \rightarrow 0$, $n_1 \rightarrow \infty$ such that $2 B_1 n_1 \rightarrow k^2$, which gives the configuration with pure non-abelian magnetic field $\beta$. From \eqref{eq:A:9}, we immediately find  $\Lambda_n^\pm \rightarrow \Lambda_\pm = k^2 + 2\beta^2 \pm 2\beta k$. An alternative and rigorous way to derive the spectrum for this case is presented below.

We may start with the gauged operator $ - D^2 = (\vec{\partial} - i \vec{A})^2$, where now $ \vec{A}_{(1)} = A^{SU(2)}_i$. We see that $[- D^2, \vec{p} \, ]= 0$, where $\vec{p} = - i \vec{\partial}$, which means that eigenvalues of $\vec{p}$, say $\vec{k}$, are good quantum numbers and we can express the eigenfunctions of $-D^2$ in the form $\psi = \phi(x_1, x_2) \, e^{i\vec{k}\cdot \vec{x}}$. On the latter, $- D^2$ takes the simple form
\begin{align}
	- D^2 = \begin{pmatrix}
		k^2 + 2\beta^2  &  -2 i \beta k_-\\
		2i \beta k_+  &  k^2 + 2\beta^2
	\end{pmatrix} \,,
\end{align}
where we have defined $k_\pm = k_1 \pm i k_2$ and $k^2 = k_1^2 +k_2^2$. Diagonalizing this matrix, we find the eigenvalues $D^2$ are
\begin{align}
	\Lambda_\pm = k^2 + 2\beta^2 \pm 2\beta k \,.
	\label{flatN0}
\end{align}
Finally, we may write the spectrum of $- D^2_{(1)}  - D^2_{(2)} + m^2$ in $\mathbb{R}^4$ as
\begin{align}
Spec(- D^2_{(1)}  - D^2_{(2)} + m^2) = k^2 + 2\beta^2 \pm 2\beta k + B_2 (2n+1) + m^2 \,.
\end{align}

\section{Spectrum of the gauged Dirac operator on $\mathbb{R}^2$}
\label{sec:B}

Here we determine the spectrum of the square of the Dirac operator introduced in \eqref{eq:2:20}. This operator is given as 
\begin{equation}
\slashed{D}= \gamma_i (\partial^i - i A^i) \,,
\end{equation}
where we have dropped the subscripts $(1)$ in order not to clutter the notation and the $2 \times 2$ $\gamma$-matrices are given as $\gamma_1 = \tau_1$ and $\gamma_2 = \tau_2$, where $\tau_1\,, \tau_2$ are the Pauli matrices.  $\slashed{D}$ can be expressed in the $2 \times 2$ block matrix form as 
\begin{align}
        \label{eq:B:4}
        \slashed{D} = 
        \scalemath{0.75}{
        \begin{pmatrix}
            0 & (\partial_1  - i \partial_2) + \frac{B}{2}(- x_1 + i x_2) + \beta (- \sigma_1 + i \sigma_2) \\
            (\partial_1 + i \partial_2) + \frac{B}{2}(x_1 + i x_2) + \beta (\sigma_1 + i \sigma_2) & 0
        \end{pmatrix}}.
    \end{align}
Using the notation already introduced in the previous section and \eqref{eq:A:6} in $\slashed{D}$, we can write it in the form 
\begin{align}
        \label{eq:B:6}
        \slashed{D} =  - \sqrt{2B}
        \begin{pmatrix}
            0 & a^{\dagger} + 2 \beta^\prime \sigma_-  \\
           - a  - 2 \beta^\prime \sigma_+ & 0
        \end{pmatrix},
    \end{align}
Squaring we find
    \begin{align}
    \label{eq:B:8}
        - \slashed{D}^2 = 2B 
         \scalemath{0.85}{
        \begin{pmatrix}
           a^{\dagger} a + \sqrt{2} \beta^\prime (a\sigma_- + a^{\dagger} \sigma_+) + 2\beta^{\prime 2} \sigma_- \sigma_+ & 0 \\
           0 & a a^{\dagger} + \sqrt{2} \beta^\prime (a\sigma_- + a^{\dagger} \sigma_+) + 2\beta^{\prime 2} \sigma_+ \sigma_-
        \end{pmatrix}} \,.
    \end{align}
Expanding the $2\times2$ blocks in \eqref{eq:B:8}, we can cast $- \slashed{D}^2$ in the form 
\begin{align}
    \label{eq:B:9}
    - \slashed{D}^2 = 2 B
    \begin{pmatrix}
        a^{\dagger}a & \sqrt{2} \beta^\prime a^{\dagger} & 0 & 0 \\
        \sqrt{2} \beta^\prime a & a^{\dagger}a + 2\beta^{\prime 2} & 0 & 0 \\
        0 & 0 & a a^{\dagger} + 2\beta^{\prime 2}  & \sqrt{2} \beta^\prime a^{\dagger} \\
        0 & 0 & \sqrt{2} \beta^\prime a & a a^{\dagger}
    \end{pmatrix}.
\end{align}

Alternatively, we may also note that 
\beqa
-\slashed{D}^2 & = & -\gamma_i \gamma_j D_i D_j \,, \nonumber \\
& =& - D^2  \mathbbm{1}_2 + B_1 (\tau_3 \otimes \mathbbm{1}_2) + 2 \beta^2 (\tau_3 \otimes \sigma_3) \,, 
\label{squaring1}
\eeqa
and this immediately yields \eqref{eq:B:9} upon using \eqref{eq:A:7}.

We may write the eigenvalue equation in the form $- \slashed{D}^2 \Phi = \lambda \Phi$, with $\Phi \equiv (\phi_1\,,\phi_2\,,\phi_3\,,\phi_4)^T$, with $T$ standing for transpose. This leads to the coupled set of operator equations, which can be explicitly written as
\begin{align}
	\label{eq:B:11}
        \omega_c (a^{\dagger} a \phi_1 + \sqrt{2} \beta^\prime a^{\dagger} \phi_2) &= \lambda \phi_1, \nonumber \\
        \omega_c (\sqrt{2}\beta^\prime a \phi_1 +a^{\dagger} a \phi_2 + 2\beta^{\prime 2} \phi_2) &= \lambda \phi_2, \nonumber \\
        \omega_c (a a^{\dagger} \phi_3 + 2\beta^{\prime 2} \phi_3 + \sqrt{2}\beta^\prime a^{\dagger} \phi_4) &= \lambda \phi_3,  \nonumber \\
        \omega_c (\sqrt{2}\beta^\prime a \phi_3 + a a^{\dagger} \phi_4) &= \lambda \phi_4.
    \end{align}

We observe that the eigenkets of the operator $- \slashed{D}^2$ can easily be given in the tensor product space $\mathcal{H} = \mathcal{F} \times \mathbb{C}^2 \otimes \mathbb{C}^2 \equiv \mathcal{F} \otimes \mathbb{C}^4 $. Here, the first copy of $\mathbb{C}^2$ stands for the spin, and the second copy for the isospin space and the Fock space is spanned by the eigenstates of the number operator $N = a^\dagger a$ as before. In order to solve $\eqref{eq:B:11}$, it is sufficient to consider the subspace $\mathcal{H}_n \subset \mathcal{H}$ spanned by the states $\lbrace \ket{n+1,+,+}\,, \ket{n,+,-} \,,\ket{n,-,+}\,, \ket{n-1,-,-} \rbrace$, where $n=0,1,2,\cdots$, except for the last ket, for which $n=1,2,\cdots$. We may write these kets in the form 
\begin{align}
 \label{eq:B:12}
    \ket{n+1,+,+}  \equiv 
    \left(\begin{smallmatrix}
      \ket{n+1} \\
                0 \\
                0 \\
                0
    \end{smallmatrix}\right) \,, 
    \ket{n,+,-} \equiv
    \left(\begin{smallmatrix}
        0 \\
        \ket{n} \\
        0 \\
        0
    \end{smallmatrix}\right) \,, 
    \ket{n,-,+} \equiv
    \left(\begin{smallmatrix}
        0 \\
        0 \\
        \ket{n} \\
        0
    \end{smallmatrix}\right) \,, 
    \ket{n-1,-,-} \equiv
    \left(\begin{smallmatrix}
        0 \\
        0 \\
        0 \\
        \ket{n-1}
    \end{smallmatrix}\right).
\end{align}
In this subspace we easily find that
 \begin{align}
  \label{eq:B:13}
  - \slashed{D}^2_{(n)} = 2 B
   \begin{pmatrix}
            n+1 & \sqrt{2} \beta^\prime \sqrt{n+1} & 0 & 0 \\
            \sqrt{2} \beta^\prime \sqrt{n+1} & n + 2\beta^{\prime 2} & 0 & 0 \\
            0 & 0 & n + 1 + 2\beta^{\prime 2}  & \sqrt{2} \beta^\prime \sqrt{n} \\
            0 & 0 & \sqrt{2} \beta^\prime \sqrt{n} & n
        \end{pmatrix} \,,
    \end{align}
and the eigenvalues of $\slashed{D}^2$ can are then readily computed to be 
    \begin{align}
       \lambda^{\pm}_{n} = B_1 \left ( 1+2n+2\beta^{\prime 2} \pm \sqrt{1 + 4\beta^{\prime 2}(1+2n+\beta^{\prime 2})} \right ) \,,
       \label{fdirac}
    \end{align}
with each eigenvalue occurring with multiplicity $2$. Let us note that it is legitimate to take $n=0$ in this expression. This yields $\lambda^{+}_{0} = 2 B_1 (1 + 2 \beta^{\prime 2})$, and $\lambda^{-}_{0} = 0$. Corresponding eigenkets can be determined with some care. Eigenkets of these states belong to the subspace spanned by the set $\lbrace \ket{0,+,+}\,, \ket{1,+,+}\,, \ket{0,+,-} \,,\ket{0,-,+} \rbrace$, where we note that the state $\ket{0,+,+}$ is not covered by the notation for $\mathcal{H}_n$ introduced above, but clearly belongs to the Hilbert space $\mathcal{H}$.  We find that the zero mode solutions are 
\begin{align}
	\label{eq:B:17}
	\ket{0,+,+} = 
	\begin{pmatrix}
	\ket{0}\\
		0\\
		0\\
		0\\
	\end{pmatrix} \,, \quad 
	\frac{(-\sqrt{2} \beta^\prime \ket{1,+,+} + \ket{0,+,-})}{\sqrt{1 + 2 \beta^{\prime 2}}}
	=\frac{1}{\sqrt{1 + 2 \beta^{\prime 2}}}
	\begin{pmatrix}
		-\sqrt{2} \beta^\prime \ket{1}\\
		\ket{0} \\
		0\\
		0\\
	\end{pmatrix}\,,
\end{align}
while the eigenkets for the $\lambda^+_0$ eigenvalue are 
\begin{align}
	\label{eq:B:18}
	\ket{0,+,+} = 
	\begin{pmatrix}
		0\\
		0\\
		\ket{0}\\
		0\\
	\end{pmatrix} \,, \quad 
	\frac{( \ket{1,+,+} + \sqrt{2} \beta^\prime \ket{0,+,-})}{\sqrt{1 + 2 \beta^{\prime 2}}}
	=\frac{1}{\sqrt{1 + 2 \beta^{\prime 2}}}
	\begin{pmatrix}
	 \ket{1}\\
	\sqrt{2} \beta^\prime \ket{0} \\
		0\\
		0\\
	\end{pmatrix}.
\end{align}

For $n \geq 1$, we can write the corresponding orthonormal eigenvectors associated with the eigenvalues \eqref{fdirac} in the generic form 
\begin{align}
  \label{eq:B:16}
  |\psi^{a_{\pm}}\rangle_n &= \frac{1}{\sqrt{{a}_\pm^2 +1}}\left ( a_\pm  \ket{n+1,+,+}  +  \ket{n,+,-} \right )  \,, \nonumber \\
  |\psi^{b_{\pm}}\rangle_n &= \frac{1}{\sqrt{{b}_\pm^2 +1}}\left ( b_\pm  \ket{n,-,+}  +  \ket{n-1,-,-} \right ) \,, 
\end{align}
where    	
\begin{align}
\label{eq:B:15}
  a_{\pm} &= \frac{1-2\beta^{\prime 2} \pm \sqrt{1 + 4\beta^{\prime 2}(1+2n+\beta^{\prime 2})}}{2\sqrt{2n+2} \beta^\prime} \,, \nonumber \\
  b_{\pm} &= \frac{1+2\beta^{\prime 2} \pm \sqrt{1 + 4\beta^{\prime 2}(1+2n+\beta^{\prime 2})}}{2\sqrt{2n} \beta^\prime} \,.
\end{align}
Here, $\psi^{a_{\pm}}_n$ are the eigenvectors corresponding to the eigenvalues $\lambda^+_n$ and $\psi^{b_{\pm}}_n$ are those corresponding to $\lambda^-_n$. Let us also note that these states are also simultaneous eigenstates of spin, since $\slashed{D}^2$ commutes with the spin operator $\gamma_3 \otimes \mathbbm{1}_2 = - i \gamma_1 \gamma_2 \otimes \mathbbm{1}_2 = \tau_3 \otimes \mathbbm{1}_2$. $|\psi^{a_{\pm}}\rangle_n$ correspond to spin up and $|\psi^{b_{\pm}}\rangle_n$ to spin down (Note that $\pm$ signs are not indicating spin direction). These are not eigenstates of isopin though, since $D^2$ and hence $\slashed{D}^2$ does not commute with $\mathbbm{1}_2 \otimes \sigma_3$. We may construct a generalized number operator commuting with both $\slashed{D}^2$ and the isospin operators, but as this is not necessary for our purposes, we do not pursue it here.

To obtain the eigenvalues of $- \slashed{D}^2$ in the case of pure non-abelian magnetic field configuration, we may proceed as follows. Setting $B_1=0$ in \eqref{squaring1} yields $- \slashed{D}^2 = - D^2 + 2 \beta^2 \tau_3 \otimes \sigma_3$. Since $\slashed{D}$ commutes with $\vec{p}$, eigenvalues of the latter are good quantum numbers. As before, we may denote these eigenvalues as $\vec{k}$. Thus $-\slashed{D}^2$ can be written in matrix form as 
\begin{align}
-\slashed{D}^2 
&=\begin{pmatrix}
	k^2 & -2i\beta k_-   &  0  &  0\\
	2i\beta k_+ & k^2 + 4\beta^2 &  0  &  0 \\
	0  &  0 & k^2+4\beta^2 & -2i\beta k_- \\
	0  &  0 &  2i\beta k_+  & k^2
\end{pmatrix} \,.
\end{align}
Diagonalization gives the eigenvalues 
\be
\lambda_\pm =  k^2 + 2\beta^2 \pm 2\beta \sqrt{k^2 +\beta^2} \,,
\label{conspecDirac1}
\ee
with each eigenvalue occurring with multiplicity 2. We remark that, in this case too, $\slashed{D}^2$ commutes with the spin operator $\gamma_3 \otimes \mathbbm{1}_2 = - i \gamma_1 \gamma_2 \otimes \mathbbm{1}_2 = \tau_3 \otimes \mathbbm{1}_2$, but do not commute with the isospin. Therefore the eigenstates of $\slashed{D}^2$ are simultaneous eigenstates of $\gamma_3$ but not $\sigma_3$. Observe also that the $\pm$ signs in the eigenvalues are not related with the direction of the spin, in fact, for each sign $\pm$ in $\lambda_\pm$ there is a state with spin up and a state with spin down. Let also note that \eqref{conspecDirac1} can also be obtained from \eqref{fdirac} by taking the limit $B_1 \rightarrow 0$, $n \rightarrow \infty$ such that $2 B_1 n \rightarrow k^2$. 

We can write the spectrum of $- \slashed{D}^2$ on $\mathbb{R}^4$. Together with the abelian magnetic field $F_{34} =B_2$ on the second $\mathbb{R}^2$ copy, we have, 
\begin{align}
Spec(- \slashed{D}^2 + m^2) = 
\begin{cases}
	k^2 + 2\beta^2 \pm 2\beta  \sqrt{k^2 + \beta^2}+ 2 n B_2 \,, \\
	k^2 + 2\beta^2 \pm 2\beta  \sqrt{k^2 +\beta^2} + (2n+2) B_2 \,.
\end{cases}
\end{align}

\section{Spectrum of the gauged Laplacian on $S^2$}
\label{sec:C}
Here, we outline the result obtained already in \cite{Estienne_2011}. We consider the following gauged Laplace operator on $S^2$ with radius $a$
\begin{align}
	\label{eq:C:3}
	D^2 = \frac{\Vec{\Lambda}^2}{a^2}\,,
\end{align}
where $\Vec{\Lambda}$ is given as
\begin{align}
\label{eq:C:4}
\Vec{\Lambda} \equiv \Vec{r} \times (\Vec{p} - \Vec{A}).
\end{align}
and the $SU(2) \times U(1)$ gauge field $\vec{A}$ is explicitly written as
    \begin{align}
    \label{eq:C:1}
        \Vec{A} = \Vec{A}_{abelian} + \vec{A}_{non-abelian}\,, \quad \vec{A}_{non-abelian} := \alpha \frac{\Vec{r}\times \Vec{\sigma}}{a^2} \,. 
    \end{align}
In \eqref{eq:C:1}, $\vec{A}_{abelian}$ stands for the gauge potential of a Dirac monopole with magnetic charge $N/2$, $N \in \mathbb{Z}$, and $\vec{\sigma}$ are the Pauli matrices spanning the "isospin" $SU(2)$ gauge symmetry. Associated field strength is computed via $\vec{B} = \vec{\nabla} \times \vec{A}$ and yields a radial magnetic field; which takes the form \footnote{Note that the choice of gauge for $\vec{A}_{abelian}$ is immaterial for our purposes.}
\begin{align}
 \label{eq:C:2}
 B = \frac{N}{2a^2} + \left (2 \left ( \alpha -\frac{1}{2}\right )^2 -\frac{1}{2} \right) \frac{\Vec{\sigma} \cdot \hat{r}}{a^2} \,.
 \end{align}
From \eqref{eq:C:2} it is manifestly seen that $B$ is symmetric under $\alpha \rightarrow (\alpha-1)$, which is a direct consequence of the gauge transformation $U B \, (\alpha) U = B \, (\alpha-1) $, with $U := \Vec{\sigma} \cdot \hat{r}$, $U^\dagger = U$, $U^2 = \mathbbm{1}_2$.

We may write, in the same manner as in the Landau problem on $S^2$ \cite{Haldane}
\begin{align}
	\label{eq:D:3}
	\vec{\Lambda} &= \vec{L} + \frac{N}{2} \hat{r} + \alpha(\vec{\sigma} -  (\Vec{\sigma} \cdot \hat{r}) \, \hat{r}) \,, \nonumber \\
	 & =  \vec{\Lambda}_{abelian} + \alpha(\vec{\sigma} -  (\Vec{\sigma} \cdot \hat{r}) \, \hat{r}) \,.
\end{align}
where $\vec{L} = \vec{\Lambda}_{abelian} -  \frac{N}{2}\hat{r} $ is the angular momentum solely generated by the charge-Dirac monopole system. The total angular momentum operator is found by adding the contribution of the isospin: 
\begin{align}
	\label{eq:C:6}
	\Vec{J} &= \vec{L} + \frac{\Vec{\sigma}}{2} = \Vec{r} \times (\Vec{p} - \Vec{A}_{abelian}) - \frac{N}{2}\hat{r}  + \frac{\Vec{\sigma}}{2} \,, \nonumber \\
	& = \Vec{\Lambda}_{abelian}  - \frac{N}{2}\hat{r} + \frac{\Vec{\sigma}}{2} \,, \nonumber \\
	& = \Vec{\Lambda} -  \alpha(\vec{\sigma} -  (\Vec{\sigma} \cdot \hat{r}) \, \hat{r}) - \frac{N}{2}\hat{r} + \frac{\Vec{\sigma}}{2} \,.
\end{align}
We have,
\be
\left ( \Vec{J} + (\alpha -\frac{1}{2}) \, \Vec{\sigma} \right )^2 =  \left ( \Vec{\Lambda} + \alpha (\Vec{\sigma} \cdot \hat{r} - \frac{N}{2}) \hat{r} \right )^2 \,. 
\ee
Upon using $\Vec{\Lambda}  \cdot \hat{r} = 0 = \hat{r} \cdot \Vec{\Lambda}$ and rearranging the terms this yields
\begin{multline}
 \label{eq:C:5}
 D^2  = \frac{1}{a^2} \Big (\Vec{J}^2 + \frac{1}{4} -  \frac{N^2}{4} + 2 \left ( \alpha -\frac{1}{2} \right )^2 -\frac{1}{2} + 2 \left (\alpha - \frac{1}{2} \right) ( \Vec{J}\cdot \Vec{\sigma} -\frac{1}{2} + \frac{N}{2} \Vec{\sigma} \cdot \hat{r} )+ \frac{N}{2} \Vec{\sigma} \cdot \hat{r} \Big ) \,.
 \end{multline}
It is useful to define $X := 2 \left ( \alpha - \frac{1}{2} \right) \left (\Vec{J}\cdot \Vec{\sigma} -\frac{1}{2} + \frac{N}{2} \Vec{\sigma} \cdot \hat{r} \right )+ \frac{N}{2} \Vec{\sigma} \cdot \hat{r}$. It squares to 
\be
X^2 = 4 \left (\alpha -\frac{1}{2} \right )^2 (\Vec{J\,}^2 + \frac{1}{4}) - \left ( \left ( \alpha -\frac{1}{2} \right )^2 -\frac{1}{4} \right ) N^2 \,,  
\ee
which is diagonal in the total angular momentum basis. The latter can take on the values given by the tensor product $j \equiv  l \otimes \frac{1}{2} = (l - 1/2) \oplus (l + 1/2)$, where $l$ is the angular momentum of the charge monopole system. Setting $l = n_1 + \frac{N}{2}$, possible values of $j$ are $j = n_1 + \frac{N-1}{2}$ and $j = n_1 + \frac{N+1}{2}$ with $n_1 =0,1,2,\dots$ and $N =1,2,\dots$. To be more precise, the Hilbert space becomes block diagonal in the total angular momentum basis and splits into the direct sum of irreducible representations
\be 
\left ( \frac{N-1}{2} \right ) \oplus \bm{2} \left ( \frac{N+1}{2} \right ) \oplus \bm{2} \left ( \frac{N+3}{2} \right ) \oplus \cdots \,,
\ee
where the coefficients written in bold typeface denote the multiplicities of the respective IRRs. Except the first IRR, which corresponds to the ground state with $n_1=0$, each IRR occurs twice. In the $j = n_1 + \frac{N+1}{2}$ branch we may shift $n_1 \rightarrow n_1 - 1$ and write the spectrum of $D^2$ as 
\begin{align}
 \label{eq:C:7}
 \Lambda^\pm_{n_1}(\alpha) =  \frac{1}{a^2} \left ( n_1 (N + n_1) + 2 \left ( \alpha -\frac{1}{2} \right )^2 -\frac{1}{2}  \pm \sqrt{4 \left (\alpha- \frac{1}{2} \right)^2 (n_1 + N)n_1 + \frac{N^2}{4}} \right ) \,.
 \end{align}
 We observe that $\Lambda^\pm_{n_1}(\alpha) = \Lambda^\pm_{n_1}(1-\alpha)$. Ground state energy is given by $ \Lambda^+_{0}=  \frac{1}{a^2} \Big(2 \left ( \alpha -\frac{1}{2} \right )^2$ $-\frac{1}{2} + \frac{N}{2} \Big)$, where we have taken $n_1=0$ and the $+$ sign in front of the square root term. The latter follows from the continuity of the energy spectrum as $\alpha \rightarrow 0$ matching the ground state energy of the Landau problem on the sphere \cite{Haldane}. Hence, we have $n_1 = 0,1,2,\dots$ for $\Lambda^+_{n_1}$ and $n_1 = 1,2,\dots$ for $\Lambda^-_{n_1}$. Let us immediately note that taking the limit $N\rightarrow \infty$, $a \rightarrow \infty$, $\alpha \rightarrow \infty$, such that $B_1 = \frac{N}{2 a^2}$, $\beta^2= \frac{\alpha^2}{a^2}$ fixed and $\beta^{\prime 2} = \frac{\beta^2}{B_1}$, yields the spectrum of $D^2$ on ${\mathbb R^2}$ given in \eqref{eq:A:9} as expected.  

It is straightforward to see that $D^2$ and $\vec{J}$ commute. We may note that the only nontrivial commutators are$ \lbrack J_i, \Vec{\sigma} \cdot \hat{r} \rbrack $ and $\lbrack J_i , \vec{J}\cdot \vec{\sigma} \rbrack$. These vanish as the following calculations demonstrate:
\beqa
\label{eq:C:9}
[J_i, \Vec{\sigma} \cdot \hat{r}] & =& \Big[\epsilon_{ijk} r_j   (p_k-A_k) -\frac{N}{2}\frac{r_i}{\sqrt{r_l r_l}}+ \frac{\sigma_i}{2}, \frac{\sigma_n  r_n}{\sqrt{r_m r_m}}\Big], \nonumber\\
&=& \epsilon_{ijk}\Big[r_j (p_k-A_k),\frac{\sigma_n r_n}{\sqrt{r_m r_m}}\Big] - \frac{N}{2}\Big[\frac{r_i}{\sqrt{r_l r_l}},\frac{\sigma_n r_n}{\sqrt{r_m r_m}}\Big] + \frac{r_n}{2r} [\sigma_i, \sigma_n], \nonumber\\
&=& \epsilon_{ijk}r_j\Big[(p_k-A_k),\frac{\sigma_n r_n}{\sqrt{r_m r_m}}\Big] + \epsilon_{ijk}\Big[r_j,\frac{\sigma_n r_n}{\sqrt{r_m r_m}}\Big](p_k-A_k) + \frac{r_n}{2r} [\sigma_i, \sigma_n], \nonumber\\
&=& \epsilon_{ijk} r_j\Big[p_k,\frac{\sigma_n r_n}{\sqrt{r_m r_m}}\Big] + \frac{r_n}{2r} [\sigma_i, \sigma_n] \,, \nn \\
& = & -i\epsilon_{ijk} r_j \sigma_n \frac{r^2 \delta_{nk}-r_n r_k}{r^3} + \frac{r_n}{2r} [\sigma_i, \sigma_n], \nonumber \\
&=& -i\epsilon_{ijk} \frac{r_j}{r} \sigma_n \delta_{nk} + \frac{r_n}{2r} [\sigma_i, \sigma_n] \nn \\
& = & -i\epsilon_{ijk} \frac{r_j}{r} \sigma_k +  \frac{r_n}{2r} 2 i\epsilon_{ink} \sigma_k \,, \nn \\
& =& 0 \,.
\eeqa
\beqa
 \label{eq:C:8}
        [J_i, J_j \sigma_j] &=& [L_i + \frac{1}{2} \sigma_i, (L_j + \frac{1}{2}\sigma_j) \sigma_j] \,, \nonumber \\
                            &=& [L_i,L_j]\sigma_j + L_j[L_i,\sigma_j] + \frac{1}{2} [\sigma_i,L_j]\sigma_j + \frac{1}{2} L_j[\sigma_i,\sigma_j], \nonumber \\
                            &=& [L_i,L_j]\sigma_j + \frac{1}{2} L_j[\sigma_i,\sigma_j] \nn \\
                            &=& i\epsilon_{ijk}(L_k \sigma_j + L_j \sigma_k) \,, \nn \\
                            & =& 0 \,.
\eeqa
Therefore, we conclude that each energy level in \eqref{eq:C:7} is $(2j+1)$-fold degenerate. The degeneracy of each branch of these energy levels is thus the same as that of the Landau problem on the sphere \cite{Haldane}. In particular, the ground level with energy $\Lambda^+_{0}$ is $N$-fold degenerate.

In the absence of the abelian magnetic field, i.e. setting $N=0$, the operator $D^2$ takes the form 
\begin{align}\label{eq:d2p}
D^2= \frac{1}{a^2} \left (J^2 + \frac{1}{4} + 2 \left ( \alpha -\frac{1}{2} \right )^2 -\frac{1}{2} + 2 \left (\alpha-\frac{1}{2} \right) (\vec{J}\cdot \vec{\sigma} -1/2) \right ) \,.
\end{align}
while $X = 2 \left ( \alpha - \frac{1}{2} \right)  (\vec{J}\cdot \vec{\sigma} -1/2) = 2 \left ( \alpha - \frac{1}{2} \right) (\vec{J}^2 -\vec{L}^2 + 1/4)$ with the eigenvalues  $2 \left ( \alpha - \frac{1}{2} \right) (l+1)$ for the IRR $j = l + 1/2$, ($l=0,1,2,3,...$) and $- 2 \left ( \alpha - \frac{1}{2} \right) l $ for the IRR $j = l - 1/2$, ($l=1,2,3,...$). Thus, the Hilbert space splits into the direct sum
\be
\bm{2} \left ( \frac{1}{2} \right ) \oplus \bm{2} \left ( \frac{3}{2} \right ) \oplus \cdots \,.
\ee
Spectrum of $D^2$ can be written as
\begin{subequations}
  \begin{align}
    \Lambda_{l, N=0}^+ &= \frac{1}{a^2} \left ( (l+1)^2 + 2 \left ( \alpha -\frac{1}{2} \right )^2 -\frac{1}{2} + 2 \left ( \alpha - \frac{1}{2} \right) (l+1) \right) \,, \quad l =0,1,2\cdots \,, \\ 
    \Lambda_{l, N=0}^- &= \frac{1}{a^2} \left ( l^2 + 2 \left ( \alpha -\frac{1}{2} \right )^2 -\frac{1}{2} - 2 \left ( \alpha - \frac{1}{2} \right) \, l \right) \,, \quad l =1,2\cdots \,.
  \end{align}
  \label{eq:C:16}
\end{subequations}
Note that as $\alpha \rightarrow 0$, we have that $D^2 = \vec{L}^2/a^2$, which has the spectrum $\frac{1}{a^2} l (l+1)$, $l =0,1,2,\cdots$ and $\Lambda_{l, N=0}^+$ gives a zero mode at $l=0$. We remark that, in this limit, $D^2$ becomes independent of $\vec{J}^2$ as expected and not only in $\Lambda_{l, N=0}^+$ but also in $\Lambda_{l, N=0}^-$ we have $l=0,1,2,\cdots$.  Shifting $l \rightarrow l-1 $ in $\Lambda_{l, N=0}^+$, we may write the spectrum more compactly as $\Lambda_{l, N=0}^\pm (\alpha) = \frac{1}{a^2} \left ( l^2 + 2 \left ( \alpha -\frac{1}{2} \right )^2 -\frac{1}{2} \pm 2 \left ( \alpha - \frac{1}{2} \right) \, l \right)$ with $l =1,2\cdots$ for $\alpha \neq 0$, while for $\alpha =0$ only, $l = 0,1,\cdots$ for the lower sign. Let us also remark that $\Lambda_{l, N=0}^\pm (\alpha) = \Lambda_{l, N=0}^\mp (1- \alpha)$ indicates that the spectrum remains the same under $\alpha \leftrightarrow 1 - \alpha$. As consequence, at $\alpha = 1$, $\Lambda_{l, N=0}^-$ is a zero mode at $l = 1$.

Taking $l \rightarrow \infty$, $a \rightarrow \infty$, $\gamma \rightarrow \infty$,  such that $\frac{l}{a} \rightarrow k$, $\frac{\alpha^2}{a^2}\rightarrow \beta$ remain finite, we obtain the spectrum on ${\mathbb R^2}$ given in \eqref{flatN0}.
 
\section{Spectrum of the gauged Dirac operator on $S^2$}
\label{sec:D}
We consider the Dirac operator in the background of the total magnetic field introduced in \eqref{eq:C:1}. This Dirac operator can be written as $\slashed{D} = \frac{1}{a} (\vec{\tau} \cdot \vec{\Lambda} +1) $ where $\vec{\tau}$ are the Pauli matrices, spanning the Clifford algebra $\{\tau_i \,, \tau_j \} = 2 \delta_{ij}$, and $\vec{\Lambda}$ is defined as previously in \eqref{eq:C:4}. For the square of the Dirac operator we have, 
        \begin{align}
        \label{eq:D:4}
            a^2 \slashed{D}^2 &= (\vec{\tau} \cdot \vec{\Lambda} +1)^2, \nonumber  \\ 
            			  &= (\vec{\tau} \cdot \vec{\Lambda})^2 + 2 \vec{\tau} \cdot \vec{\Lambda} +1 \nonumber \,,  \\ 
                          &= \tau_i \tau_j \Lambda_i \Lambda_j = (\delta_{ij} + i \epsilon_{ijk} \tau_k)\Lambda_i \Lambda_j + 2 \vec{\tau} \cdot \vec{\Lambda} +1 \,, \nonumber \\
                          &= \Lambda^2 + \frac{i}{2}  \epsilon_{ijk} [\Lambda_i,\Lambda_j] \tau_k +2 \vec{\tau} \cdot \vec{\Lambda} +1\,, \nonumber\\
                          &= \Lambda^2 + \frac{i}{2} \epsilon_{ijk}[L_i + \frac{N}{2} \hat{r}_i + \alpha(\sigma_i - \sigma_n \hat{r}_n \hat{r}_i),L_j + \frac{N}{2} \hat{r}_j + \alpha(\sigma_j - \sigma_m \hat{r}_m \hat{r}_j)]\tau_k + 2 \vec{\tau} \cdot \vec{\Lambda} +1 \,,\nonumber \\ 
                          &= \Lambda^2 + \frac{i}{2} \epsilon_{ijk}\Big([L_i,L_j] + \frac{N}{2}[L_i, \hat{r}_j] + \frac{N}{2}[ \hat{r}_i , L_j] - \alpha [L_i, \sigma_m \hat{r}_m \hat{r}_j] - \alpha [\sigma_n \hat{r}_n \hat{r}_i,L_j ] \,,\nonumber\\ 
                          & -\alpha^2 [\sigma_i,\sigma_m \hat{r}_m \hat{r}_j]-\alpha^2 [\sigma_n \hat{r}_n \hat{r}_i, \sigma_j] + \alpha^2 [\sigma_i, \sigma_j] + \alpha^2  [\sigma_n \hat{r}_n \hat{r}_i,\sigma_m \hat{r}_m \hat{r}_j]\Big)\tau_k + 2 \vec{\tau} \cdot \vec{\Lambda} +1 \,,\nonumber\\
                          &= \Lambda^2 + \frac{i}{2} \epsilon_{ijk}\Big([L_i,L_j] + \frac{N}{2}[L_i, \hat{r}_j] + \frac{N}{2}[ \hat{r}_i , L_j] - \alpha \sigma_m  [L_i, \hat{r}_m \hat{r}_j] - \alpha\sigma_n[ \hat{r}_n \hat{r}_i,L_j ] \,, \nonumber\\ 
                          & -\alpha^2 \hat{r}_m \hat{r}_j [\sigma_i,\sigma_m ]-\alpha^2 \hat{r}_n \hat{r}_i [\sigma_n, \sigma_j] + \alpha^2 [\sigma_i, \sigma_j] + \alpha^2 \hat{r}_n \hat{r}_i  \hat{r}_m \hat{r}_j [\sigma_n ,\sigma_m]\Big)\tau_k + 2 \vec{\tau} \cdot \vec{\Lambda} +1 \,, \nonumber\\
                          &= \Lambda^2 + \vec{\tau} \cdot \vec{\Lambda} - \frac{N}{2} \vec{\tau} \cdot \hat{r}  + 1 - 2 \left ( \left (\alpha -\frac{1}{2} \right)^2 -\frac{1}{4} \right) (\vec{\tau}\cdot \hat{r})(\vec{\sigma}\cdot \hat{r})\,.
        \end{align}
Using $\vec{J} = \vec{L} + \frac{\vec{\sigma}}{2}$, we have the intermediate expression
        \begin{multline}
        \label{eq:D:5}
           a^2 \slashed{D}^2 = J^2 + 2 \left(\alpha - \frac{1}{2}\right) \vec{J}\cdot\vec{\sigma} + \frac{3}{4} - 3\alpha + 2\alpha^2 + N \, \alpha \, \vec{\sigma}\cdot \hat{r} - \frac{N^2}{4} + \vec{\tau} \cdot \vec{J}  \\
                            + \left(\alpha - \frac{1}{2}\right) \vec{\tau} \cdot \vec{\sigma} - 2 \alpha  \left(\alpha - \frac{1}{2}\right)  (\vec{\sigma} \cdot \hat{r})(\vec{\tau} \cdot \hat{r}) + 1 \,.
        \end{multline}
At this stage, we may introduce the total angular momentum operator $\vec{K}$ and the operator $\vec{\tau} \cdot \vec{J}$ as
\begin{subequations}
  \label{eq:D:6}
  \begin{align}
    \label{eq:D:6:a}
    \vec{K} &= \vec{J} + \frac{\vec{\tau}}{2} = \vec{L} + \frac{\vec{\sigma}}{2} + \frac{\vec{\tau}}{2}, \\
    \label{eq:D:6:b}
    \vec{\tau} \cdot \vec{J} &= K^2 - J^2 - \frac{\tau^2}{4}\,.
  \end{align}
\end{subequations}
These allow us to express $\slashed{D}^2$ in the form
\begin{multline}
\label{eq:D:7}
           a^2 \slashed{D}^2 = K^2 - \left(\frac{N^2}{4} - 2 \left (\alpha - \frac{1}{2} \right)^2 \right) \\
                            + \left [ 2 \left(\alpha - \frac{1}{2}\right ) (\vec{K}\cdot\vec{\sigma}-\frac{1}{2}) + N \alpha \, \vec{\sigma} \cdot \hat{r} - 2 \alpha  \left(\alpha - \frac{1}{2}\right ) (\vec{\sigma}\cdot\hat{r}) (\vec{\tau}\cdot\hat{r}) \right ] \,.
        \end{multline}
In this expression first two terms are already diagonal, but we have to diagonalize the operator in the brackets $\chi : = 2 \left(\alpha - \frac{1}{2}\right ) (\vec{K} \cdot \vec{\sigma} - \frac{1}{2}) + N\, \alpha \, \vec{\sigma} \cdot \hat{r} -2 \, \alpha  \left(\alpha - \frac{1}{2}\right) (\vec{\sigma} \cdot \hat{r})(\vec{\tau} \cdot \hat{r})$. Squaring it, we find
\begin{align}
        \label{eq:D:9}
            \chi^2 = 4 \left(\alpha - \frac{1}{2}\right)^2 \left( K^2 + \frac{1}{4} \right) - \left ( \left (\alpha -\frac{1}{2} \right)^2 -\frac{1}{4} \right) \left (N^2 - 4\left(\alpha - \frac{1}{2}\right )^2 \right ) \,,
        \end{align}
which is diagonal in the total angular momentum basis. We let $l$ represent the angular momentum of the charge- Dirac monopole system as before. Then, the total angular momentum $K$ could take on the possible values given by the tensor product
\begin{align}
	\label{eq:D:11}
	k \equiv  & \, l \otimes \frac{1}{2} \otimes \frac{1}{2} \nonumber \\
	\equiv &  \, (l+1) \oplus \pmb{2} l \oplus (l-1) \nonumber \\
	= &\,  (n_1 + \frac{N}{2} + 1) \oplus \pmb{2} (n_1 + \frac{N}{2}) \oplus (n_1 + \frac{N}{2} - 1) \,,
\end{align}
where in the last line we have used $l = n_1 + \frac{N}{2}$, with $N =1,2,\dots$ and $n_1 =0,1,2,\cdots$ except in the last direct summand at $N=1$ for which $n_1 =1,2,\cdots$.  Thus, at a given orbital angular momentum spectrum of $\slashed{D}^2$ has four distinct eigenvalues and we may express the spectrum as
\begin{subequations}
    \label{eq:D:12}
    \begin{align}
        \label{eq:D:12:a}
        \lambda_{n_1 + 1}(\alpha) =\, & \frac{1}{a^2} \left ( \xi_{n_1+1} + \left(\alpha - \frac{1}{2}\right)^2 - \sqrt{4 \left(\alpha - \frac{1}{2}\right)^2 \xi_{n_1+1} + \frac{N^2}{4} } \right)\,, \\
        \label{eq:D:12:b}
        \lambda^\pm_{n_1}(\alpha) =\, &  \frac{1}{a^2} \left (  \xi_{n_1} + \left(\alpha - \frac{1}{2}\right)^2 \pm \sqrt{4 \left(\alpha - \frac{1}{2}\right)^2 \xi_{n_1} + \frac{N^2}{4}} \right )\,, \\
        \label{eq:D:12:c}
        \lambda_{n_1-1}(\alpha) = \, & \frac{1}{a^2} \left ( \xi_{n_1-1} + \left(\alpha - \frac{1}{2}\right)^2 + \sqrt{4 \left(\alpha - \frac{1}{2}\right)^2 \xi_{n_1-1} +\frac{N^2}{4}} \right ) \,,
    \end{align}
\end{subequations}
where
\begin{align}
\label{eq:D:13}
\xi_{n_1}(\alpha) = \left ( n_1 + \frac{1}{2} \right)^2 + N  \left( n_1 + \frac{1}{2} \right) + \left (\left(\alpha - \frac{1}{2}\right)^2 - \frac{1}{4} \right) \,.
\end{align}
and each eigenvalue being $(2k +1)$-fold degenerate with $k =l+1 \,, l \,, l-1$ as given in \eqref{eq:D:11}. Clearly the spectrum is symmetric under $\alpha \leftrightarrow 1-\alpha$. For $N \geq 2$, this spectrum has zero modes at the branches $\lambda_{n_1}^-$ and $\lambda_{n_1-1}$ for $n_1 = 0$. For  $N=1$, we have from \eqref{eq:D:12:c} that the set of eigenvalues $\lambda_{n_1-1}$ start with $n_1 = 1$, since the IRR $(n_1 + \frac{N}{2} - 1)$ does not exist at $N=1 \,, n_1 = 0$, meaning that this branch does not include a zero mode, while the spectrum of $\slashed{D}^2$ retains the zero mode from the branch $\lambda_{n_1}^-$. Let us recall that at $N=0$ and $\alpha =0$, i.e. in the absence of the entire magnetic background, the Dirac operator on $S^2$ has no zero modes. In a similar manner, for the present problem, absence of the zero mode in the branch $\lambda_{n_1-1}$ at $N = 1$ can therefore be understood as the insufficient contribution of the Dirac monopole flux to the total angular momentum to compensate the total spin and isospin down contributions to the latter.

To make these results manifestly clear, firstly, let us note that as $\alpha \rightarrow 0$, we have
\begin{align}
\label{eq:D:14}
\chi {\big |}_{\alpha = 0} = : \chi_0 = - (\vec{K} \cdot \vec{\sigma} - 1/2) = - (\vec{K}^2 - \vec{M}^2 + 1/4) \,,
\end{align}
where $\vec{M} = \vec{L} + \frac{\vec{\tau}}{2}$ and from \eqref{eq:D:7} we immediately recover $a^2 \slashed{D}^2 = \vec{M}^2 - \frac{N^2}{4} + \frac{1}{4}$, which is the standard form of the Dirac operator on $S^2$ in the presence of the Dirac monopole \cite{FuzzySUSY} and manifestly independent of $\vec{K}^2$ and depends on $\vec{M}^2$. Its spectrum can easily be written and matches with that obtained from \eqref{eq:D:12} by taking $\alpha \rightarrow 0$. This yields one copy of the set of eigenvalues $(n_1+1)^2 + (n_1+1) N$ and $n_1^2 + n_1 N$ for isospin up and one for isospin down with $n_1=0,1,\dots$ and $N \geq1$, with the zero modes at $n_1=0$ from the latter set for both the isospin up and down configurations.

In the present case, for the branch $\lambda_{n_1-1}$ at $n_1 =0$ and $N \geq 2$, we have to be careful about the square root term in the r.h.s. of \eqref{eq:D:12:c}. In this case, we find that $\chi^2$ has the eigenvalue $(\frac{N}{2} - 2 (\alpha - \frac{1}{2})^2)^2$. Above, we have already seen how the spectrum and in particular the zero modes of $\slashed{D}^2$ are obtained in the $\alpha \rightarrow 0$ limit. By continuity in $\alpha$, we conclude that the correct eigenvalue of $\chi$ is $\frac{N}{2} - 2 (\alpha - \frac{1}{2})^2$ (which could be positive or negative depending on the values of $N$ and $\alpha$) and this yields the zero mode in the indicated branch. In other words, for $n_1 =0$ and $N \geq 2$, $\sqrt{4 \left(\alpha - \frac{1}{2}\right)^2 \xi_{n_1-1} +N^2/4}$ in \eqref{eq:D:12:c} is replaced with $\frac{N}{2} - 2 (\alpha - \frac{1}{2})^2$.

We may note that, taking the limit $N\rightarrow \infty$, $a \rightarrow \infty$, $\alpha \rightarrow \infty$, such that
$B_1 = \frac{N}{2 a^2}$, $\beta^2 = \frac{\alpha^2}{a^2}$ kept fixed and $\beta^{\prime 2} = \frac{\beta^2}{B_1}$,
yields the spectrum of $\slashed{D}^2 $ on ${\mathbb R^2}$ given in \eqref{fdirac} as expected.

In the absence of the abelian magnetic field, $\slashed{D}^2$ takes the form
\begin{align}
		\slashed{D}^2 = K^2 + 2 \left(\alpha - \frac{1}{2}\right)^2 + \chi \,,
		\label{D2alpha0}
\end{align}
and $\chi = 2\left(\alpha - \frac{1}{2}\right) \Big(\vec{K}\cdot \vec{\sigma}-\frac{1}{2}\Big) -  2 \left (\left (\alpha -\frac{1}{2} \right)^2 -\frac{1}{4} \right) (\vec{\sigma} \cdot \hat{r}) \, ( \vec{\tau} \cdot \hat{r})$, with the  eigenvalues $\pm 2 \left | \alpha - \frac{1}{2}\right | \left ( (k+\frac{1}{2})^2+ \left (\alpha -\frac{1}{2} \right)^2 -\frac{1}{4} \right)$ as easily seen from \eqref{eq:D:9} after setting $N=0$ in that expression. Total angular momentum $\vec{K}$ could carry the irreducible representations: $(l+1), l, l, (l-1)$ with $l=0,1,\cdots$ for the first two of the IRRs and  $l=1,2,\cdots$ for the remaining two.  Spectrum of $\slashed{D}^2$ becomes
\begin{subequations}
  \begin{align}
    \label{alpha0}
    \lambda_{l+1}(\alpha) =\, & \frac{1}{a^2} \left( l^2+3 l + 2 +  2 \left(\alpha - \frac{1}{2}\right)^2 - 2 \left | \alpha - \frac{1}{2}\right | \sqrt{(l+3/2)^2 + \left ( \left (\alpha -\frac{1}{2} \right)^2 -\frac{1}{4} \right)} \right ) \,, \\
    \lambda^\pm_{l}(\alpha) =\, & \frac{1}{a^2} \left(l^2+ l +  2 \left(\alpha - \frac{1}{2}\right)^2 \pm 2 \left | \alpha - \frac{1}{2}\right | \sqrt{(l+1/2)^2+\left ( \left (\alpha -\frac{1}{2} \right)^2 -\frac{1}{4} \right ) } \right ) \,, \\
    \lambda_{l-1}(\alpha) = \, & \frac{1}{a^2} \left(l^2- l +  2 \left(\alpha - \frac{1}{2}\right)^2 +2 \left | \alpha - \frac{1}{2}\right | \sqrt{(l-1/2)^2+\left ( \left (\alpha -\frac{1}{2} \right)^2 -\frac{1}{4} \right ) }\right ) \,,
  \end{align}
\end{subequations}
where $l=0,1,\cdots$ for $\lambda_{l+1}$ and $\lambda_l^+$ and $l=1,2,\cdots$ for $\lambda_l^-$ and $\lambda_{l-1}$. Quite interestingly, we notice that $\lambda^+_{l}\big|_{l=0} = \lambda_{l-1}\big|_{l=1} = 4 (\alpha - \frac{1}{2})^2$, from which we make the observation that these states produce zero modes at the special value $\alpha = \frac{1}{2}$ of the non-abelian gauge coupling. Taking $l \rightarrow l +1$ in $\lambda_{l-1}$ yields the same as $\lambda_l^+$ and $l \rightarrow l - 1$ in $\lambda_{l+1}$ yields the same as $\lambda_l^-$. Thus, we may write the spectrum in \eqref{alpha0} as
\be
\lambda^\pm_{l}(\alpha) =\, \frac{1}{a^2} \left(l^2+ l +  2 \left(\alpha - \frac{1}{2}\right)^2 \pm 2 \left | \alpha - \frac{1}{2}\right | \sqrt{(l+1/2)^2+\left ( \left (\alpha -\frac{1}{2} \right)^2 -\frac{1}{4} \right ) } \right ) \,,
\label{alpha0short}
\ee
with $l =0,1,2,\cdots$ for the upper and  $l = 1,2,\cdots$ for the lower sign and each eigenvalue occurring with multiplicity $2$. We see that $\lambda^\pm_{l}(\alpha) = \lambda^\pm_{l}(1-\alpha)$, i.e. the spectrum remains the same under $\alpha \leftrightarrow 1- \alpha$. Taking $\alpha \rightarrow 0$, we obtain from \eqref{alpha0} the eigenvalues $(l+1)^2$ with $l=0,1,\cdots$ and this matches with the spectrum of $\slashed{D}^2$ on $S^2$. These two facts uniquely fix the sign choices in front of the square root term in \eqref{alpha0short}.

Finally, we note that, with $l \rightarrow \infty$, $a \rightarrow \infty$, $\alpha \rightarrow \infty$,  such that $\frac{l}{a} \rightarrow k$, $\frac{\alpha^2}{a^2} \rightarrow \beta^2$ remaining finite, we obtain the spectrum on ${\mathbb R^2}$ given in \eqref{conspecDirac1}.

\end{document}